\documentclass[a4paper,longauth]{aa}
\usepackage[dvipsnames,usenames]{xcolor}
\usepackage{txfonts}
\usepackage{etoolbox}
\usepackage{graphicx}
\usepackage{amsmath}
\usepackage{hyperref}
\hypersetup{colorlinks=true,linkcolor=blue,citecolor=blue,filecolor=blue,urlcolor=blue}
\begin{document}
\title{The LOFAR Two-Metre Sky Survey (LoTSS)}
\titlerunning{Optical IDs for LoTSS DR2}
\subtitle{VI. Optical identifications for the second data release\thanks{The catalogues described in this paper are are only available
  in electronic form, via the LOFAR surveys project website at
  \url{https://lofar-surveys.org/dr2_release.html}, 
at the CDS via anonymous ftp to cdsarc.cds.unistra.fr (130.79.128.5),
or via \url{https://cdsarc.cds.unistra.fr/cgi-bin/qcat?J/A+A/}.}
}
\author{M.J.~Hardcastle\inst{1}\thanks{\email{m.j.hardcastle@herts.ac.uk}}
  \and M.A.~Horton\inst{1,2} \and W.L.~Williams\inst{3} \and
  K.J.~Duncan\inst{4}\and L.~Alegre\inst{4} \and B.~Barkus\inst{5} \and J.H.~Croston\inst{5} \and
  H.~Dickinson\inst{5} \and E.~Osinga\inst{6} \and
  H.J.A.~R\"ottgering\inst{6} \and
  J.~Sabater\inst{4} \and T.W.~Shimwell\inst{7} \and
  D.J.B.~Smith\inst{1} \and
  P.N.~Best\inst{4} \and A.~Botteon\inst{16} \and M.~Br\"uggen\inst{17} \and
  A.~Drabent\inst{10} \and F.~de~Gasperin\inst{16,17} \and
  G.~G\"urkan\inst{1,10,20} \and M.~Hajduk\inst{15} \and
  C.L.~Hale\inst{4} \and
  M.~Hoeft\inst{10} \and
  M.~Jamrozy\inst{8} \and M.~Kunert-Bajraszewska\inst{14} \and
  R.~Kondapally\inst{4} \and
  M.~Magliocchetti\inst{12} \and V.H.~Mahatma\inst{10} \and
  R.I.J.~Mostert\inst{6,7} \and
  S.P.~O'Sullivan\inst{21} \and
  U.~Pajdosz-\'Smierciak\inst{8} \and
  J.~Petley\inst{13} \and
  J.C.S.~Pierce\inst{1} \and I.~Prandoni\inst{16} \and D.J. Schwarz\inst{11} \and
  A.~Shulewski\inst{7} \and
  T.M. Siewert\inst{11} \and J.P.~Stott\inst{19} \and
  H.\ Tang\inst{22} \and
  M.\ Vaccari\inst{23,24,16} \and
  X.~Zheng\inst{6,18}\and
  T.~Bailey\inst{25} \and S.~Desbled\inst{25} \and A.~Goyal\inst{7} \and
  V.~Gonano\inst{25} \and M.~Hanset\inst{25} \and W.~Kurtz\inst{25}
  \and S.M.~Lim\inst{25} \and L. Mielle\inst{25} \and C.S.~Molloy\inst{25} \and
  R.~Roth\inst{25} \and I.A.~Terentev\inst{25} \and M.~Torres\inst{9} }
\authorrunning{M.J. Hardcastle et al.}
\institute{Centre for Astrophysics Research, University of Hertfordshire, College Lane, Hatfield
  AL10 9AB, UK
  \and
Cavendish Astrophysics, University of Cambridge, Cavendish Laboratory, JJ Thomson Avenue Cambridge CB3 0HE, UK
\and
SKA Observatory, Jodrell Bank, Lower Withington, Macclesfield, SK11 9FT, UK
\and
Institute for Astronomy, University of Edinburgh, Royal Observatory, Blackford Hill, Edinburgh, EH9 3HJ, UK
\and
School of Physical Sciences, The Open University, Walton Hall,
Milton Keynes, MK7 6AA, UK
\and
Leiden Observatory, Leiden University, PO Box 9513, NL-2300 RA
Leiden, the Netherlands
\and
ASTRON, the Netherlands Institute for Radio Astronomy, Postbus 2, 7990
AA Dwingeloo, the Netherlands
\and
Astronomical Observatory of the Jagiellonian University, ul. Orla 171, 30-244 Krakow, Poland
\and
Universidad Nacional Autónoma de México (UNAM), Avenida Insurgentes
Sur 3000, Mexico City, Mexico
\and
Th\"uringer Landessternwarte, Sternwarte 5, D-07778 Tautenburg, Germany
\and
Fakult\"at f\"ur Physik, Universit\"at Bielefeld, Postfach 100131, 33501 Bielefeld, Germany 
\and
INAF-IAPS, Via Fosso del Cavaliere 100, 00133, Rome, Italy
\and
Centre for Extragalactic Astronomy, Department of Physics, Durham
University, Durham, DH1 3LE, UK
\and
Institute of Astronomy, Faculty of Physics, Astronomy and Informatics,
NCU, Grudziadzka 5, 87-100 Toru\'n, Poland
\and
Space Radio-Diagnostics Research Centre, University of Warmia and
Mazury, ul.Oczapowskiego 2, 10-719 Olsztyn, Poland
\and
INAF - Istituto di Radioastronomia, via P. Gobetti 101, 40129,
Bologna, Italy
\and
Hamburger Sternwarte, Universit\"at Hamburg, Gojenbergsweg 112, 21029,
Hamburg, Germany
\and
Key Laboratory for Research in Galaxies and Cosmology, Shanghai
Astronomical Observatory, Chinese Academy of Sciences, 80 Nandan Road,
Shanghai 200030, China
\and
Department of Physics, Lancaster University, Lancaster, LA1 4YB, UK 
\and
CSIRO Space and Astronomy, ATNF, PO Box 1130, Bentley WA 6102,
Australia
\and
Departamento de Física de la Tierra y Astrofísica, Universidad Complutense de Madrid, 28040 Madrid, Spain
\and
Department of Astronomy, Tsinghua University, Beijing 100084, China
\and
Inter-University Institute for Data Intensive Astronomy, Department of
Astronomy, University of Cape Town, 7701 Rondebosch, Cape Town, South
Africa
\and
Inter-University Institute for Data Intensive Astronomy, Department of
Physics and Astronomy, University of the Western Cape, 7535 Bellville,
Cape Town, South Africa
\and
Citizen Scientist}
\abstract{The second data release of the LOFAR Two-Metre Sky Survey
  (LoTSS) covers 27\% of the northern sky, with a total area of $\sim 5,700$
  deg$^2$. The high angular resolution of LOFAR with Dutch baselines
  (6 arcsec) allows us to carry out optical identifications of a large
  fraction of the detected radio sources without further radio
  followup; however, the process is made more challenging by the many
  extended radio sources found in LOFAR images as a result of its
  excellent sensitivity to extended structure. In this paper we
  present source associations and identifications for sources in the
  second data release based on optical and near-infrared data, using a
  combination of a likelihood-ratio cross-match method developed for
  our first data release, our citizen science project Radio Galaxy
  Zoo: LOFAR, and new approaches to algorithmic optical identification,
  together with extensive visual inspection by astronomers. We also present
  spectroscopic or photometric redshifts for a large fraction of the
  optical identifications. In total 4,116,934 radio sources lie in the
  area with good optical data, of which 85\% have an optical or infrared
  identification and 58\% have a good redshift estimate. We
  demonstrate the quality of the dataset by comparing it with earlier
  optically identified radio surveys. This is by far the largest ever
  optically identified radio catalogue, and will permit robust
  statistical studies of star-forming and radio-loud active galaxies.}

\maketitle
\section{Introduction}
\label{sec:intro}

The LOFAR Two-Metre Sky Survey\footnote{See
\url{http://lofar-surveys.org/}} (LoTSS: \citealt{Shimwell+17}) aims
to survey the entire northern sky using the Low-Frequency Array
(LOFAR: \citealt{vanHaarlem+13}) at a central frequency of 144 MHz.
The survey, which already covers a significant amount of the
extragalactic northern sky, will provide an unrivalled resource for
wide-area low-frequency selection of extragalactic samples, both of
star-forming galaxies (hereafter SFG) and of radio-loud active
galactic nuclei (hereafter
RLAGN). In addition to the wide-field component, LoTSS has several
deep fields with published and publicly available images and catalogues, including the Lockman Hole, Bo\"otes \citep{Tasse+21}, and
ELAIS-N1 \citep{Sabater+21} fields. There is also a counterpart survey at
lower LOFAR frequencies, the LOFAR Low-Band Antenna Sky Survey (LoLSS:
\citealt{deGasperin+21}). Key to the science goals of the project is
accurate redshift information for the host galaxies of the radio
sources. This information will be provided in part by more than one
million optical spectra that will be obtained using the William
Herschel Telescope Enhanced Area Velocity Explorer (WEAVE) instrument
\citep{Jin+23} as part of the WEAVE-LOFAR project \citep{Smith+16}, by
the Sloan Digital Sky Survey \citep{Blanton+17} and other ongoing and future large-scale
spectroscopic campaigns such as the Dark Energy Spectroscopic
Instrument (DESI: \citealt{Levi+13}) or the {\it Euclid} Wide Survey \citep{Euclid+20}, and for the remaining LOFAR sources by
state-of-the-art photometric redshifts already in hand \citep{Duncan+21}.

\begin{figure*}
  \includegraphics[width=1.0\linewidth]{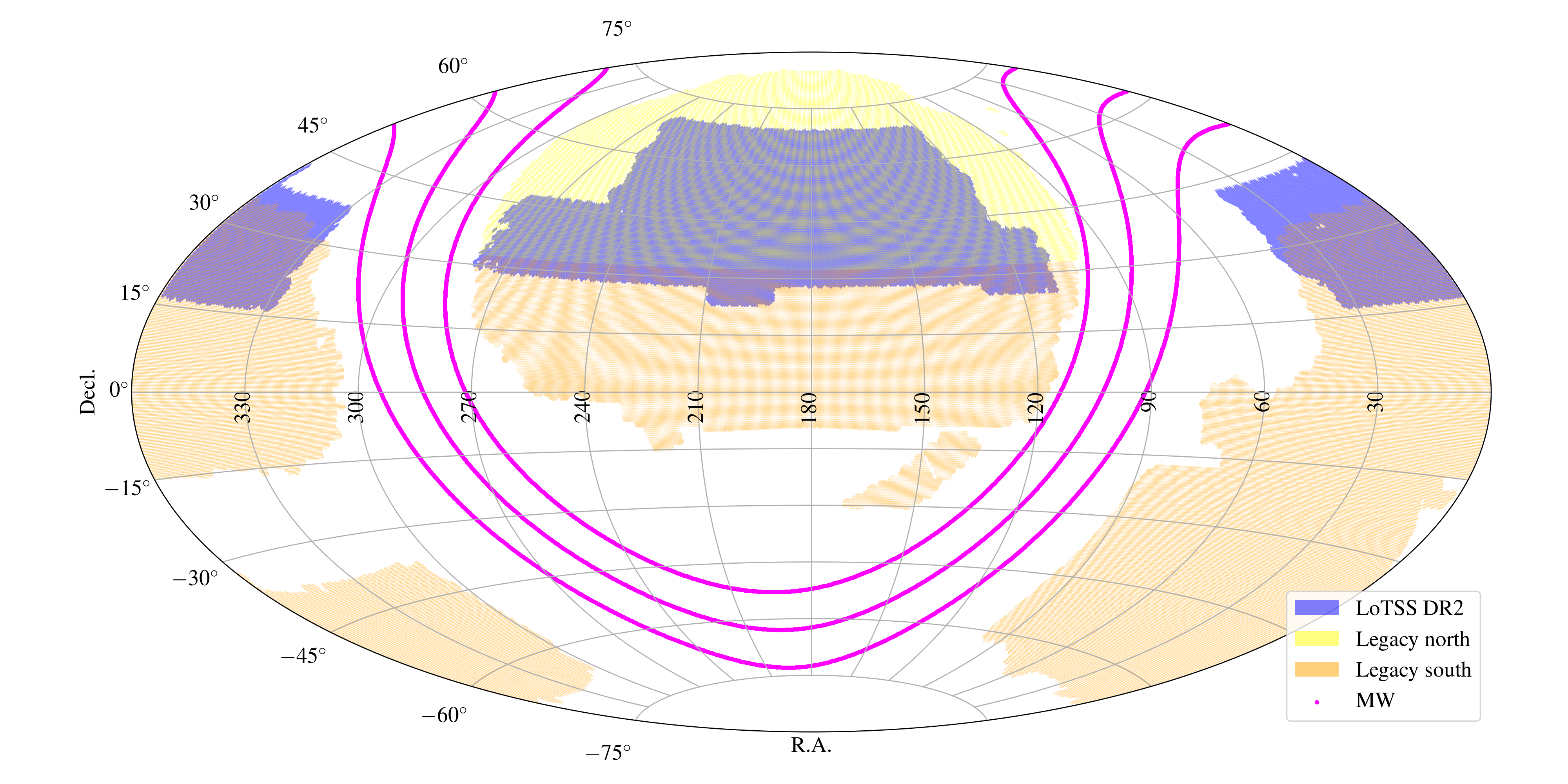}
  \caption{Sky coverage of the LoTSS DR2 (blue) and the Legacy DR9 (yellow and
    orange) optical surveys.
    The purple lines (`MW') show the Galactic plane and lines of
    $|b|=10^\circ$. As described in the text, `Legacy North' data is made up of BASS and MzLS
    data, `Legacy South' data are from DECaLS. }
  \label{fig:coverage}
\end{figure*}

In order to exploit the full potential of deep extragalactic radio
surveys, we need optical identifications, and the photometric and/or
spectroscopic redshifts that they make possible. Spectroscopic
followup projects such as WEAVE-LOFAR also rely, where possible, on
accurate optical positions of target sources. Historically, radio
continuum surveys have produced catalogues of radio sources for others
to follow up with further radio or optical observations: for example,
the highly influential revised Third Cambridge Revised (3CR) sample of the brightest
extragalactic low-frequency radio sources in the northern sky (3CRR:
\citealt{Laing+83}), itself based on radio data taken in the 1960s
\citep{Bennett62,Gower+67}, only received its final optical
identification in 1996 \citep{Rawlings+96}. The radio survey that was
the largest in terms of numbers of sources detected until very
recently, the NRAO Very Large Array (VLA) Sky Survey (NVSS:
\citealt{Condon+98}), which covers the whole sky above declination
$-40^\circ$, has never had anything approaching a full optical
identification catalogue, partly because of the lack of any
appropriate counterpart optical catalogue but also because its low
resolution (45 arcsec) precludes reliable matching of the radio
sources with deep optical data. Higher-resolution large-area surveys, such
as Faint Images of the Radio Sky at Twenty-Centimeters (FIRST:
\citealt{Becker+95}) are more easily matched to optical data, but
high-resolution surveys with the VLA are insensitive to large-scale
structure due to a lack of short interferometric
baselines\footnote{In addition to this problem, wide-area
  high-resolution surveys
  with the VLA, such as FIRST, are also necessarily strongly
  surface-brightness limited because of the small VLA field of view,
  which means that short observations are required in order to cover wide areas.
  In the case of some low-surface brightness structures, such as
  moderately resolved star-forming galaxies, it is this surface
  brightness limit that prevents FIRST from
  seeing all of their emission rather than missing short baselines;
  here a VLA survey with a larger beam, such as NVSS, can perform much better \citep{Condon+02}.}, and so
obtaining a catalogue that is both optically identified and
flux-complete in the radio has historically involved labour-intensive
combination of multiple radio catalogues with the optical data
\citep[e.g.][]{Gendre+Wall08,Best+Heckman12}. While the VLA Sky Survey
(VLASS: \citealt{Lacy+20}), now in progress, will have excellent
angular resolution and improved image fidelity compared to FIRST, it
will still be insensitive to structures on scales larger than 30
arcsec.

A major complication of the process of optical identification of radio
sources is due to the fact that radio structures, if properly imaged,
can be physically large, with complex, resolved structure extending to
much larger scales than those of the host galaxy observed in the
optical. In extreme (but far from uncommon) cases, the catalogued
positions of the two lobes of a double RLAGN may both lie arcminutes
away from the true optical host and from each other
\citep[e.g.][]{Oei+23}. In situations like this two operations are
required --- the radio components must be `associated', that is
they must be recognised as a single physical source, and the source
must be `identified', that is an optical counterpart must be found.
In general it is easier to do these two operations together and, at
present, visual inspection remains the best way of doing so --- a
human being with a small amount of training can efficiently pick out
radio structures that look like an extended radio galaxy and
simultaneously select the best optical counterpart for the candidate
radio source. For the very large surveys being generated by the
current generation of radio telescopes, though, visual inspection is
extremely expensive in terms of time. \cite{Banfield+15} describe
`Radio Galaxy Zoo', the first citizen-science project to aim
specifically at providing associations and optical identifications for
extended radio sources. Radio Galaxy Zoo involved the inspection by
citizen scientists of $\sim 100,000$ radio sources, mostly from FIRST,
and obtained infrared (IR) IDs from the {\it Wide-field Infrared
  Survey Explorer} ({\it WISE}) catalogue for a large
fraction of them (56\% in Data Release (DR) 1: \citealt{Wong+23}), demonstrating the
applicability of citizen science methods to such very large datasets.

The LoTSS surveys, because of the wide range of baselines provided by
even the Dutch subset of LOFAR antennas, have the capability to detect
extended emission on scales up to $\sim 1^\circ$ while also having
resolution good enough (6 arcsec) for unambiguous identification of a
large fraction of the detected radio sources and sensitivity nearly an
order of magnitude higher than FIRST for sources of typical radio
spectra, $\alpha \sim 0.7$. It has always been the goal of the LoTSS
project not only to produce the surveys, but also to provide the
ancillary data needed for their scientific exploitation. In the first
LoTSS data release, DR1, which covered 424 deg$^2$ in a region of the
Northern sky matched to the coverage of the Hobby-Eberly Telescope
Dark Energy Experiment (HETDEX: \citealt{Gebhardt+21}), we were able
to generate an optically identified catalogue \citep{Williams+19} by
combining the LoTSS data with Panoramic Survey Telescope and Rapid
Response System (PanSTARRS) DR1 \citep{Chambers+16} and
All{\it WISE} data \citep{Wright+10,Mainzer+11}, a process that
generated a value-added catalogue of 318,520 radio sources, with
plausible optical and/or IR counterparts for 73\% of them. We developed an
algorithm for deciding whether a particular radio source needed visual
inspection for association and identification, described in detail by
\cite{Williams+19}. When required, we used a private Zooniverse
project, `LOFAR Galaxy Zoo', (hereafter LGZ) based on the approach of
Radio Galaxy Zoo (RGZ), as a platform for distributing and collating
the effort of inspection. This visual classification was largely done
by members of the Surveys Key Science Project. The resulting optical
identifications enabled a range of science including the study of
RLAGN \citep{Sabater+19,Hardcastle+19,Mingo+19}, their environments
\citep{Croston+19} and their host galaxies \citep{Zheng+20}, giant
radio galaxies \citep{Dabhade+20}, quasars
\citep{Gurkan+19,Morabito+19,Rankine+21}, star-forming galaxies \citep{Wang+19},
and the search for extra-terrestrial intelligence
\citep{Chen+Garrett21}. The process that we developed for DR1 was
adapted to provide the optical identifications for the first release
of the LoTSS deep fields \citep{Kondapally+21}, where an
identification rate close to 100\% was achieved thanks to the
excellent optical data available in those fields.

The second wide-area data release, DR2, of LoTSS \citep{Shimwell+22}
covers 27\% of the northern sky, but specifically targets areas at
high Galactic latitude with good optical coverage for extragalactic
sources. It has a total sky coverage of 5,700 deg$^2$, provided by 841
LOFAR pointings, and is split between two regions: the RA-13
(`Spring') region centred at approximately 12h45m00s $+44^\circ
30'00''$ and the RA-1 (`Fall') region centred at 1h00m00s $+28^\circ
00'00''$. The DR2 sky coverage (Fig. \ref{fig:coverage}) reflects the
contiguous sky area that the survey had built up at the start of the
DR2 processing run in 2019, but excludes both the Galactic plane and
also low-declination regions where the sensitivity of LOFAR is reduced
due to geometrical effects; in total DR2 covers 46\% of the
extragalactic Northern sky with $|b|>10^\circ$ and $\delta >
15^\circ$. DR2 contains 4.4 million catalogued sources, the largest
radio source catalogue released so far, and so the required effort for
optical identification and source association was over an order of
magnitude larger than for DR1. We took an early decision to involve
citizen scientists in the optical identifications for DR2 through a
successor project to Radio Galaxy Zoo, which we named Radio Galaxy
Zoo: LOFAR. For the remainder of this paper, this public project is
referred to as RGZ(L) to make clear the distinction between it, the
original Radio Galaxy Zoo (RGZ), and our previous internal platform,
LGZ.


In this paper we describe the process of deriving optical
identifications for LoTSS DR2 targets. Section \ref{sec:optdata}
describes the datasets that we use for the optical counterpart
catalogue and Section \ref{sec:ml} describes the approach to
likelihood-ratio cross-matching that we adopt for these datasets.
Section \ref{sec:algorithm} describes the choices made to decide
whether likelihood-ratio matches should be used for a given source or
whether visual inspection is needed for optical identification and/or
association. Section \ref{sec:zooniverse} describes our public
Zooniverse project, `Radio Galaxy Zoo: LOFAR' and its outputs. We
discuss the post-processing of the Zooniverse and likelihood-ratio
identifications and associations in Section \ref{sec:postproc}, source
angular sizes are discussed in Section \ref{sec:angsize}, and our methods for
estimating photometric redshifts, galaxy masses and other physical
quantities are briefly
summarized in Section \ref{sec:photoz}. The final catalogue is
described in Section \ref{sec:catdesc}. We discuss some properties of
the sources in the resulting catalogue in Section \ref{sec:properties}
and summarize our results in Section \ref{sec:summary}.

Throughout this paper we use a cosmology in which $H_0 = 70$ km
s$^{-1}$ Mpc$^{-1}$, $\Omega_{\rm m} = 0.3$, and $\Omega_\Lambda =
0.7$. Radio flux density is quoted in Jy: 1 Jy is $10^{-26}$ W
Hz$^{-1}$ m$^{-2}$. The radio spectral index $\alpha$ is defined in
the sense $S_\nu \propto \nu^{-\alpha}$. Optical and IR
magnitudes used are in the AB system unless stated otherwise. Code
used for the operations described in this paper is available for
download and modification online\footnote{See
\url{https://github.com/mhardcastle/lotss-catalogue/}.}.

\section{The input data}
\label{sec:optdata}

For radio data, our starting point is the DR2 images and combined
catalogue described by \cite{Shimwell+22}. The images used are the mosaiced
images described in that paper, which have the greatest depth at any
position in DR2. The catalogue is a radio catalogue generated by
combining runs of the Python Blob Detector and Source Finder ({\sc pybdsf}: \citealt{Mohan+Rafferty15}) over all the
mosaics, and so is the result of decomposing the image of the
sky into many Gaussian components. For our purposes the key elements
of the catalogue are, for each source: position, total flux density,
major and minor full width at half-maximum (FWHM) and position angle of the fitted Gaussian,
and the deconvolved versions of the last three quantities (i.e. after
correcting for the 6-arcsec restoring beam). For the cataloguing parameters that we use, {\sc pybdsf} can sometimes combine the originally detected
Gaussians into composite sources, and so for some purposes
(discussed further below) we use the original Gaussian catalogue as
well as the DR2 source catalogue. Since the latter is the starting point for our later efforts to associate components together into sources, we refer to it as the component
catalogue in what follows.

Optical data for the identification effort are provided by the DESI
Legacy Imaging Surveys, hereafter the Legacy
Survey\footnote{\url{https://legacysurvey.org/}} \citep{Dey+19}. This
combines three optical surveys of the sky away from the Galactic
plane: the Dark Energy Camera Legacy Survey (DECaLS), covering mostly southern
declinations, and the Beijing-Arizona Sky Survey (BASS) and Mayall
$z$-band Legacy Survey (M$z$LS), covering the northern sky. The
coverage of the Legacy survey is shown in relation to LoTSS DR2 in
Fig.\ \ref{fig:coverage}. As can be seen in that figure, the bulk of
our sky coverage in the RA-13 region is from BASS and M$z$LS, which
reach typical point-source depths of 24.3, 23.7, and 23.3 mag in the $g$,
$r$ and $z$ bands respectively. The coverage available in the RA-1
region, and a small amount to the south of the RA-13 region, is from
the deeper DeCALS which reaches mean depths of 24.8, 24.2, and 23.3 mag
in the northern sky, with the extinction-corrected depth being more or
less constant over the areas of interest to LOFAR. Even the northern
parts of the survey are 1.0 mag deeper in $g$ and $z$, and 0.5 mag
deeper in $r$, than PanSTARRS DR1, which provided the optical data for
our DR1 optical cross-matching effort.

As can be seen in Fig.\ \ref{fig:coverage}, there is an area of DR2 to the
north of the RA-1 field that does not have Legacy Survey coverage,
amounting to 48 LOFAR pointings or a little over 300 deg$^2$ of our
area. For simplicity this area is omitted from our analysis and from
the value-added catalogues, which reduces the number of radio sources
that can be optically identified to $\sim 4.1$ million. For our
likelihood-ratio cross-matching, as discussed below, we combined the
Legacy DR9 `sweep' catalogues, joining North and South at a
declination of 32.375$^\circ$. To obtain FITS images for visual
inspection (Sections \ref{sec:zooniverse} and \ref{sec:postproc}) we
used the publicly available survey web-based APIs to download {\it
  WISE} band 1 and $grz$ Legacy image cubes. Around 2,600 $4096 \times
4096$ {\it WISE} images and 295,000 $1000\times1000\times 3$ Legacy
cubes, totalling $\sim 3$ TB, were downloaded.

\section{Likelihood-ratio cross-matching}
\label{sec:ml}

We cross-matched radio sources to their optical and/or IR counterparts 
using a likelihood-ratio (LR) method \citep{Sutherland+Saunders92}.
First, we cross-matched the Legacy Survey data with the un{\it WISE} data \citep{Schlafly+19} to
create a combined optical and IR catalogue. We used a simple nearest
neighbour match limited to a maximum radius of 2.0 arcsec to match 
optical to IR sources. This value for the radius was empirically 
found to be optimal to provide actual matches. Unmatched sources were 
added to the final combined catalogue without corresponding {\it WISE}
or Legacy photometry. The combined optical and IR catalogue was then
cross-matched to the 
LoTSS DR2 radio sources using the LR method presented by 
\citet{Williams+19}, which uses both optical magnitude and colour as
an input. This LR method is a statistical technique to 
match counterparts of the same source observed at different wavelengths.
We considered ten colour ($r$-band to un{\it WISE} W1) bins plus two bins for 
objects with only un{\it WISE} data: one for objects with W1 and W2 
magnitudes, and one for objects with only W2 magnitudes.

The cross-match was done separately for three different regions: a) the
RA-1 (`Fall') region which is covered by the Legacy South survey; b)
the RA-13 (`Spring') region covered by the Legacy South survey; and,
c) the RA-13 (`Spring') region covered by the Legacy North survey. We
did this to take into account the different locations on the sky and
the possible differences in the optical survey properties. Within each
of these regions we computed the $Q_0$ values (where, as described by
\cite{Williams+19}, $Q_0$ represents the fraction of sources that have
an optical counterpart down to the magnitude limit of the survey) in
different areas where the optical and IR coverage was complete. The
values of $Q_0$ for those different areas within a region were similar
within the errors. This suggests that the range of declinations did
not generate any significant biases for the LR method. The LR cutoff
thresholds for the different regions are slightly different for the
different regions, as expected. As a result of the LR matching, every
source either had a best-match LR candidate ID, or no potential
counterpart above the LR threshold.

\section{The decision tree}
\label{sec:algorithm}

The decision tree used for selecting which radio sources to accept
their statistical LR identification (or lack thereof, see
Section\ \ref{sec:ml}) and which sources to further process visually
through the public RGZ(L) Zooniverse project (described in
Section\ \ref{sec:zooniverse}) was very similar to that used by
\cite{Williams+19} for LoTSS DR1. This decision tree aims to identify
{\sc pybdsf} sources that are components of physical radio
sources and that therefore need to be associated before the
optical and IR cross-identification is made, together with other
sources that are not suitable targets for the LR method. Here we give only
a brief summary and highlight any changes to the process used for
DR1. Fig.~\ref{fig:cat:flowchart} shows the modified decision
tree used in this work, along with the numbers and fractions of
sources at each outcome. Key parameters used for the decisions
are defined in Table\ \ref{tab:cat:flowkey}. A separate decision
process is followed within the decision tree for {\sc pybdsf}
sources that are composed of multiple Gaussians. The decision tree used
for this was essentially identical to that used for DR1 and is
described by \cite{Williams+19}.

\begin{figure*}
  \includegraphics[width=0.9\linewidth]{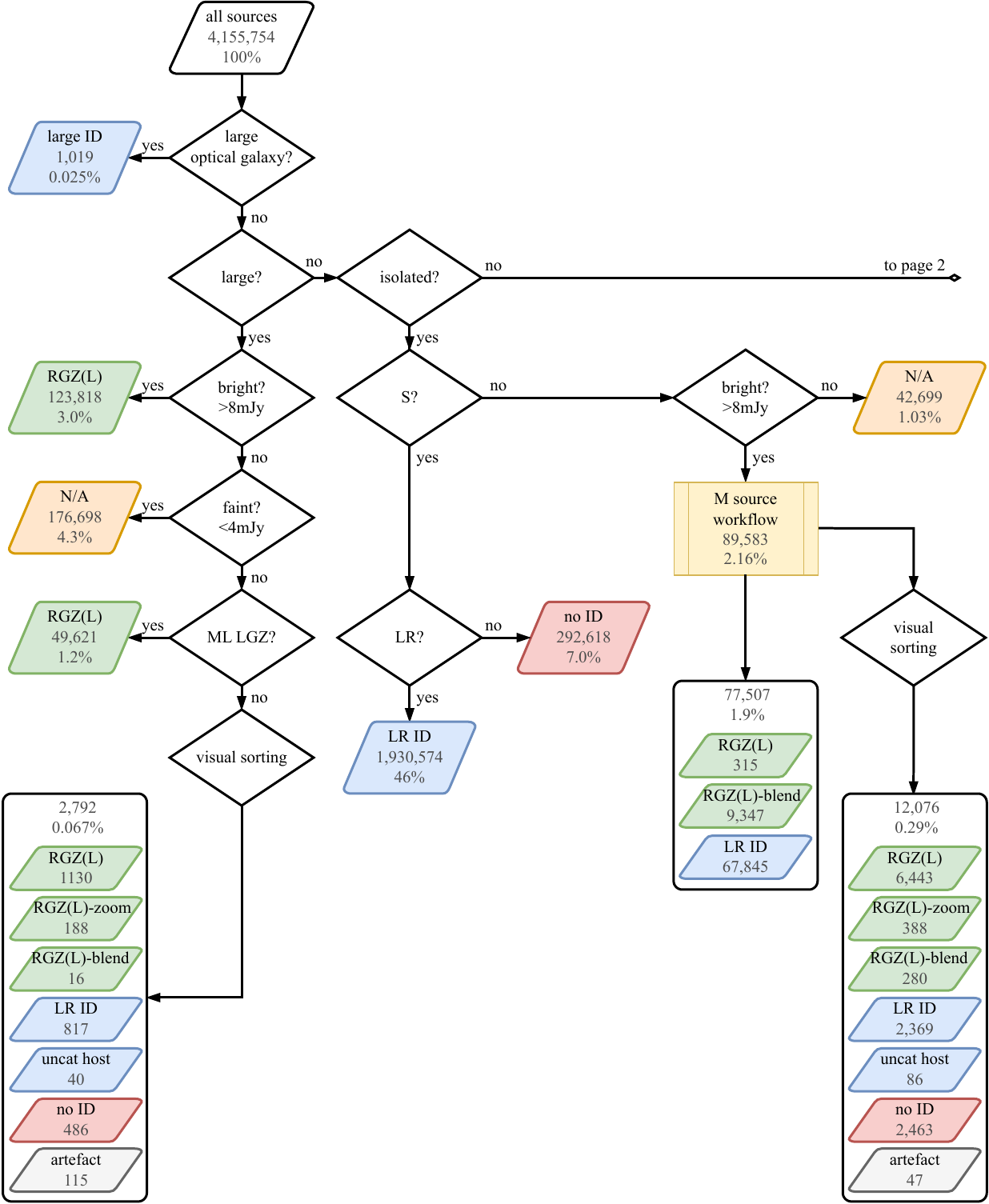}
\caption{Representation of the decision tree used to process all entries in 
the {\sc pybdsf} catalogue lying in the Legacy Survey sky area. Following this workflow a decision is made for each 
source whether to: (i) make the optical and IR identification, or lack thereof, 
through the LR method (blue and red outcomes respectively); (ii) process the 
source in RGZ(L)  (green outcomes, including direct RGZ(L) post-processing); (iii) reject the source as an artefact (grey outcomes). The key parameters are defined in 
Table~\ref{tab:cat:flowkey}. The number and percentage of {\sc pybdsf}
sources in each final bin are shown for each final outcome. Some faint
sources are not processed further (orange outcomes); these are
discussed in the text.}
\label{fig:cat:flowchart}
\end{figure*}
\addtocounter{figure}{-1}
\begin{figure*}
  \includegraphics[width=0.9\linewidth]{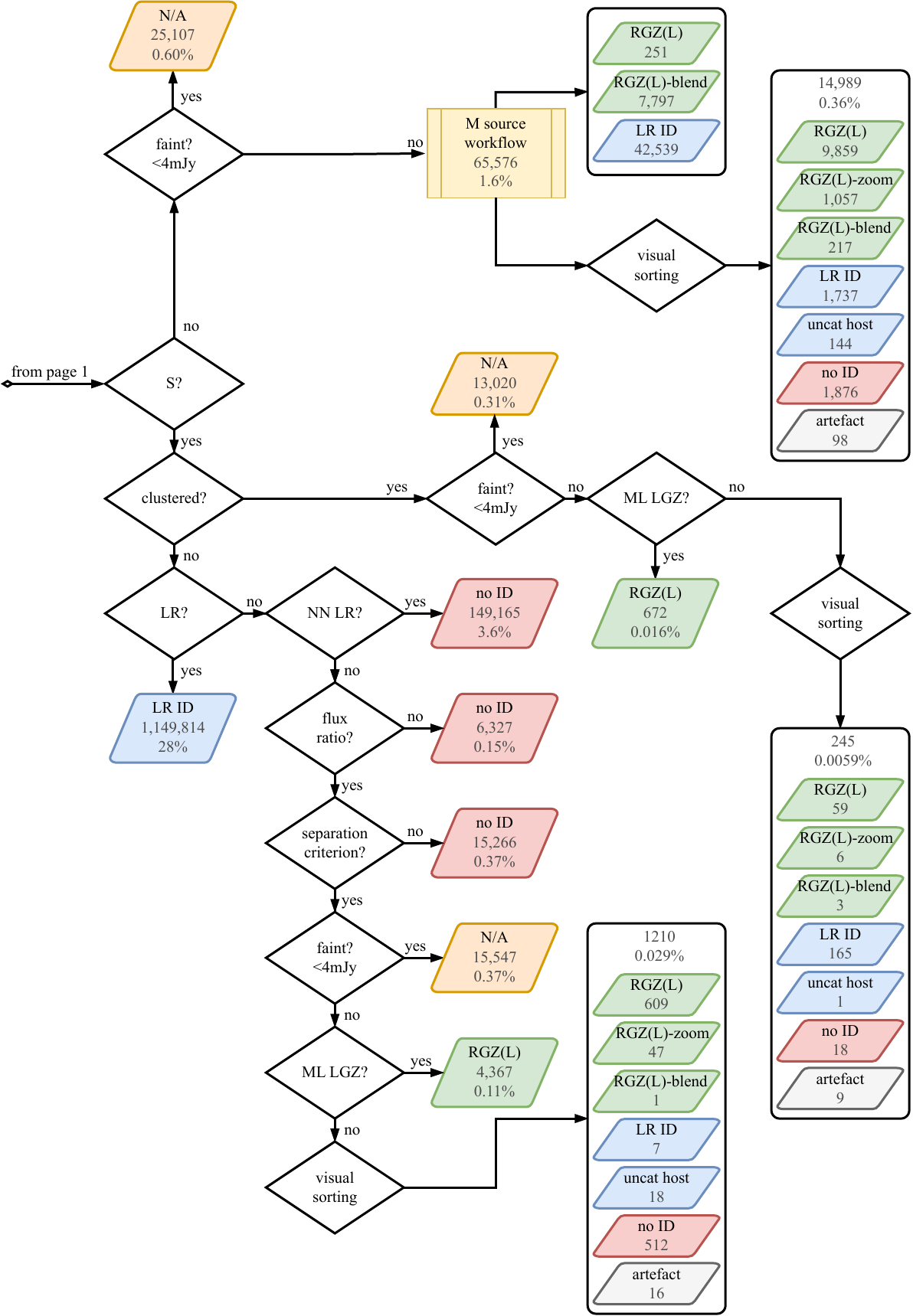}
\caption{continued} 
\end{figure*}

\begin{table*}[htp]
 \centering
 \caption{Definition of the parameters used in the main decision tree in 
Fig.~\ref{fig:cat:flowchart}.}
 \small
 \label{tab:cat:flowkey}
\begin{tabular}{ll}
\hline
\hline
Parameter & definition \\
\hline
Large optical galaxy & 2MASX size ($r_{\mathrm{ext}}$) $\geq 60${\arcsec} \\
Large & {\sc pybdsf} major axis $>15${\arcsec} \\
Bright & total flux density $>8$ mJy \\
Isolated &  distance to nearest {\sc pybdsf} neighbour (NN) $>45${\arcsec} \\
S & single Gaussian component within an island \\
LR &  $LR > LR_{\mathrm{thresh}}$\\
ML & machine-learning classification \\
Clustered & distance to fourth nearest {\sc pybdsf} neighbour $<45${\arcsec} \\
NN LR & $LR_{\mathrm{NN}} > LR_{\mathrm{thresh}}$\\
Flux ratio & $S/S_{\mathrm{NN}} < 10 $\\
Separation criterion & $S + S_{\mathrm{NN}} \leq 50 (d_{\mathrm{NN}} / 
100{\arcsec})^2$ mJy\\
\hline
\end{tabular}
\end{table*}

The input parameters to the decision tree are the {\sc pybdsf} source
size (taken to be the major axis), source flux density, and number of
fitted Gaussian components, as well as the calculated distances to the
nearest neighbour (NN) and to the fourth closest neighbour (NN4).
Further inputs are the likelihood ratios for sources smaller than
$30$ arcsec as well as for individual Gaussian components smaller than
$30$ arcsec. The outcomes of the decision tree are labels for each
{\sc pybdsf} source which determine how it should be treated subsequently. Some of these are derived directly from the
source properties, but, as for DR1, some outputs of the decision tree required `visual sorting' or filtering done by a small number of experienced people.
This rapid process, performed using a simple {\sc Python} interface to
view the RGZ(L) images and categorise the sources, was done to
avoid overpopulating the RGZ(L) sample with sources that would not
benefit from citizen science inspection.

A key difference with the DR1 flowchart was that we did not attempt to
include faint sources, below a total flux density of 4 mJy, in the
list of objects sent to RGZ(L) for visual sorting. The reason for this
was twofold: firstly, experience from DR1 shows that these faint
objects are often extremely difficult to associate and identify,
especially for large sources; secondly, these sources are very
numerous and would overwhelm the capacity of the Zooniverse project.
The level of the limit was selected because we were aiming to produce an almost
complete sample of physical radio sources for the WEAVE-LOFAR project,
which will target all LoTSS sources brighter than 8 mJy for
spectroscopic followup. In almost all cases we used a limit of 4 mJy,
as these {\sc pybdsf} sources might be components of an 8-mJy physical
source and need to be associated, thereby ensuring greater
completeness for the WEAVE 8-mJy flux-density selection criterion.
Only in the branch of the decision tree addressing small, isolated,
multiple-Gaussian component sources did we use a different limit of 8
mJy since, given their isolation, these sources are unlikely to be
components of another source. Within this category of faint sources,
all except the largest sources ($>15$ arcsec) will have LR
determinations available, and the identification (or lack thereof)
from these has been adopted for the catalogue; these can be used with
the caveat that they may be wrong if the source is actually a
component of a larger physical radio source. However,
\cite{Williams+19} showed that not many sources in this flux range
benefited from visual inspection.

A second key change to the decision tree from DR1 was the inclusion of
the machine-learning (ML) classifications developed by
\cite{Alegre+22}. This gradient-booster classifier, whose features are
similar to the parameters used in the decision tree here, was trained
using the final outcomes from the DR1 processing, that is, whether a {\sc
  pybdsf} source needed to be associated or deblended or had a
different identification to that provided by LR, and therefore needed
to be processed with LGZ, and used to predict the same for the DR2
{\sc pybdsf} sources. While these ML classifications were not used to
fully replace the decision tree, they were used to reduce the number
of sources requiring visual sorting in several branches of the
decision tree. Firstly, for large ($>15$ arcsec) and intermediate flux
density sources ($4<S<8$ mJy), instead of visually sorting all
sources, we used the ML classifications to select most (95 per cent)
for direct processing in RGZ(L), while only the remaining 5 per cent
were visually sorted. Roughly half of the latter category were
selected for RGZ(L) after the visual inspection process. Secondly, the
ML classifications were also used for clustered sources. Faint sources
($<4$ mJy) were not processed, while the brighter sources with ML
RGZ(L) classifications were processed directly in RGZ(L) and the
remainder through visual sorting. Finally, the non-isolated sources
without LR identifications that did not meet either the flux density
or separation criteria to identify possible double sources were
selected either for RGZ(L) or visual sorting based on the ML
classification after excluding the faintest ($<4$ mJy) sources. We are
confident that this ML approach did not prevent unusual sources from
being inspected through the RGZ(L) platform, as (a) the training data
from DR1 are very well matched to the type of data used in DR2 and (b)
the training set size from DR1 was close to 10\% of the total size of
DR2, meaning that all source types seen in DR2 are likely to be well
represented in the training set.

The final outputs of the decision tree, combining algorithmic,
machine-learning, and visual inspection outcomes, are flags indicating
which of several post-processing steps are required. These outcomes
are summarized in Table \ref{tab:id_flags_final} along with the number
of {\sc pybdsf} sources within each category. Similar to the approach
of \citet{Williams+19}, the visual sorting used in several branches of
the decision tree identifies some sources directly for the post
processing which is normally applied to sources that have passed
through the RGZ(L) project, either through the deblending or
too-zoomed-in workflows (described in Section\ \ref{sec:zooniverse}).
 
\begin{table*}
  \centering
  \caption{Summary of the decision tree outcomes}
  \label{tab:id_flags_final}
  \begin{tabular}{lll}
    \hline
    ID\_Flag & Meaning & Number\\
    \hline
    0 & No identification after prefilter & 5,355\\
    1 & LR (including no counterpart above threshold) & 3,659,243 \\
    2 & Large optical galaxy & 1,019 \\
    3 & Send to RGZ(L) & 197,144\\
    4 & Artefact after prefilter & 285 \\
    5 & N/A (full identification not attempted) & 273,071 \\
    6 & Send to deblend workflow & 17,682 \\
    7 & Send to too zoomed in workflow after prefilter & 1,686\\
    8 & Uncatalogued host after prefilter & 268\\
    \hline
    &Total&4,155,468\\
    \hline
  \end{tabular}
  \end{table*}

\section{Zooniverse visual inspection}
\label{sec:zooniverse}

\begin{figure*}
  \includegraphics[width=0.33\linewidth]{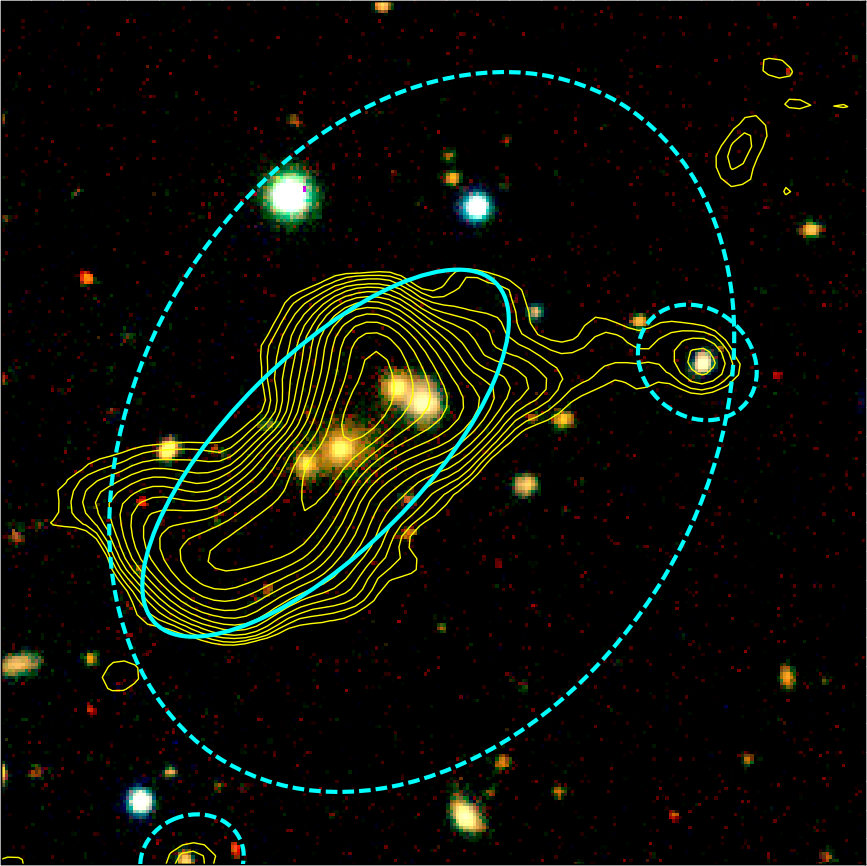}
  \includegraphics[width=0.33\linewidth]{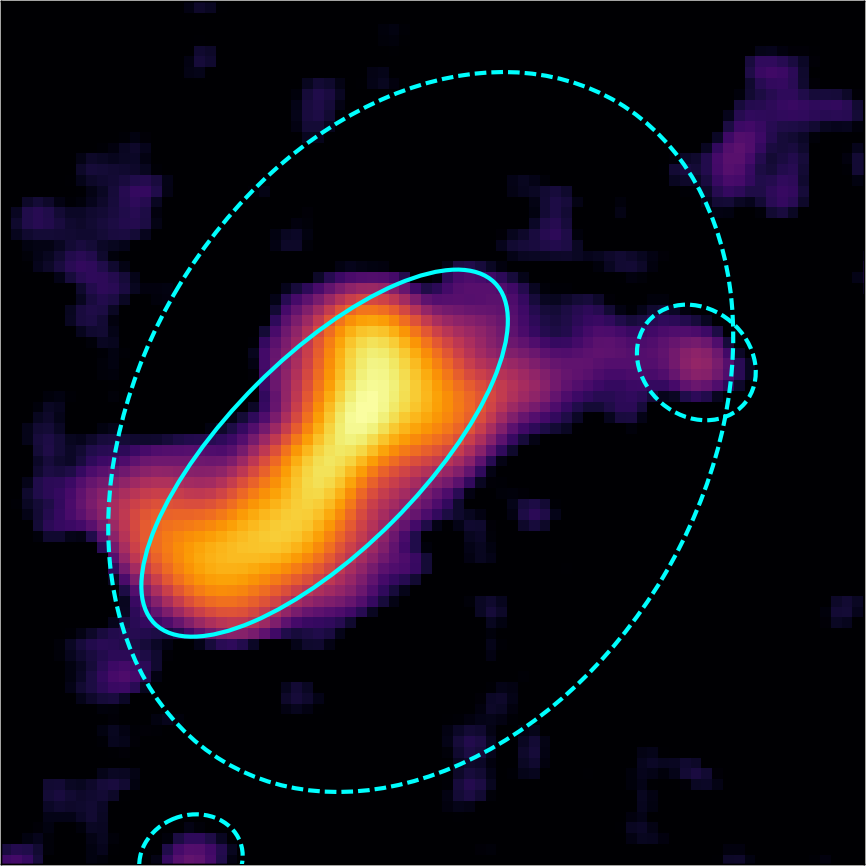}
  \includegraphics[width=0.33\linewidth]{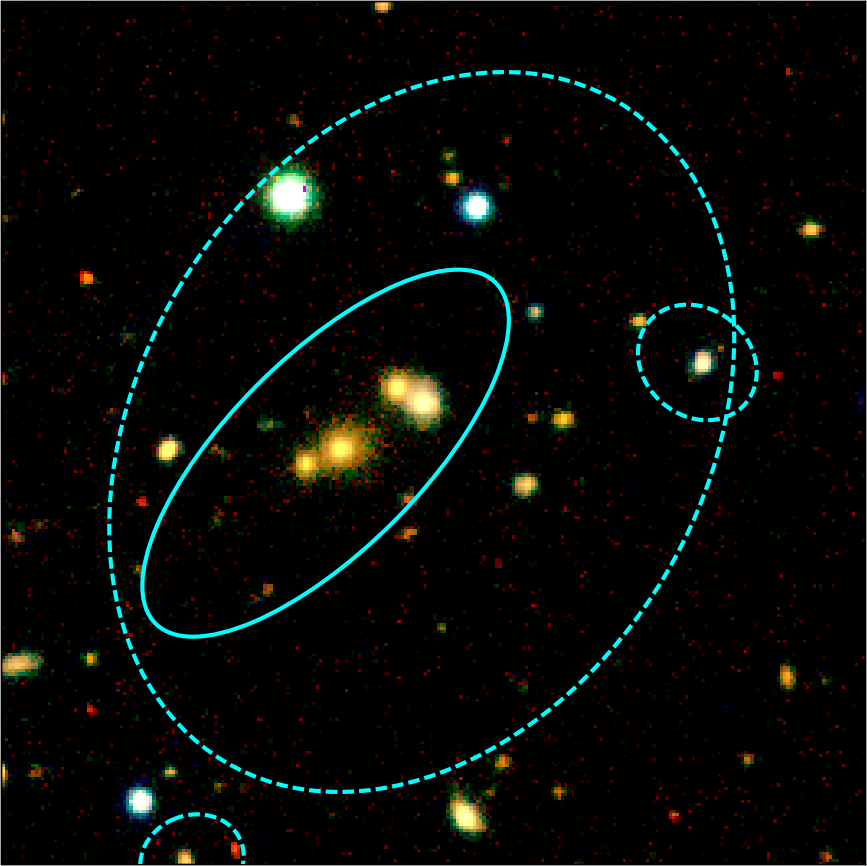}
\caption{Example of the three images presented to citizen scientists
  for one catalogued LOFAR radio source (ILTJ093236.46+602825.5). Left
  panel, the default view: radio contours from the LOFAR data
  (logarithmically increasing by a factor two at each interval from five
  times the local noise level) are superposed on the Legacy
  three-colour image. Cyan ellipses denote catalogued radio sources,
  with sizing as described in the text; the solid ellipse is the one
  under study and dotted ellipses represent other sources in the radio
  catalogue. Middle panel: the colour scale shows the LOFAR radio data
  only. Right panel: a view of the optical sky only. This image is 2
  arcmin on a side.}
\label{fig:zooniverse-images}
\end{figure*}
 
Almost all of the objects selected above as requiring visual
inspection were sent to citizen scientists\footnote{A small number of
sources, just over 4,000 in total, were classified through a test
version of the same interface by members of the collaboration before
the launch of the public project. These classifications are merged in
with the citizen science classifications in the final analysis.}
participating in the RGZ(L) project through the Zooniverse web
interface\footnote{\url{http://lofargalaxyzoo.nl/}}. The basic process
for generating these images was very similar to that described by
\cite{Williams+19}, with radio and optical images again being
generated using {\sc aplpy}, but was modified to present citizen
scientists with a simpler and more attractive view of the targets.
Fig.\ \ref{fig:zooniverse-images} shows an example of the three views
provided to Zooniverse volunteers for one randomly chosen LOFAR source
from the `large, bright' category, where the user can flip between all
three views at any time. The main differences in this interface
compared to the LGZ interface used for DR1 was the inclusion of a
multi-colour optical image, a colourmap version of the radio image (to
enhance accessibility), and the exclusion of the {\it WISE} image. The
latter choice was made to simplify the interface at the cost of losing
a small number of distant RLAGN which are easy to spot in
near-IR, since many of those sources were recoverable using the steps
described below.

The field of view presented to the user for each catalogued radio
source was chosen algorithmically with the aim of maximizing the
probability of seeing all of a large, multi-component source.
Initially the field of view was taken to encompass all of the target
source itself, where a catalogued component from the DR2 catalogue
with deconvolved FWHM values $\theta_{\rm maj}$ and $\theta_{\rm min}$
is represented by an ellipse with semi-major axis $\theta_{\rm maj}$
and semi-minor axis $\theta_{\rm min}$. It was then extended
iteratively to cover any other overlapping elliptical components; this
helps to ensure that complex contiguous sources, where possible, are
represented in the image sent to Zooniverse. Next, nearby resolved
neighbour objects from the component catalogue with total flux density
similar to (no more than a factor three less than) the target source and
an offset of no more than 3 arcmin from the field centre were
iteratively added to the field of view -- once a nearest neighbour was
added, the mean positional centroid of all the sources selected so far
was calculated and the process repeated until convergence. This
approach was intended to pick up, for example, lobes of a double
source that might have similar total flux density but did not appear
to overlap on the sky. Finally, the centroid and bounding box of the
resulting set of components were computed. If the bounding box was
larger than 5 arcmin, then only the size of the original component was
used. This prevented very large fields being sent for inspection, as
those would present the user with too large a field of view to
reliably select components and optical counterparts. A minimum field
of view of 1 arcmin (ten times the FWHM of the LOFAR restoring beam)
was also imposed to ensure that at least some neighbouring sources and
galaxies would be visible. Finally, the field of view used was rounded
to the nearest 10 arcsec (this allows for simple formatting of the
number when the data are uploaded to Zooniverse in ASCII format) and
the three images were generated. As illustrated in
Fig.\ \ref{fig:zooniverse-images}, ellipses mark the positions of all
catalogued radio sources in the field of view, with a solid ellipse
indicating the `current' source and dashed ellipses indicating others
that might potentially be associated with it.

\begin{figure*}
  \includegraphics[width=0.50\linewidth]{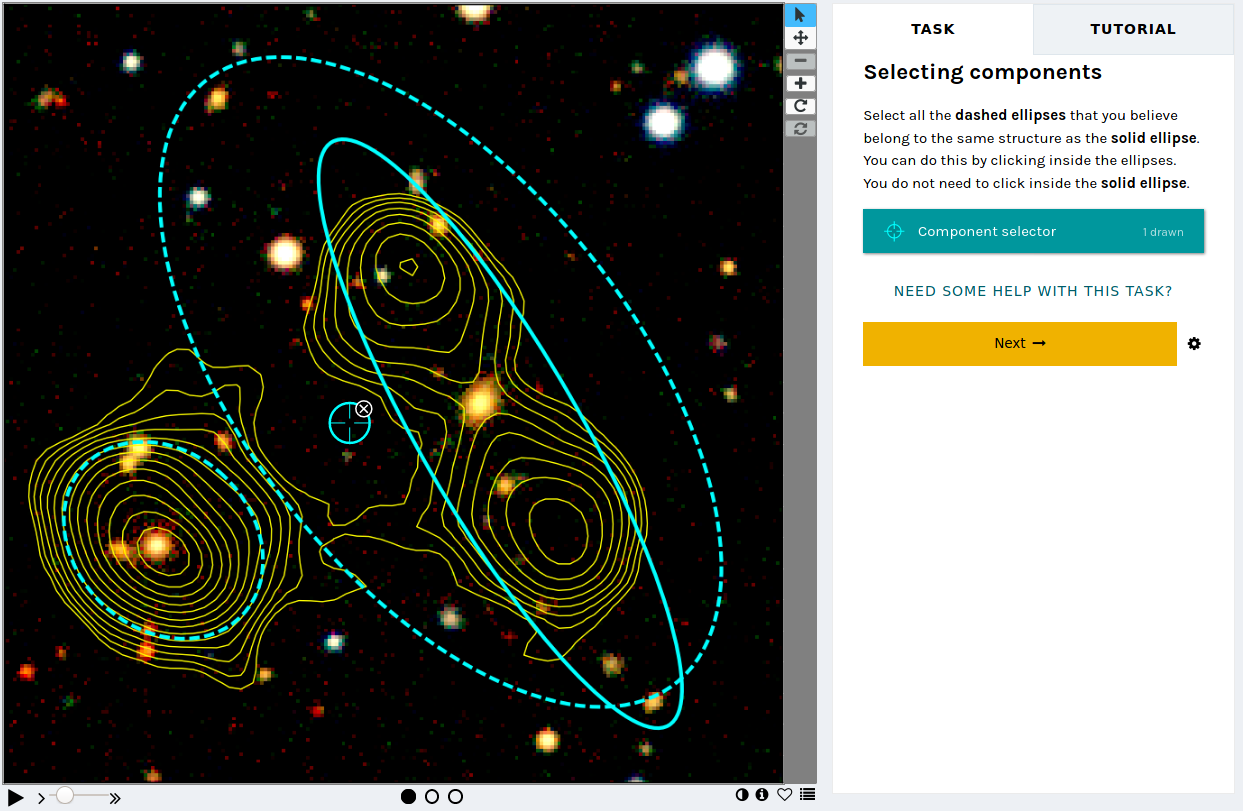}
  \includegraphics[width=0.50\linewidth]{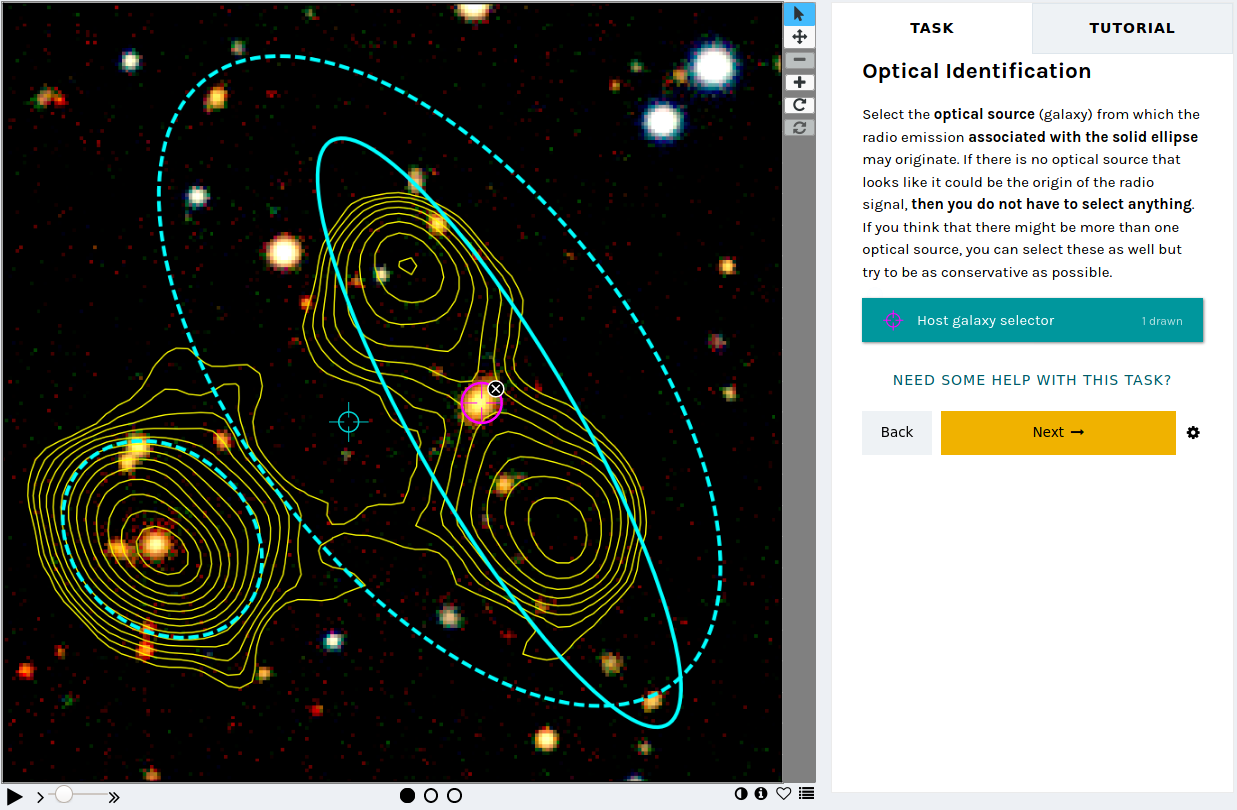}
  \includegraphics[width=0.50\linewidth]{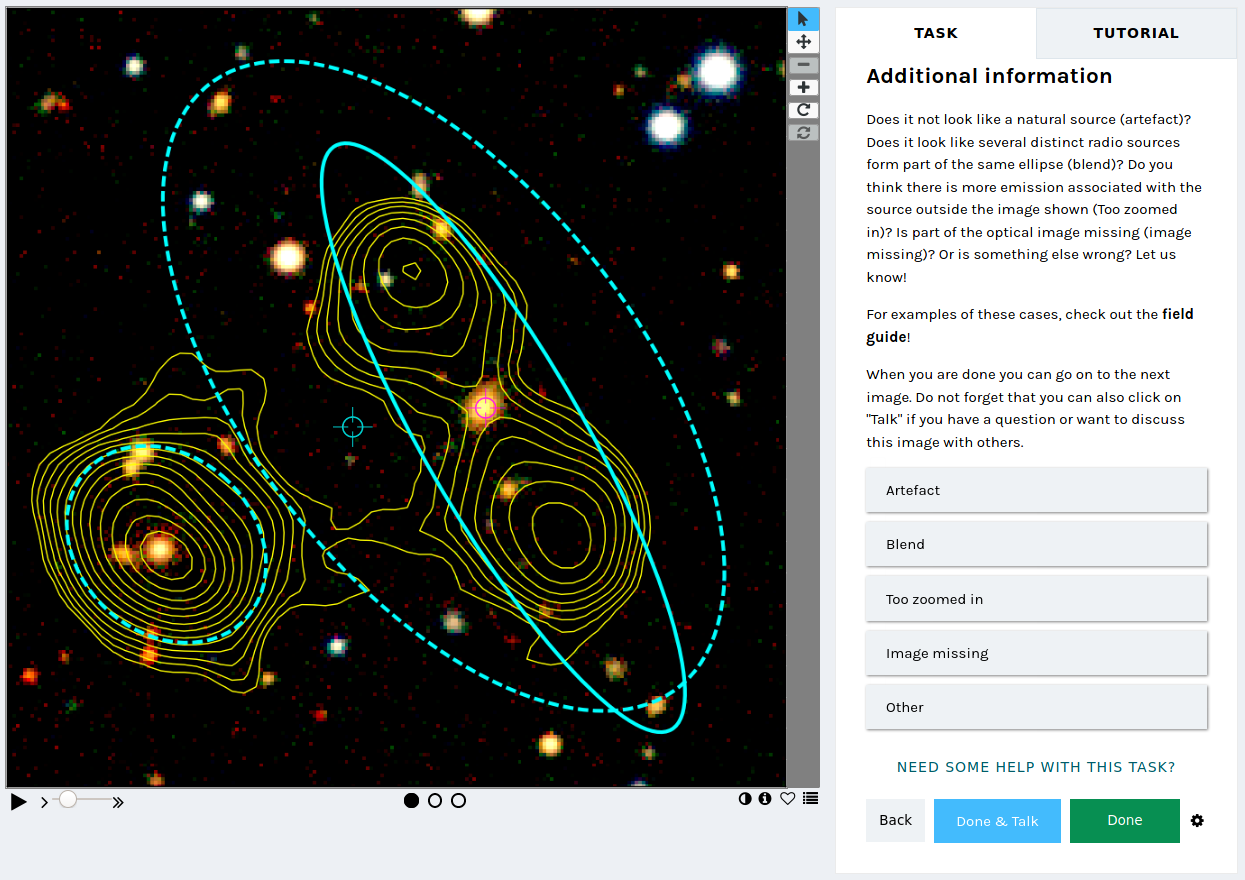}
  \caption{Images from the classification section of the Zooniverse interface.
    This shows the three task screens presented for one catalogued source,
    ILTJ172125.82+370417.2, seen by the user in the order top left,
    top right, bottom left panels. The image shown here is 100 arcsec on
  a side. All three views here show the standard image (Legacy colour
  scale, radio contours, and ellipses to represent catalogued
  Gaussians). In the first panel, the marker for an associated
  component can be seen; in the second panel, the user has also marked
  an optical identification for the radio source. In the third panel,
  the user has the opportunity to apply various flags to the source or
  to discuss it on the talk pages. Note the
  unassociated radio source to the southeast (bottom left). The toolbar below
  the image allows the user to switch images, to get information on
  the source, to invert the colour map, or to add the source to a list of favourites. Additionally, the user has the option to zoom, pan, and rotate the image using the buttons on the right. }
  \label{fig:zooniverse-interface}
\end{figure*}

Citizen scientists were asked to go through a three-stage process for
each source sent to Zooniverse, illustrated in
Fig.\ \ref{fig:zooniverse-interface}. These can be summarized as
`association', `identification', and `commenting'. In the first step,
volunteers were asked to select any radio sources in the field of view
that were physically associated with the object of interest (indicated
with a solid ellipse) by clicking on the image. Next, they were asked
to select one or more potential optical identifications for the
associated source in the same way. In the final screen they could
select one or more flags to indicate potential problems with the
source, and/or choose to leave comments on the object on the
Zooniverse talk page. Problems that could be flagged up included
stating that the source was an artefact (i.e. not a physical source),
that the source combined emission from two or more separate sources (a
blend), that it was too zoomed in (i.e. there might be associated
components outside the field of view), that one or other of the
required images was missing, or some other general problem with the
image (for example a bright star preventing the optical
identification). Volunteers were also encouraged to tag the objects
with descriptive but consistently used words (`hashtags': cf. \citealt{Rudnick21}) which could be recovered in
processing. No previously defined hashtags were supplied, so the
consistent use of these relied on communication between participants
on the Zooniverse forums.

To guide and train the citizen scientists in the process, various
resources were made available. The first time a user started
classifying, a text-based tutorial appeared on the screen which
explained the interface, the radio-optical overlay and the
association, identification and commenting tasks. Additionally, we
provided a tutorial video which explained the process with ten examples
of common radio sources. Finally, a separate interactive training
workflow was set up where volunteers could practice on those ten
example radio sources and receive feedback interactively after
clicking on the images. The project and text based tutorials were made
available in eight languages\footnote{In order of the volume of use by
volunteers these were English, French, German, Italian, Polish, Dutch,
Swedish, and Chinese.}, while the tutorial video was made in four
different languages, plus an additional version using closed captions.

Following the approach of \cite{Williams+19}, we required a minimum of five
classifications for each catalogued source, but large complex physical
sources are often broken down into smaller sub-components in {\sc
  pybdsf}, so that many more individual classifications can contribute
to the interpretation of a complex source. A refinement added part-way
through the process was to `retire' after only three views a source
that no user had classified in any way at that point. This avoids wasting user time on
sources where volunteers have nothing to say (i.e. compact sources with no
optical identifications).

A total of 189,375 sources (4\% of the total source count in the
survey) were sent to RGZ(L): this includes 104,582 large,
bright sources (where we selected sources with flux density $>8$ mJy
and size $>15$ arcsec but also a peak flux $>2$ times the local rms
noise)\footnote{6,978 sources that failed the rms criterion could not
be sent to RGZ(L) as they could not be visualized using contour maps. Some of these were
deleted as artefacts in subsequent processing, and a few were included
in RGZ(L) or post-processing outputs, but many simply end up with a
likelihood-ratio ID. In the final catalogue these objects can be selected by requiring {\tt Total\_flux} $>8$ mJy, {\tt DC\_Maj} $>15$ arcsec and {\tt Peak\_flux} $<2 \times${\tt Isl\_rms}. They should be treated with caution in
the final catalogue.}, 64,835
sources with flux density $>4$ mJy selected directly from decision
tree endpoints, and 19,958 sources pre-filtered from decision tree
endpoints by visual inspection from members of the project team. Results from RGZ(L) were initially processed in the manner described by
\cite{Williams+19}. User `clicks' were provided in the JSON-format
Zooniverse output, and these were matched to the radio and
optical and {\it WISE} catalogues. Once clicks had been matched to the catalogue, quality factors for the association and identification of the sources were calculated based purely on the fraction of Zooniverse volunteers who had picked any particular identification or association. Overall, the whole process differed from the approach taken with our internal LGZ platform used for DR1 only because we used a magnitude-size relation for galaxies to give more leeway to the optical identifications with bright, nearby, extended galaxies. The default maximum circular offset threshold was 3 arcsec but it could be extended up to $\sim 25$ arcsec for the brightest galaxies.

\begin{figure*}
  \includegraphics[width=\columnwidth]{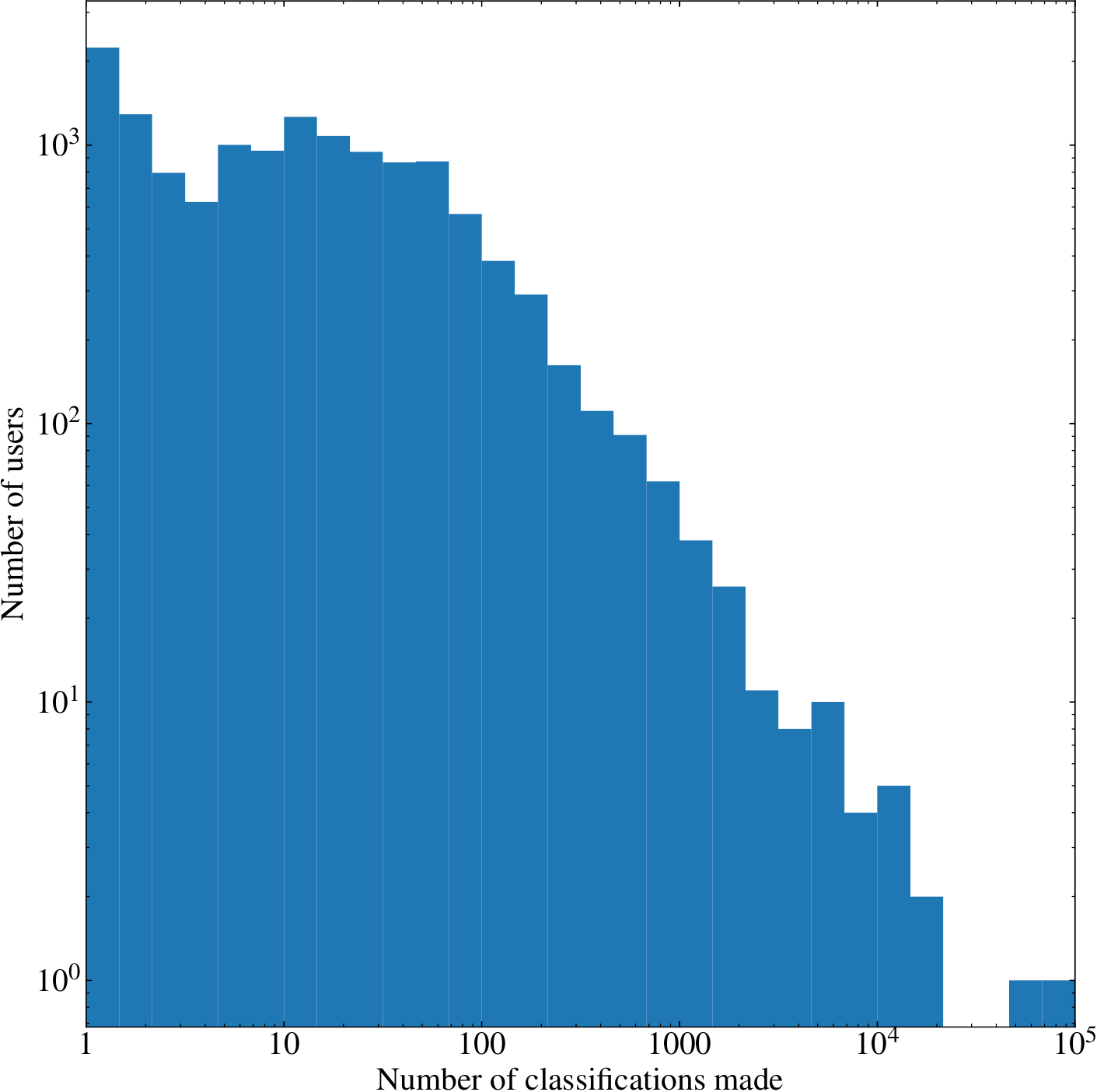}
  \includegraphics[width=\columnwidth]{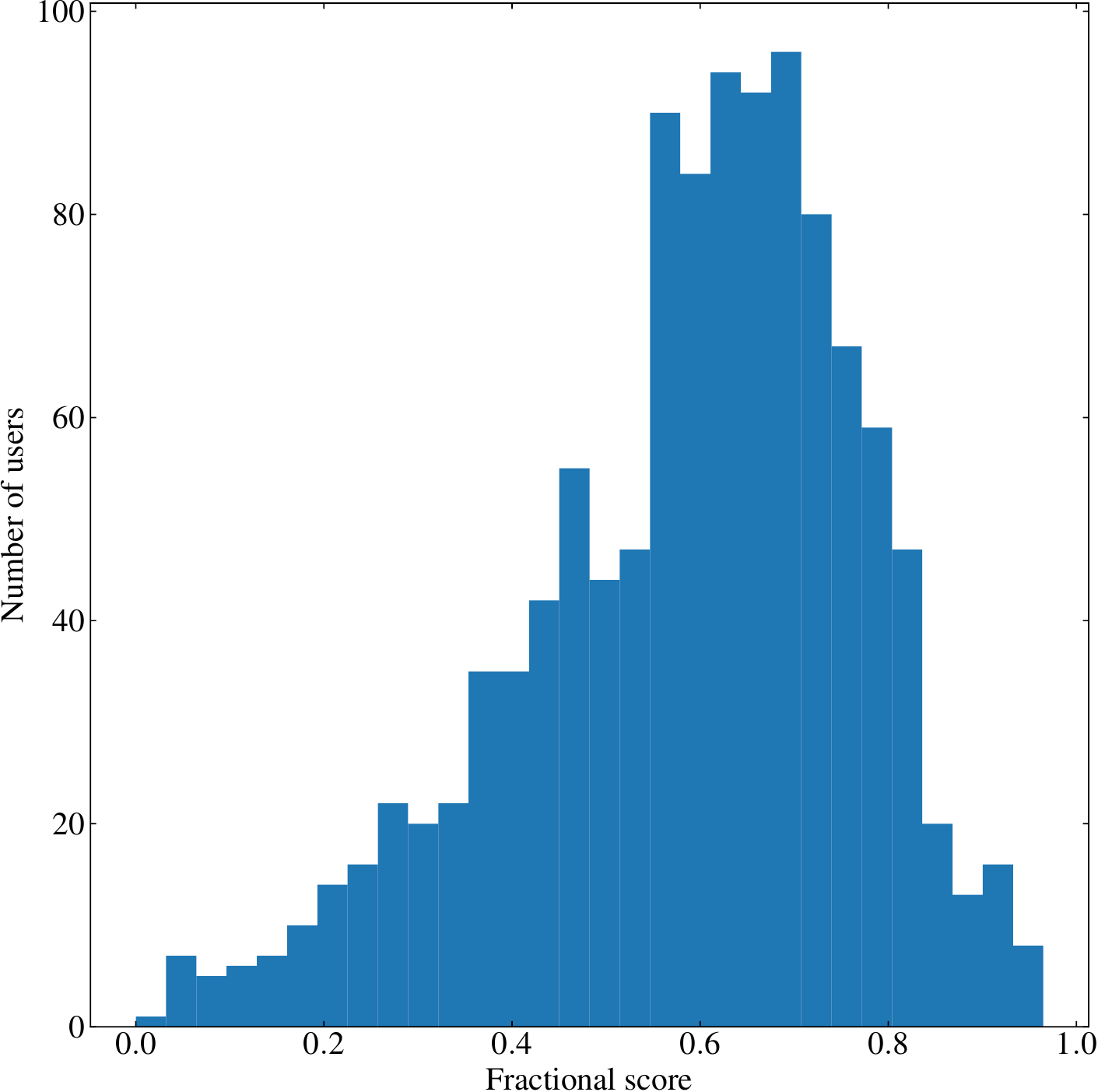}
  \caption{Statistics of the Zooniverse volunteer population. Left: histogram showing the numbers of Zooniverse volunteers who made a certain
    number of classifications.
    On a log scale the rough power-law distribution of classification
    numbers is apparent, with a slope $\approx -1$. Right: histogram of the distribution of optical ID consensus
    scores for volunteers with more than 100 classifications.}
  \label{fig:user_classifications}
  \label{fig:consensus_scores}
\end{figure*}

A total of 957,374 classifications were made through the Zooniverse
system by 13,711 distinct users, including users who were not logged in
to the platform. Of these, only $\sim 100$ made more than
1,000 classifications --- the most prolific $\sim 125$ volunteers
contributed half the total classifications. The distribution of user
classification numbers is plotted in
Fig.\ \ref{fig:user_classifications}. It can be seen that several
thousand volunteers tried classifying just once or twice before
disengaging with the project --- this may be a reflection of the
comparative difficulty of the combined radio and optical classifications.
However, the numbers level off above a few tens of classifications and
show a rough power-law form between 100 and $\sim 2000$ classifications. This
type of distribution is not uncommon, in some parts of the
range, for measures of `scientific productivity', loosely defined
\citep{Lotka26}. For projects like this one it means that many of the
classifications will be contributed by volunteers who have had the
opportunity to develop expertise in source classification. Volunteers with more
than 2,000 classifications were offered co-authorship on this
paper and personally contacted for assistance in finishing off the
later parts of the project, and this may account for the change in the
slope of the histogram at this point.

\begin{figure*}
  \begin{center}
  \includegraphics[width=0.8\columnwidth]{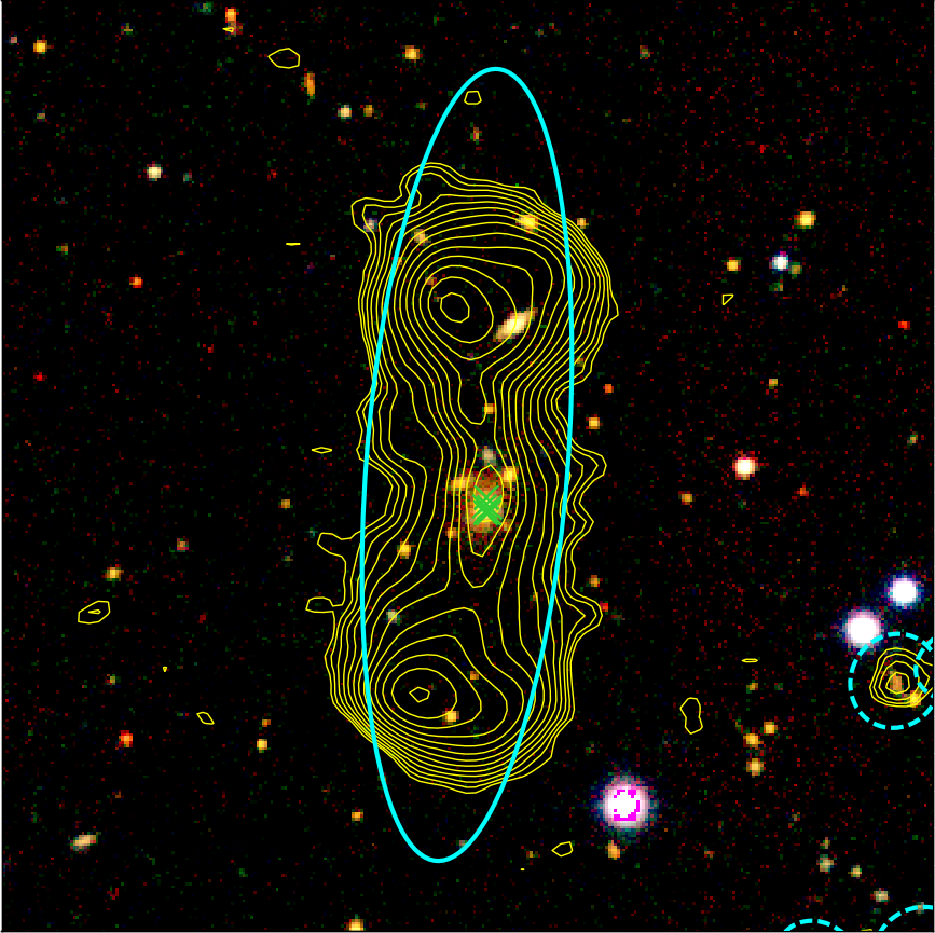}
  \includegraphics[width=0.8\columnwidth]{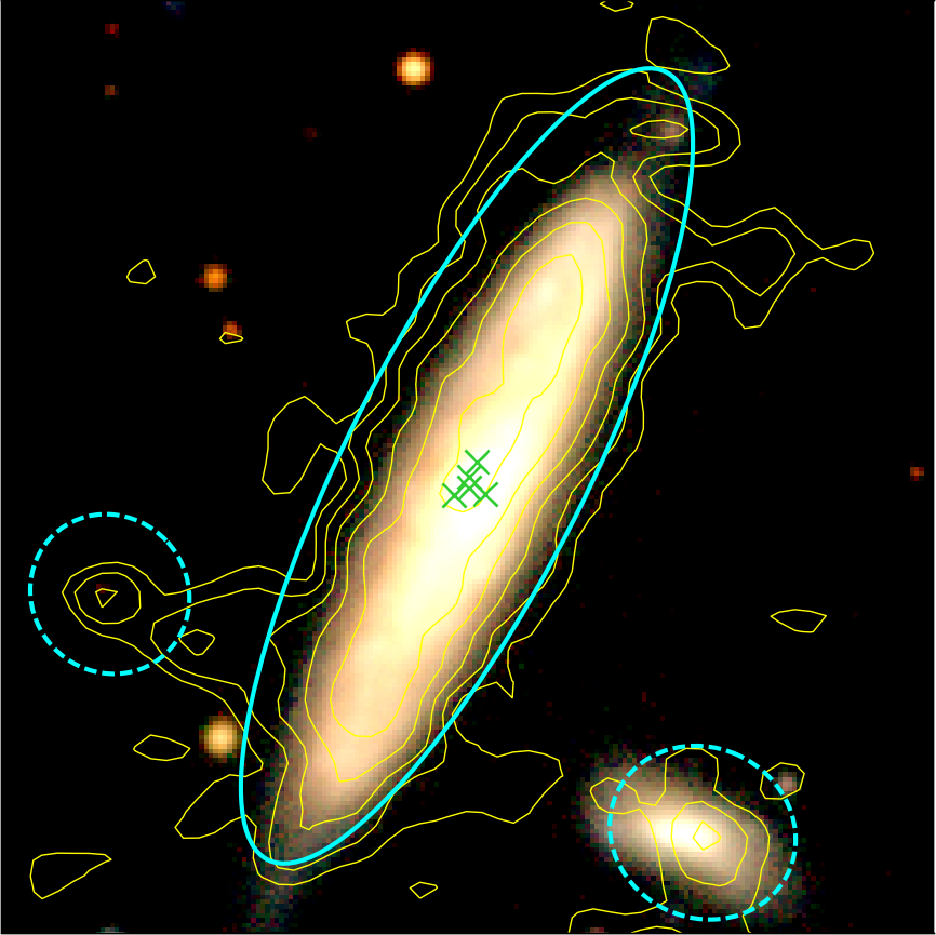}
  
  \includegraphics[width=0.8\columnwidth]{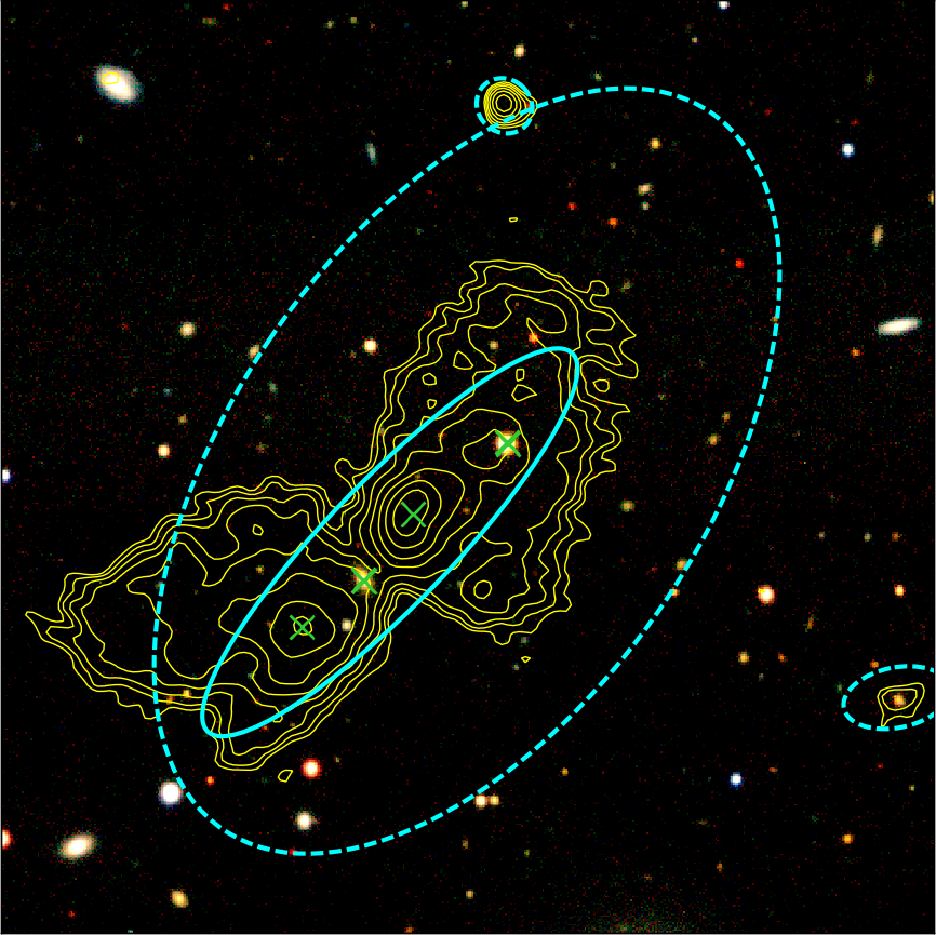}
  \includegraphics[width=0.8\columnwidth]{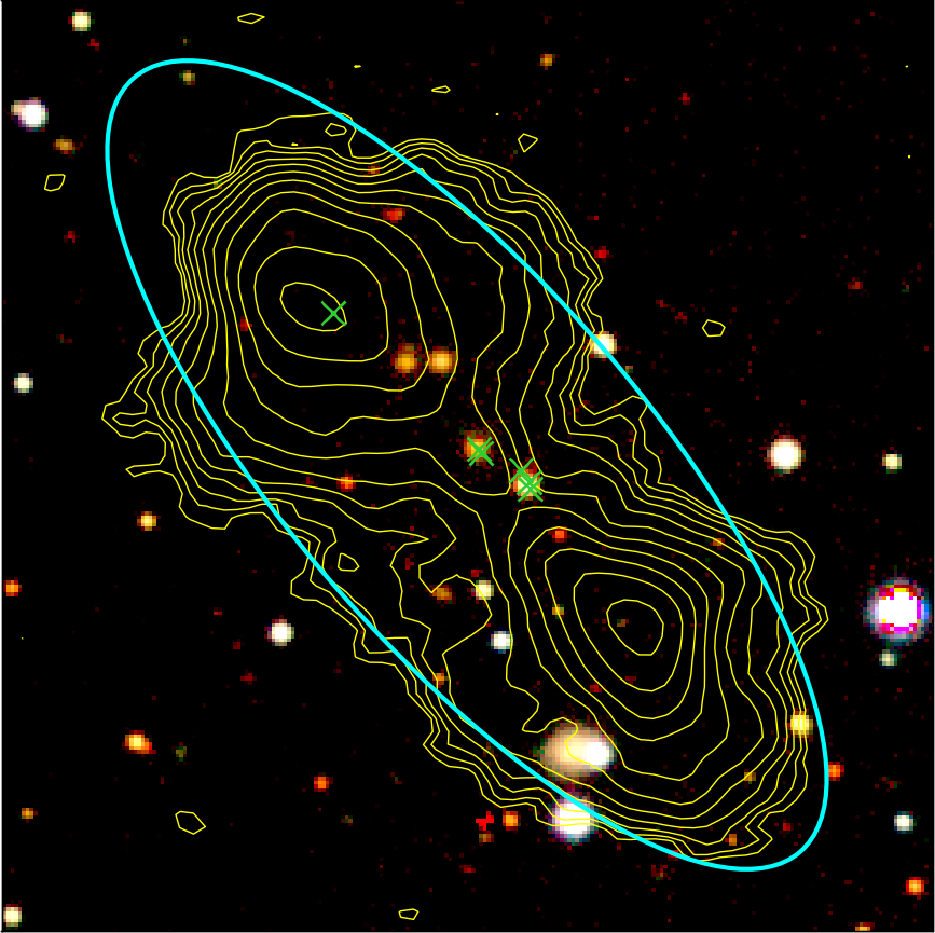}
  \end{center}
\caption{Examples of RGZ(L) subjects with the optical and radio
  contour image seen by Zooniverse platform volunteers overplotted with the optical IDs selected by
  the volunteers, marked as green crosses. All sources have five or more optical ID selections
  (volunteers could optionally select more than one possible ID). The top
  row shows examples where a consensus was achieved and the correct
  optical ID selected, the bottom row
  ones where no consensus was found and no optical ID returned from
  RGZ(L).}
\label{fig:consensus_example}
\end{figure*}
  
Interestingly, the raw rate of optical identification from the
Zooniverse project was low. Only 27\% of all sources sent to be viewed
by volunteers returned with a consensus optical ID (that is, one where
more than 2/3 of the votes on a given target agreed on the best associated
optical object: examples of sources where this is and is not the case
are shown in Fig. \ref{fig:consensus_example}). This contrasts with 51\% for the internal
classifications through the same interface, and illustrates the
difficulty of selecting the right optical object for relatively
untrained volunteers. By contrast, the fraction of radio sources associated
with others (around 18\%) is similar for astronomers and Zooniverse
volunteers as a whole. Objects with no consensus optical ID may still have associations and simply propagate through to the next stages of processing with no ID. On visual inspection of a randomly selected subsample of the RGZ(L) optical IDs by two independent astronomers, the error rate was found to be $\sim 3$\%; in other words, the RGZ(L) optical ID process is conservative and probably does not assign an ID to every source that should have one, but where an ID is assigned, it is almost always correct.


As we have no `gold standard' sources, we have no means of assessing
the quality of individual volunteers' classifications as objectively good
or bad. What we can do instead is to assess the extent to which volunteers
tend to agree with others. To do this, for optical IDs, we considered
the final RGZ(L) source catalogue, and compared all optical ID
classifications made by volunteers to it. If the final catalogue
contained no ID for the source, each user who selected no ID for that
particular source scored one point, and all volunteers who selected any ID
scored no points. If the final catalogue did contain
an optical ID, volunteers who had selected an ID positionally matched to the one in
the catalogue scored one point, and all others scored no points. Dividing the points scored by the number
of sources classified by each user gives a per-user `consensus score'
which must lie between 0 and 1, and the histogram of this (for all
volunteers with more than 100 classifications, to give adequate statistics)
is shown in Fig.\ \ref{fig:consensus_scores}. Since a selected
optical ID requires more than 3/5 classifiers to agree on it, we
expect this score to exceed 0.6 in general --- that is, for any
finally catalogued optical ID, at least 3/5 volunteers should score points.
Consistent with this, the median of the consensus score is almost
exactly 0.6. Volunteers who had a consensus score much lower than this were
consistently disagreeing with other volunteers, and this suggests that
they were not interpreting the images in the same way. Over 116,000 classifications were made by volunteers whose
consensus score was less than 0.3. The histogram also shows that a few
volunteers, generally with quite small numbers of classifications, have
consensus scores approaching 1.0. Since this degree of consensus would be
quite hard to achieve by other means, we suspect that these are volunteers
who declined to classify (by hitting reload) all sources where the
optical ID was not obvious, but this hypothesis cannot be confirmed from the available data on user interactions with the Zooniverse platform, which does not list classifications that were started but not completed.

Given the wide range of consensus scores for optical IDs and the low
optical ID fraction, we elected to rerun the processing code with
volunteers' optical ID votes (only) reweighted by their consensus scores as
shown in Fig.\ \ref{fig:consensus_scores}: volunteers who had not
classified more than 100 objects were given a weighting of 0.6, the
median value. This gave a modest improvement in the optical ID fraction from the RGZ(L) volunteers, increasing it to 31\%, and so it is
these consensus optical IDs which are fed to the next stages of the
process.

Hashtags assigned by volunteers to each source were added to a supplementary
catalogue file made available as a JSON dictionary. This will allow
catalogue users to search easily for objects which have been tagged in
a particular way. Widely used tags are listed in Table \ref{tab:tags},
and include a number which could give morphological information on the
resulting source. However, it is worth noting that these tags were not
consistently applied and should not be used to try to derive complete
samples. Some morphological structures are labelled more reliably than
others; for example, there are a reasonable amount of wide angle
tailed sources labelled as WATs, but very few of the sources tagged as NATs
have narrow angle tails, even though both tags have been applied a
similar number of times. In general around 10-40\% of tagged sources
appear to be clearly described by their morphological tags. Additionally, only a small percentage of objects of any given kind were tagged to begin with.

\begin{table}
\caption{Tags applied by RGZ(L) volunteers to 50 or more sources. Italics indicate tags that are descriptive of the images seen by the volunteers or the processes they followed rather than the sources themselves.}
    \label{tab:tags}
    \centering
    \begin{tabular}{rlrl}
      \hline
    Rank&Tag&Rank&Tag\\
    \hline
3082 & {\it solid-ellipse} & 131 & stretched \\
1963 & core-jet & 120 & one-sided \\
1958 & doublelobe & 112 & ddrg \\
1503 & compact & 101 & x-shaped \\
1252 & triple & 96 & disk \\
1164 & diffuse & 89 & jets \\
1092 & compacts & 85 & galaxycluster \\
967 & hourglass & 81 & diffuseradiosources \\
433 & {\it submitted} & 79 & interesting \\
409 & hybrid & 75 & s-shaped \\
381 & core-jets & 75 & complex \\
354 & nat & 72 & corejet \\
348 & blend & 67 & dashed-ellipses \\
325 & bent & 63 & artefact \\
287 & wat & 62 & orc \\
282 & sdragn & 58 & unusual \\
273 & extended & 57 & nascent-doublelobe \\
269 & {\it too-zoomed-in} & 56 & tail \\
265 & galaxy & 56 & spiral \\
234 & clumpy & 55 & v-shaped \\
219 & {\it overedge} & 55 & hybrid-doublelobe \\
206 & no\_clear\_source & 55 & doublelobes \\
196 & no-optical-source & 53 & double-lobe \\
191 & stretched-compact & 52 & star \\
187 & possible\_jets & 52 & cluster \\
179 & double & 50 & noise \\
164 & restarted & 50 & hybrid-feature \\
158 & {\it no-dashed-ellipses} & 50 & difficult \\
148 & {\it toozoomedin} & 46 & diffuse-clumpy \\
 \hline
    \end{tabular}
\end{table}

\section{Catalogue generation, further visual inspection and processing}
\label{sec:postproc}

Once the RGZ(L) outputs were processed, a first catalogue was
created by merging the decision tree results (including a decision
on whether or not to accept a likelihood-ratio optical ID for a given
source) with the radio and optical catalogue generated by the process described in the
previous section. For this we adapted the code written for the LoTSS Deep
Fields analysis \citep{Kondapally+21} which keeps track of the
provenance of all finally generated sources, their components and
their optical identifications. The output of this combination was (i)
an initial catalogue of associated sources that combines the basic
{\sc pybdsf}, optical, and RGZ(L) catalogues into one, along with provenance
information, and (ii) a component catalogue that allows the final
state of each {\sc pybdsf} source (whether as a catalogued source in its own
right or as a component of an associated source) to be looked up.
Objects flagged by a majority of Zooniverse volunteers as artefacts (or
flagged as artefacts in the pre-filtering process discussed in the
previous section) were removed from the catalogue at this point, and
the catalogue generation process also generates derived table entries
for quantities like the total flux density of a composite source from
RGZ or the maximum size of the convex hull enclosing all of its
components ({\tt Composite\_Size}).

\begin{figure}[t]
    \centering
    \includegraphics[width=1.0\linewidth]{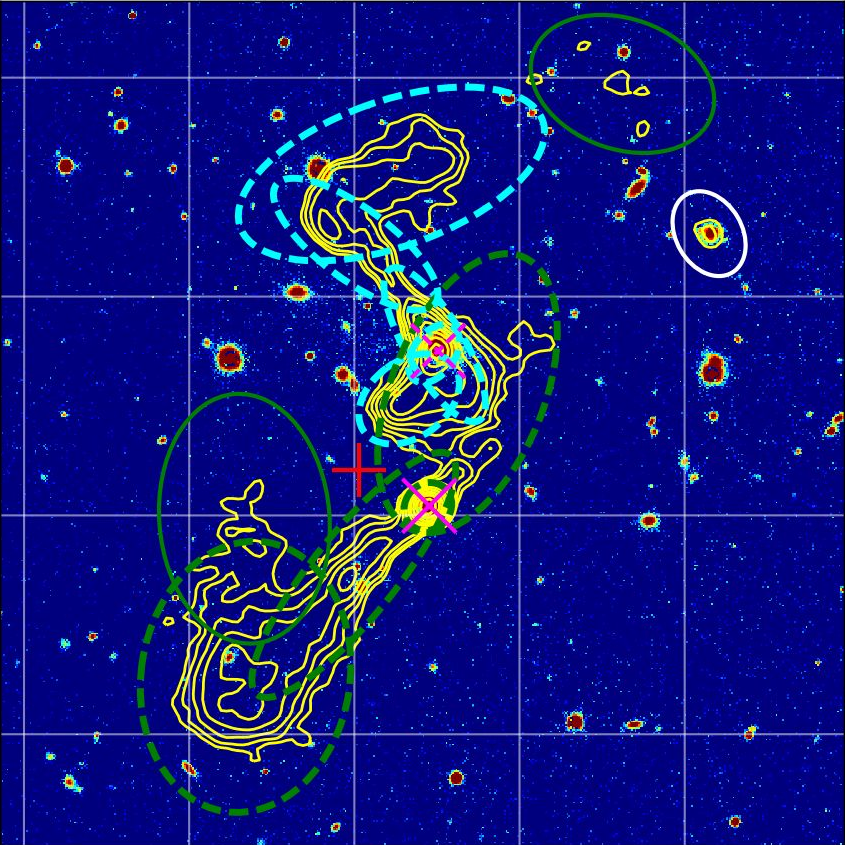}
    \caption{Example user interface for the `Second deblend' workflow.
      In an interactive Matplotlib window the expert classifier has
      separated the emission from two extended sources that had been
      combined in {\sc pybdsf}, seen in green and cyan, and has selected
      optical IDs for both. An unrelated source marked in white has
      been left unchanged. The new source is a mixture of {\sc pybdsf} components (solid lines) and Gaussians (dashed lines).}
    \label{fig:new_blend}
\end{figure}

Further visual inspection was needed for a small minority of sources
after this was done, with the aim being to ensure that the catalogue was
as accurate as possible for extended, complex radio sources. This was
done using six workflows carried out by astronomers on the LoTSS
team, all of which involved an expert classifier editing either
or both the association or identification of the catalogued source.
These ran roughly in the following order:

\begin{enumerate}
\item `First deblend': in this workflow {\sc pybdsf} components of a single
  composite source were broken down into their component Gaussians in
  order to allow a finer-grained allocation of radio sources to
  optical counterparts. This was particularly important in the case of
  two close but physically distinct radio sources that were merged
  into one {\sc pybdsf} source. Sources flagged as blends by more than half
  of RGZ(L) volunteers or in pre-filtering were either sent to this
  workflow or to `Second deblend' (see below). Users of the workflow
  could choose to send deblended sources on to the `too zoomed in'
  workflow (see below) for further processing.
\item `Too zoomed in' (TZI): this workflow was used for sources where
  RGZ(L) volunteers flagged sources as `too zoomed in' meaning that
  there appeared to be extended structure on scales larger than was visible in the image
  presented to the user. This was also used for sources prefiltered as TZI, or sent directly there by other workflows such as `Postfilter' or `First
  deblend', or for sources that exhibited other problems after the
  initial processing of the RGZ(L) catalogue. The original {\sc pybdsf}
  component decomposition was retained and components could be added (or
  removed) from the current output of the catalogue to generate a new
  composite source. Remaining blended sources could be sent on to the
  `Second deblend' workflow and the size of a source could be recorded
  manually if the {\sc pybdsf} components did not represent this well.
\item `Deduplication': this workflow provided a simple interface for
  merging objects with duplicate optical IDs or removing one of the
  duplicates as an artefact, and was set up part-way through the
  processing to reduce the labour costs of the more time-consuming TZI
  workflow. It was applied after the production of the initial catalogue.
\item `Postfilter': this workflow involved the visual inspection of
  all sources from the Zooniverse or TZI workflows with an angular
  size ({\tt Composite\_Size}) greater than 1 arcmin in order to check the
  validity of the source association --- the `post-filtering' step.
  Around 30\% of these sources were flagged as problematic in some way
  (mostly sources that should have been flagged as `too zoomed in' by
  RGZ(L) volunteers but were not) and these were sent on to a further
  iteration of the TZI workflow. A small number were flagged as
  blended and sent to the `Second deblend' workflow. 
  \item `Blend prefilter': Later in the
  processing, prefiltering was carried out on a large number of sources flagged as
  blends by RGZ(L) volunteers or by the flowchart to check whether these were genuine
  blends (which were sent on to the `Second deblend' workflow) or
  should be dealt with in some other way, such as splitting into all individual components with IDs. This was an important step as only around 13\% of blend prefiltered sources were sent to the time-consuming `Second deblend' workflow.
\item `Second deblend': this workflow was a combination of TZI and
  deblending that allowed detailed editing of the components of
  complex sources, including the ability to include previously
  unassociated components, which was missing in `First deblend'.
  Sources flagged in Postfilter, TZI or (later in the processing) by
  RGZ(L) volunteers as blends were sent to this workflow, as shown in
  Figure~\ref{fig:new_blend}.
\end{enumerate}

Finally, a version of the ridge-line optical ID code {\sc RL-Xid} of
\cite{Barkus+21} was used on large ($>15$ arcsec) sources with flux
density above 10 mJy that did not have an optical ID assigned from
visual inspection. This code, which uses the radio morphology of
extended sources to help to select the most plausible host, allowed us
to pick up a number of {\it WISE}-only or faint optical IDs that had
been missed by RGZ(L) volunteers and/or by the expert classifiers.
Relative to the version of the code described by \cite{Barkus+21}, the
main changes were optimizations of the size measurement and flood-filling
algorithms to allow the code to run in reasonable time on the large
number of sources present in DR2. The size and flux density limits
were selected based on tests of the reliability of the ridge lines
constructed by the code.

\begin{figure*}
\includegraphics[width=\columnwidth]{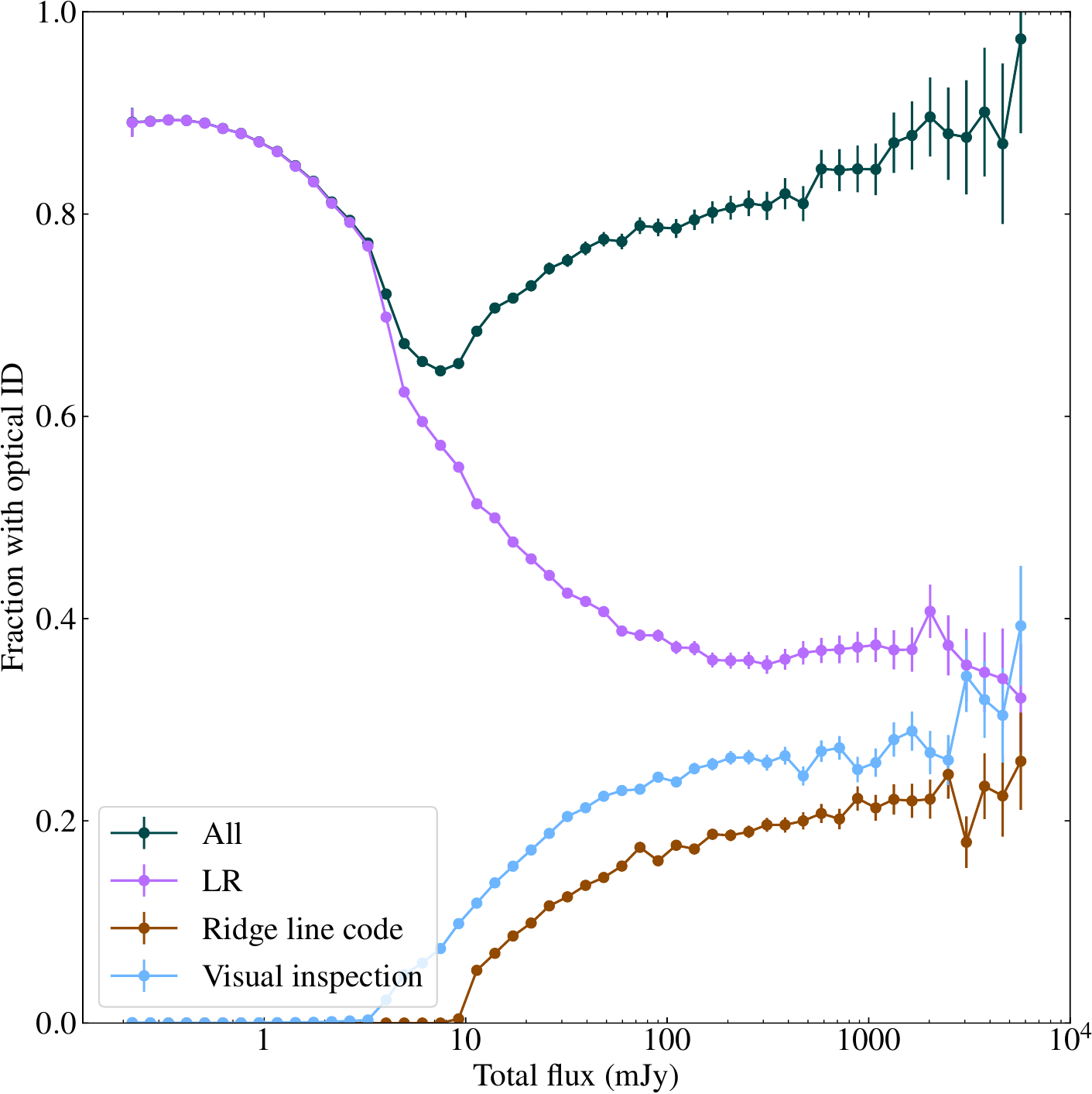}
\includegraphics[width=\columnwidth]{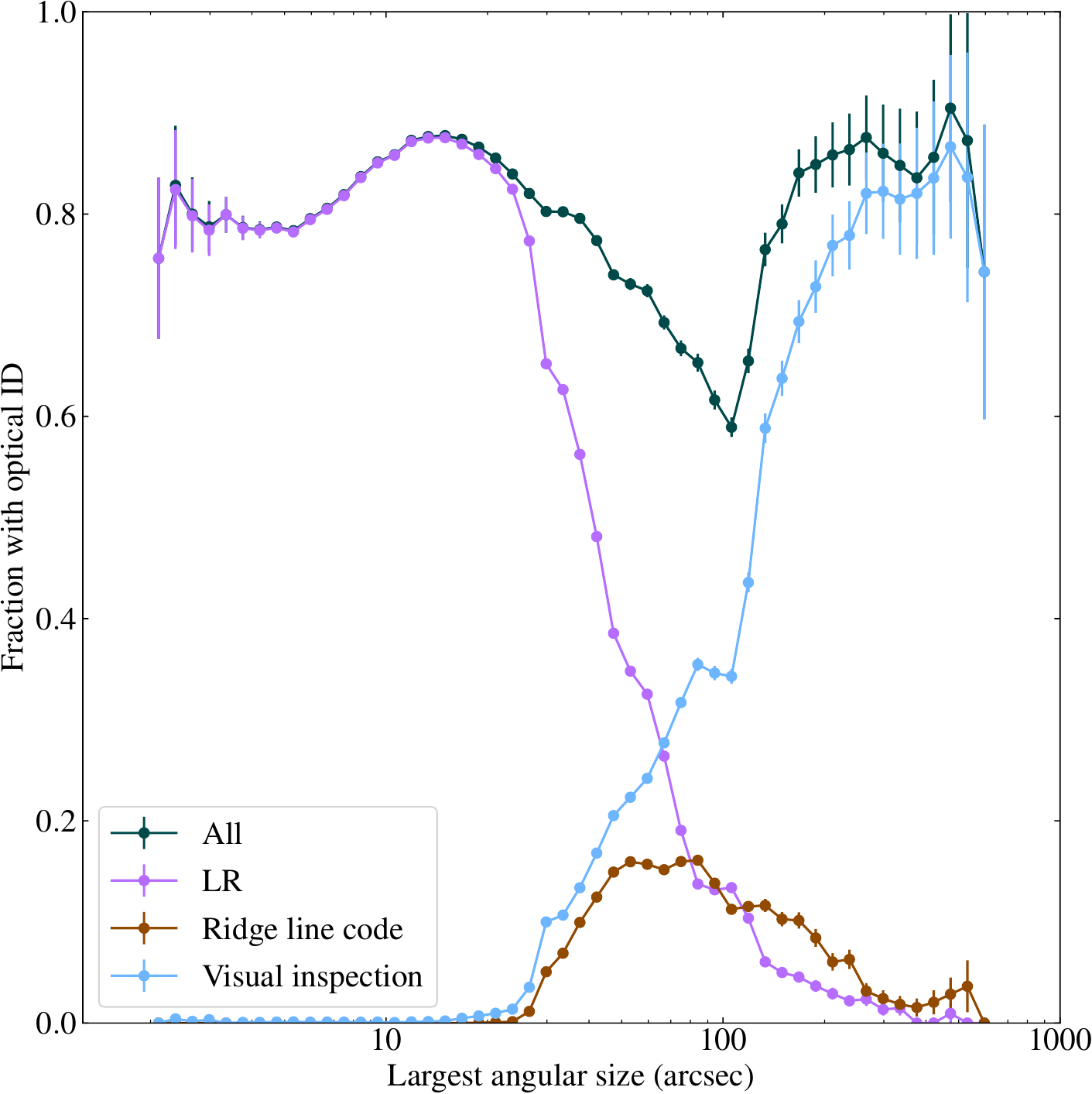}
\caption{Fractional optical IDs in the DR2 catalogue. The two plots
  show the total fraction of optically identified objects, and the breakdown by different methods of optical identification, as a function of (left) total flux density of the resulting source and (right) catalogued largest angular size.}
\label{fig:fractional_id}
\end{figure*}

Table \ref{tab:provenances} gives the recorded radio source provenance,
as recorded in the {\tt Created} column, of all sources in the final
catalogue, and the sources of optical
IDs ({\tt Position\_from}) for all objects that have them. It can be
seen that the vast majority of optical IDs (97\%) come from the
likelihood-ratio cross-matching (LR). However,
Fig.\ \ref{fig:fractional_id} shows that half of all IDs for the
brightest sources, and nearly 100\% of IDs for the largest sources,
come from visual inspection. The curves of optically identified
fraction as a function of flux density and source largest angular size
show that our methods are not uniformly good at identifying all
sources: the fact that no source with a flux density less than 4 mJy
was sent to visual inspection and only sources with fluxes $>10$ mJy went
to the ridge line code leads to a drop in the fraction of sources with
IDs between 1 and 10 mJy, while the ID fraction steadily rises above
this point. It is noteworthy that fewer than half of the sources
returned from RGZ(L) have an ID returned from visual
inspection, even after TZI processing. The sharp increase in the ID
fraction above an angular size of 2 arcmin is presumably due to the
postfilter step, and the data suggest that more IDs could be obtained
with yet more visual inspection of sources with sizes $>30$ arcsec.

More details of the different routes to optical IDs are provided in
the {\tt ID$\_$flag} column of the final catalogue, and the statistics
of this are given in Table \ref{tab:id_flags_additional}. At the end
of the processing we achieved an 85.0\% optical ID fraction for
sources in the Legacy sky coverage.

\begin{table}
  \caption{Provenances of radio sources, IDs, redshifts, and sizes in the final catalogue}
  \label{tab:provenances}
  \begin{tabular}{llr}
    \hline
    Provenance of&Origin&Number\\
    \hline
Source creation&Create initial sources&3,983,901\\
({\tt Created})&Ingest RGZ(L)&146,147\\
&Too zoomed in&21,343\\
&Process flowchart blends&6,349\\
&New$\_$blend&5,737\\
&Deduplicate&2,823\\
&Deblend&1,059\\[10pt]
Optical ID&LR&3,412,365\\
({\tt Position$\_$from})&Visual inspection&71,368\\
&Ridge line code&34,333\\[10pt]
Redshift&Photometric&2,083,466\\
({\tt z$\_$source})&SDSS&272,888\\
&DESI&33,726\\
&HETDEX&2,535\\
&High-$z$ quasar&24\\[10pt]
Angular size&Gaussian&4,079,827\\
({\tt LAS\_from})&Flood-fill&62,799\\
&Composite&24,598\\
&Manual&135\\
\hline
  \end{tabular}
  \end{table}

\begin{table}
  \caption{Final ID flag statistics for sources with optical ID}
  \label{tab:id_flags_additional}
\begin{tabular}{rlr}
  \hline
    ID\_Flag & Meaning & Number\\
    \hline
-99 & Outside Legacy optical coverage & 31,076 \\
1 & LR ID & 3,151,983 \\
2 & Match with large optical galaxy & 322 \\
3 & ID from RGZ(L) & 47,536 \\
5 & Faint source not visually inspected & 206,336 \\
6 & Sent for deblend & 11,738 \\
8 & Uncatalogued host after prefilter & 4 \\
9 & Automatic or visually selected deblend & 6,625 \\
10 & Blend workflow & 2,214 \\
11 & Second blend workflow & 5,238 \\
12 & Too zoomed in workflow & 17,888 \\
13 & Ridge line code & 34,333 \\
14 & Deduplicate workflow & 2,773 \\
\hline
\end{tabular}
\end{table}

\section{Radio source angular size estimates}
\label{sec:angsize}

As discussed above, non-composite sources have a size estimate (twice
the deconvolved major axis of the fitted Gaussian), while a rough size
estimate for composite sources can be obtained from the largest
dimension of the convex hull encompassing all of the {\sc pybdsf} components
({\tt Composite\_Size}). A small number of sources also have manual
size measurements made during the too-zoomed-in visual inspection
process. Because {\sc pybdsf} tends systematically to overestimate the size
of faint components \citep{Boyce+23}, while sometimes not detecting at
all the largest-scale parts of an extended radio source, this size
estimate is not ideal for physical size inference. As part of the {\sc
  LoMorph} (LM) code, \cite{Mingo+19} describe a method for estimating
what we here refer to as `flood-fill sizes', in which the {\sc pybdsf}
ellipses are used as the starting point for a measurement which in
principle should include only the pixels of the image of the source
that are above the local noise level. This method cannot return a size
estimate much smaller than the beam size (i.e. the beam is not
deconvolved from the size estimate) and so it is not suitable for
application to compact sources.

We applied the flood-fill method to all sources in the catalogue with total flux
density $> 5$ mJy and estimated extended size $>20$ arcsec, 147,141
sources in total. The code returns flags if the flux density in the
flood-fill source is significantly below the lower limit in the input
catalogue, or if there are too few pixels to estimate a size after
masking, and these, along with the size estimates, are included in the
catalogue (column names {\tt LM\_size}, {\tt LM\_flux}, {\tt
  Bad\_LM\_flux} and {\tt Bad\_LM\_image}).

Some heuristic is then needed to make an overall best angular size
estimate. The small number of manual size measurements in the
catalogue (which can be assumed to be accurate since they are based on
visual inspection) offer a guide: many of the flood-fill sizes are in
good agreement with the manually measured sizes but some are smaller
by a significant factor. The latter group, on inspection, are all
sources with faint extended structure which does not appear above
the noise floor in the flood-fill code. To some extent this problem
can be mitigated by requiring the flux density measured by the
flood-fill code to be close to the total catalogued flux density of
the source -- if a significant fraction of the radio emission is
missing that can be taken as an indication that the flood-fill code is
missing important structure.

To obtain an overall best size estimate (largest angular size, or LAS) we proceed as follows:
\begin{enumerate}
\item If a manual size measurement is available, we use that;
\item If not, a catalogue-based LAS is estimated by
  taking the {\tt Composite\_size} where available, and $2\times${\tt
    DC\_Maj} otherwise.
\item The flood-fill size, if one exists, is adopted as the LAS in
  preference to the catalogue-based one if
  all three of the following conditions are met:
  \begin{enumerate}
    \item No flood-fill flags are set
    \item The flood-fill flux density matches the catalogue flux
      density to within
      20\%
    \item The LAS is larger than 30 arcsec and smaller than 600
        arcsec (this avoids regions where the flood-fill code cannot
        return good results).
  \end{enumerate}
\end{enumerate}

The final LAS and, for each source, an indication of the origin of the
LAS ({\tt LAS\_from}) are given in columns in the final catalogue and
the distribution of the origins of LAS is shown in Table \ref{tab:provenances}.
Sources where the {\tt LM\_Size} is adopted even though it is
significantly different from the {\tt Composite\_Size} should be
treated with caution -- visual inspection shows that some of these
sources have genuine low-surface brightness extended structure that
was missed by the flood-fill algorithm, while others are point sources
surrounded by artefacts.

For sources where the size estimate comes from the fitted Gaussian
(the vast majority) we implement the resolution criterion of
\cite{Shimwell+22}, in the {\tt Resolved} column of the catalogue.
Size estimates should only be used where the source is flagged as
resolved. All sources with alternative size measurements are taken to
be resolved.

\section{Redshifts and physical source properties}
\subsection{Spectroscopic and photometric redshifts}
\label{sec:photoz}
Photometric redshift (photo-$z$) estimates for the LoTSS sample with
optical detections in the Legacy Surveys DR8 are taken from
\citet{Duncan2022}, where full details of the methodology, training
samples, and catalogue properties are presented.
In summary, the photo-$z$ estimation methodology was designed to
produce robust photo-$z$ predictions for a broad range of optical
populations, including active galactic nuclei (AGN). 
The method employed Gaussian mixture models (GMMs) derived from the colour, magnitude, and size properties of the observed population to divide it into different regions of parameter space for training and prediction.
The sparse Gaussian processes redshift code \textsc{GPz} \citep{Almosallam2016a,Almosallam2016b} was then used to derive photo-$z$ estimates for individual regions of observed parameter space, including cost-sensitive learning weights derived from the GMMs to mitigate against biases in the spectroscopic training sample.

\citet{Duncan2022} explored the photo-$z$ performance as a function of spectroscopic redshift, optical magnitude, and morphological type, finding that the photo-$z$ estimates offer substantially improved reliability and precision at $z > 1$, with negligible loss in accuracy for brighter, resolved populations at $z < 1$ when compared to other photo-$z$ predictions available in the literature for the same optical population.
Crucially for the LoTSS sample, the photo-$z$ predictions for the radio continuum selected population are suitable for use over a wide range in parameter space -- with low robust scatter ($\sigma_{\text{NMAD}} < 0.02 - 0.10$) and outlier fraction ($\textup{OLF}_{0.15} <10\%$)\footnote{Where $\sigma_{\textup{NMAD}} =1.48 \times \text{median} ( \left | \delta z \right | /(1+z_{\textup{spec}}))$ and the outlier fraction, $\textup{OLF}_{0.15}$, is the fraction of sources with $\left | \delta z \right | / (1+z_{\text{spec}}) > 0.15$, for $\delta z = z_{\textup{phot}} - z_{\textup{spec}}$.} at $z < 1$ across a broad range of radio continuum (and X-ray) properties.
At a given true redshift, $z_{\textup{spec}}$, there is no evidence that photo-$z$ precision or reliability exhibits any dependence on the radio continuum flux density (and hence luminosity).
The photo-$z$ quality for a given LoTSS sample will therefore largely be dictated by the associated optical properties.

In the combined value-added catalogues presented in this paper we provide the derived photo-$z$ columns presented in Table~3 of \citet{Duncan2022}.
By construction, the \textsc{GPz} predictions are unimodal, with \texttt{zphot} representing the mean of the normally distributed photo-$z$ posterior and \texttt{zphot\_err} the corresponding standard deviation.

In addition to the photo-$z$ estimates, we also included spectroscopic
redshifts from the Sloan Digital Sky Surveys Data Release 16
\citep[SDSS DR16;][]{SDSSDR16} when available. As the LoTSS DR2 sample
contains a mixture of both galaxy and quasar type sources, we matched the
SDSS spectroscopic sample in two stages. We first matched the main DR16
spectroscopic sources with $z_{\text{spec}} < 2$ to the LoTSS sources
through a positional match between the SDSS coordinates and the
corresponding Legacy Surveys optical catalogue with a 1.5-arcsec
radius. We then matched the SDSS DR16 Quasars catalogue
\citep[][`DR16Q\_V4']{Lyke2020} sample with the same matching radius.
For the sources with matches in both samples (which should largely be quasars at
$z_{\text{spec}} < 2$), the $z_{\text{spec}}$ value is taken to be
that provided by \citet{Lyke2020}. In total, we found SDSS counterparts
for 296,921 LoTSS radio sources, of which 273,935 had spectroscopic
redshifts with no warning flags.

To these, we added spectroscopic redshifts from the early data release
of the DESI spectroscopic survey \citep{DESI+23} which covers a number of non-uniformly
distributed fields within the LoTSS DR2 area. We positionally matched
the DESI target position with the positions of LoTSS optical
counterparts within 1.5 arcsec, taking only DESI sources with {\tt
  ZWARN=0} and {\tt ZCAT\_PRIMARY=True}. This gives us 45,128
counterparts to LoTSS radio sources, although a significant fraction
of these also have SDSS redshifts.

Finally, we merged in spectroscopic redshifts from the first HETDEX
data release \citep{MentuchCooper+23}. This gave a comparatively
small number of redshifts for LoTSS optical IDs, all in the DR1 area
(3,339), and increases the available spectroscopic redshifts for the
sample by only $\sim 1$\%, but we include
them in this release of the catalogue as it is our intention to make
further releases that will include the full spectroscopic results from
HETDEX.

Redshifts $>5$ are not reliable either in the SDSS quasar catalogue or
in the photometric redshift estimates. We have therefore removed all
redshifts $z>5$ from either of these two sources from the final catalogue but have merged in the DR2
high-$z$ quasar catalogue, based on spectroscopic redshifts, from \cite{Gloudemans+22}.

\begin{figure*}
  \includegraphics[width=\columnwidth]{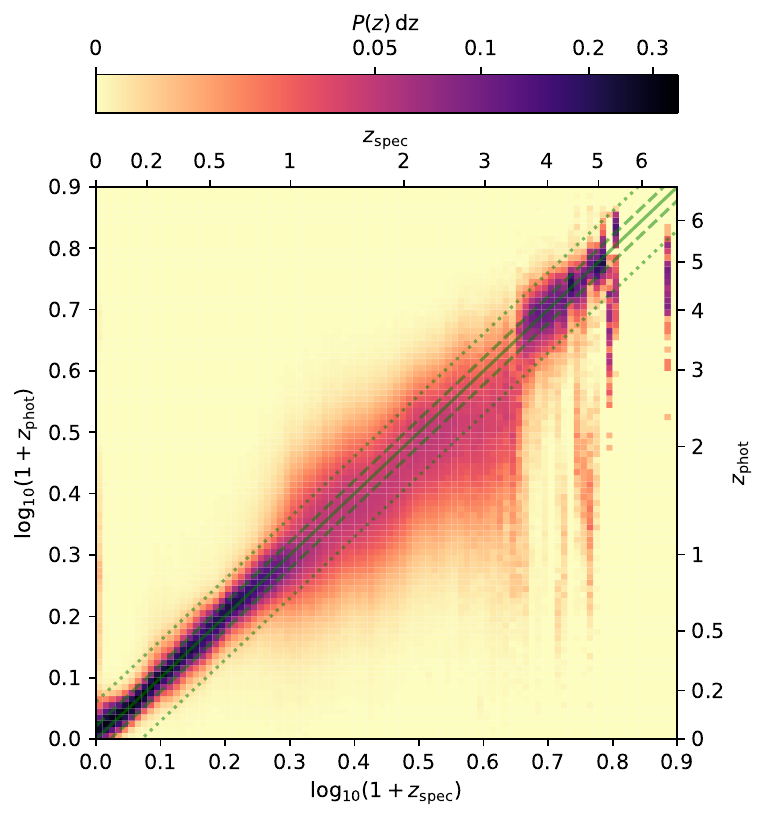}\hskip 10pt
  \includegraphics[width=\columnwidth]{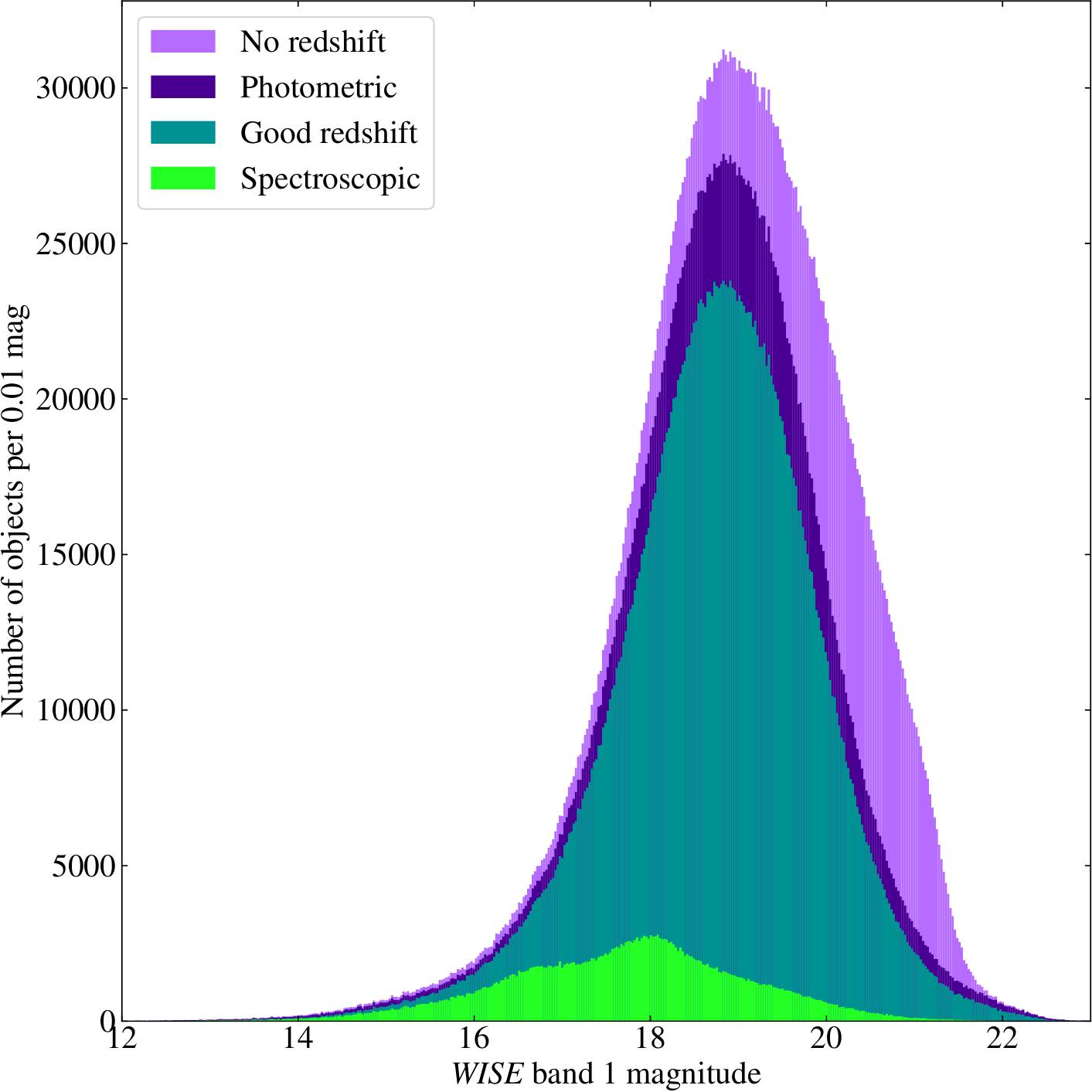}
  \caption{Statistics of the photometric and spectroscopic redshifts. Left: Photo-$z$ posterior distributions as a function of
    SDSS spectroscopic redshift for LoTSS DR2 sources with reliable
    spectroscopic redshift (\texttt{zwarning\_sdss} = 0) and photo-$z$
    estimates that pass the photo-$z$ quality selection
    (\texttt{flag\_qual} = 1). The photo-$z$ distribution is
    normalized such that the distribution for each $z_{\rm{spec}}$ bin
    integrates to unity. Dashed and dotted lines illustrate the bounds
    $z_{\rm{phot}} = z_{\rm{spec}}\pm0.05$ and $0.15\times(1+z)$
    respectively. Right: the distribution of available redshifts for
    all optically identified objects as a
    function of {\it WISE} band 1 magnitude, where a `good redshift'
    is defined in the text.}
  \label{fig:specz_photz}
\end{figure*}

In the final catalogue, we define a \texttt{z\_best} column which
contains the best estimate of the source's redshift. This is defined as follows, with
earlier redshift types taking precedence over later ones:
\begin{enumerate}
  \item The high-$z$
    quasar redshift if it exists; else
  \item the SDSS redshift
\texttt{zspec\_sdss} if there are no SDSS warnings
(\texttt{zwarning\_sdss} = 0); else
\item the DESI redshift {\tt z\_desi} if one is available; else
\item the HETDEX redshift {\tt z\_hetdex} if
  one is available; else
\item the photometric redshift
  \texttt{zphot} if the photo-$z$ quality flag \texttt{flag\_qual} = 1.
\end{enumerate}
The column is blank if there is no good-quality spectroscopic or
photometric redshift, although the original redshifts are retained in
the catalogue if they exist. A \texttt{z\_source} column in the catalogue
gives the origin of the `best' redshift and the statistics of this are given
in Table \ref{tab:provenances}. As shown in
Fig.\ \ref{fig:specz_photz}, the redshifts are dominated by
photometric redshifts above a {\it WISE} band 1 magnitude $\sim 17$, but we are close to
having complete good spectroscopic or photometric redshifts down to $W1
\sim 19$ mag. 58.0\% of sources in the Legacy Survey sky area, and 83.8\%
of sources with an ID in the Legacy catalogue, have a `good
redshift' listed in \texttt{z\_best}.

The best redshift estimate, for those sources that have it, is used
to define an estimated projected physical size ({\tt Size}) in kpc
from the largest angular size {\tt LAS} as discussed in Section \ref{sec:angsize} and an estimated radio luminosity ({\tt L\_144})
in W Hz$^{-1}$ from the total source flux density, on the assumption
of a spectral index $\alpha = 0.7$. These physical properties will in general have
significant systematic uncertainties (from the assumption of $\alpha =
0.7$ in the case of the total luminosity and from the relatively crude
size estimates in the case of the projected physical size) as well as
statistical uncertainties, which are not tabulated, in the case of the
quantities derived from photometric redshifts: however, they represent
our best estimates and should allow the initial selection of
interesting sub-populations. As noted in Section \ref{sec:angsize}, the {\tt Size} column should only be used
for sources that are flagged as {\tt Resolved}.

\subsection{Stellar mass estimates and rest-frame magnitudes}
Although the available photometry is not sufficient for detailed
spectral energy distribution (SED) modelling, the combination of
rest-frame optical colours from Legacy Survey with {\it WISE} constraints on the overall normalisation of the rest-frame near-IR make stellar mass estimates possible for the LoTSS population with SEDs dominated by host galaxy light.
We estimate stellar masses and key rest-frame magnitudes for the LoTSS
sample with optical-IDs and robust photo-$z$s following a similar
approach to that of \citet{Duncan2022}.
In summary, stellar masses are estimated using the \textsc{Python}-based SED fitting code previously used by \citet{Duncan+14} and \citet{Duncan+19}. 
Composite stellar populations are generated using the stellar
population synthesis models of \citet{Bruzual+03} for a
\citet{Chabrier+03} initial mass function (IMF), with the model SEDs
convolved with the Legacy Surveys $g$, $r$, and $z$
filters\footnote{Model grids are generated separately for the Legacy
Surveys North and South datasets separately to account for the
differing optical filters, with LoTSS sources fit to the corresponding
grid.} and {\it WISE} $W1$ and $W2$.
The assumed set of parametric star-formation histories follow those outlined by \citet{Duncan2022}, spanning a range of double power-laws.
Similarly, we assume the same dust attenuation law \citep{Charlot+00} and range of extinction values.
Due to the limited available photometry, we restrict the available metallicities to $Z \in \left \{ 0.2, 1.0 \right \} Z_{\odot}$ and fix the escape fraction of ionising photons to $f_{\textup{esc}} = 0$.

One key change from the approach taken by \citet{Duncan2022} in this analysis is the incorporation of the photo-$z$ uncertainty into the stellar mass estimates.
The SED model grid is evaluated at 100 redshift steps from $0 < z < 1.5$, with redshift steps evenly spaced in $\log_{10}(1+z)$.
When fitting the LoTSS sample, we draw 100 Monte Carlo samples from the photo-$z$ posterior and fit the observed photometry to the nearest corresponding redshift step for each draw, calculating the optimal scaling and the corresponding $\chi^{2}$ for every model in the grid \cite[see][]{Duncan+19}.
The stellar mass and associated 1-$\sigma$ uncertainties are taken to be the 50th (and 16-84th percentiles, \texttt{Mass\_median} and \texttt{Mass\_l68}/\texttt{Mass\_u68} respectively) of the likelihood weighted mass distribution from all Monte Carlo trials after marginalising over the stellar population parameters.
Additionally, we provide rest-frame magnitudes for key optical to IR bands, taken to be the median of the distribution of best-fitting templates from the Monte Carlo draws in each of the corresponding filters.
For sources with spectroscopic redshift available, we assume a small redshift uncertainty of $\sigma = 0.001\times(1+z_{\text{spec}})$.

As $z$-band is the reddest optical filter available, constraints on
the strength of the $D_{4000\text{\AA}}$ break required to constrain
the age of the stellar population (and hence mass to light ratio)
beyond $z\sim1$ will be limited. We therefore restrict stellar-mass
fitting to LoTSS sources with $z_{\text{phot}} +
\sigma_{z_{\text{phot}}} < 1.5$, or $z_{\text{spec}} < 1.5$, as well
as requiring reliable estimates and clean photometry
(\texttt{flag\_qual} = 1). In total, we fitted the SEDs of 2,193,448
sources in the LoTSS sample. However, this number includes a
significant fraction of sources for which the SED fits (and associated
stellar masses) are not expected to be reliable, primarily sources
with significant contributions to the observed SED from either
unobscured (i.e. radio-quiet or radio-loud quasar) or obscured
radiative accretion activity.

\begin{figure}
  \includegraphics[width=\columnwidth]{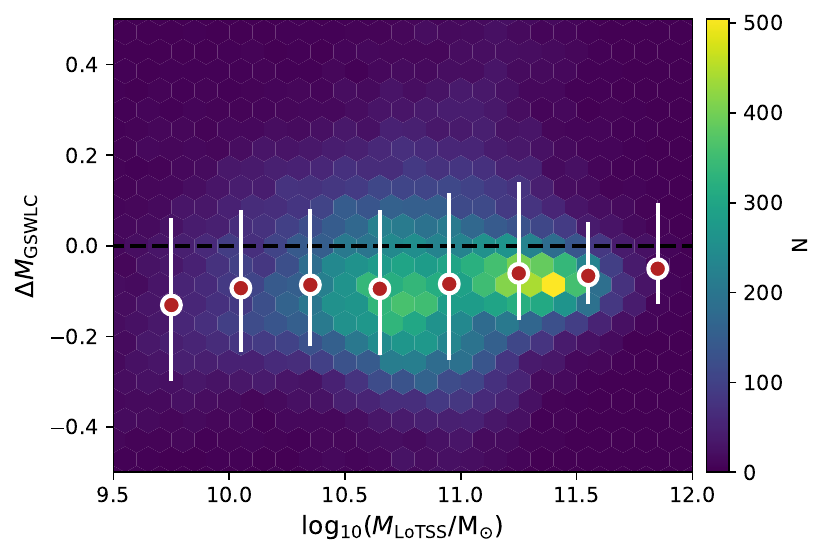}
  \includegraphics[width=\columnwidth]{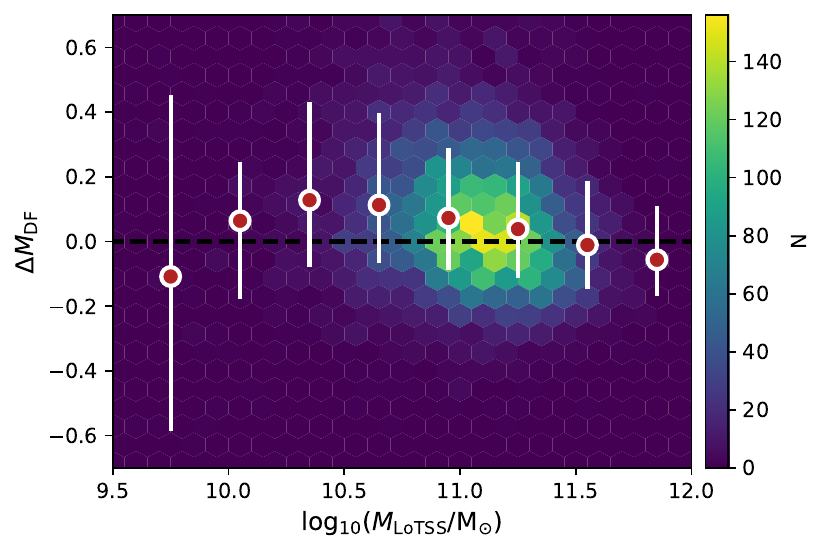}
  \caption{Distribution of estimated stellar mass differences compared to GSWLC \citep[$\Delta M_{\text{GSWLC}}$;][upper panel]{Salim2016,Salim2018} and LoTSS Deep Fields DR1 \citep[$\Delta M_{\text{DF}}$;][lower panel]{Duncan2022} for sources in common. Red circles and corresponding error bars illustrate the median and 16-84th percentile $\Delta M$ within a fixed $\log_{10}(M_{\text{LoTSS}}/\text{M}_{\odot})$ bin.}
  \label{fig:mass_comparison}
\end{figure}

To validate the precision of our stellar mass estimates, we compared our estimates to others available within the literature. 
At low redshifts, we cross-matched the LoTSS sample to the
GALEX-SDSS-{\it WISE} Legacy Catalogue \citep[GSWLC version 2:][]{Salim2016,Salim2018}, which provides stellar mass and star-formation rate estimates using the full UV to mid-IR photometry for a large sample of SDSS galaxies.
We limited the analysis to sources where the photometric redshift
\texttt{zphot} is close to the redshift assumed for the GSWLC fitting
($\delta_{z} < 0.02\times(1+z_{\text{phot}})$) and the source is not
flagged as a poor fit or an IR AGN in either GSWLC or in LoTSS DR2
\citep[based on the $C_{75}$, `75\% completeness', $W1-W2$ colour
  criteria of][]{Assef2013}. The resulting sample consists of 90,626
sources with matches within 1 arcsec separation.
The upper panel of Fig.~\ref{fig:mass_comparison} presents the difference in stellar mass estimate, 
\begin{equation}
    \Delta M_{\text{GSWLC}} = \log_{10}\left (\frac{ M_{\text{LoTSS}}}{M_{\text{GSWLC}}} \right ),
\end{equation} as a function of the stellar mass estimated in this work.
We find that the GSWLC mass estimates are consistently $\sim0.1\text{dex}$ higher than $M_{\text{LoTSS}}$ across all masses, but with a significant scatter that is equal to or greater than the systematic offset.

Extending to higher redshifts, we also compared the LoTSS DR2 stellar mass estimates for sources within the footprints of the LoTSS Deep Fields with those presented by \citet[][DF hereafter]{Duncan2022}.
As outlined above, the methodology applied here follows that of \citet{Duncan2022}; however, the DF estimates incorporate both deeper and more extensive (in wavelength range and filter coverage) photometry that should yield both more reliable estimates.
Similar to GSWLC, we limited the comparison to sources where the photo-$z$ from the Deep Fields are in good agreement ($\delta_{z} < 0.1\times(1+z_{\text{phot}})$).
Additionally, due to the different photometry measurements used for the estimates (corrected apertures versus model fluxes for Deep Fields and this work respectively), we also applied a correction based on the measured $z$-band flux, such that we define
\begin{equation}
\Delta M_{\text{DF}} = \log_{10}\left (\frac{M_{\text{LoTSS}}}{M_{\text{GSWLC}}}\times\frac{f_{z,\text{DF}}}{f_{z,\text{LoTSS}}}\right ).
\end{equation}
Similarly to our approach above, we limited the DF comparison sample
to sources with $z_{\text{phot}} > 0.3$ (where the DF aperture
corrections are appropriate) and non IR AGN, we find a total of 11,404
matches within 1 arcsec across all three DF fields).
The lower panel of Fig.~\ref{fig:mass_comparison} shows the corresponding distribution of mass offsets.
After accounting for the difference in total flux estimates (which is
a strong function of observed galaxy size and hence most severe at low
redshift), we found that our stellar mass estimates are also in good agreement with those from LoTSS DF, with masses within $\sim0.1\,\text{dex}$.
However, unlike the flat $\Delta M_{\text{GSWLC}}$$\sim0.1\text{dex}$ distribution, $\Delta M_{\text{DF}}$ shows a noticeable dependence on $M_{\text{LoTSS}}$.
Further investigation reveals that the apparent mass dependence is driven by a residual dependence on redshift (and hence likely source size), with higher $\Delta M_{\text{DF}}$ values for lower redshift sources indicating that our simple aperture corrections are insufficient.
Nevertheless, at $z_{\text{phot}} > 0.7$ where the photometry is in good agreement, our stellar mass estimates are in excellent agreement with those from the DF catalogues.

Overall, Fig.~\ref{fig:mass_comparison} demonstrates that the mass estimates presented in this work are reliable, with no significant systematic offsets resulting from the limited photometric information available.
Given the differences in photometry and assumed stellar population
properties (and associated priors), the $\sim0.1\,\text{dex}$ offsets
are consistent with those expected from, for example, different star-formation history assumptions \citep{Pacifici2023}.
However, we caution that this is only the case for sources with no significant radiative AGN contribution to the observed optical to near-IR photometry. 
We therefore provide an additional catalogue column,
\texttt{flag\_mass}, to indicate which stellar mass estimates are safe
to use.
For $\texttt{flag\_mass}$ set to {\tt True}, we require that sources
have a physically meaningful fit ($\texttt{Mass\_median} > 7.5$ and
$\texttt{Mass\_u68}-\texttt{Mass\_l68} < 2$) and are not expected to
contain a significant radiative AGN contribution \citep[$\texttt{type}
  \neq \text{PSF}$ to exclude likely quasars, and $W1_{\text{Vega}} -
  W2_{\text{Vega}} < 0.77$ to select sources not satisfying the $C_{75}$ criteria of][]{Assef2013}.

\section{Catalogue description}
\label{sec:catdesc}

The catalogues described in this paper are available for download electronically\footnote{From the LOFAR surveys website
\url{https://lofar-surveys.org/dr2_release.html} or from CDS via \url{https://cdsarc.cds.unistra.fr/cgi-bin/qcat?J/A+A/}.}. Details of the
columns are given in Appendix \ref{sec:tables}.

Our main product is a science-ready source catalogue which contains
all objects that we think are physical sources, together with their
radio properties, their
optical ID information, and their associated optical properties if available,
our best estimate of redshift combining spectroscopic and photometric
constraints, and derived physical quantities as described in the
previous section. The source names in this catalogue are the names
from the LoTSS DR2 radio source source catalogue described by
\cite{Shimwell+22}, except for composite sources, where the tabulated
RA and Dec, and therefore the name, are generated from the
flux-weighted mean position of the components that make up the source.

Accompanying the source catalogue is a component catalogue which is essentially an
annotated version of the DR2 radio source catalogue, with the
following differences:
\begin{enumerate}
  \item The name of
    entries in the catalogue is {\tt Component$\_$Name};
  \item Some entries in
the original table may have been deleted as artefacts and so will not
be present in our component table;
\item Some components are Gaussians
promoted to components as part of the deblending process, and so were
not originally present in the DR2 source catalogue: in this case there
will be an entry in the {\tt Deblended\_from} column which refers back
to the DR2 source catalogue;
\item All components have a {\tt Parent\_Source} column entry
  referring to an object in the main source table.
\end{enumerate}

Finally, as noted above, a JSON-format dictionary provides a list of
all tags for sources that were tagged by RGZ(L) volunteers. This can easily be
iterated over to generate lists of sources with a particular tag, bearing in mind the caveats given in the previous section.

\section{Properties of the final catalogue}
\label{sec:properties}

\subsection{Quality comparisons}
\label{sec:compare}

\begin{table*}
  \caption{Matches of 3CRR objects with sources in the final catalogue}
  \label{tab:3crr}
  \scriptsize
  \begin{tabular}{llrrrrllllll}
    \hline
    Name&ILT name&3C LAS&LoTSS
    LAS&3C $z$&LoTSS $z$&Source creation&LAS from&Optical ID from&Flux
    &Size&$z$\\
    &&(arcsec)&(arcsec)&&&&&&match?&match?&match\\
    \hline
3C14     &  ILTJ003606.50+183758.4 &   26.0 &   37.1  & 1.469 & 1.470  & Ingest RGZL            & Flood-fill             & Visual inspection      & Y  & N  & Y \\
3C19     &  ILTJ004054.99+331007.2 &    6.2 &   15.9  & 0.482 & 0.420* & Create initial sources & Gaussian               & LR                     & Y  & N  & N \\
3C28     &  ILTJ005550.31+262434.4 &   45.6 &   55.1  & 0.195 & 0.195  & Ingest RGZL            & Flood-fill             & Visual inspection      & Y  & Y  & Y \\
3C31     &  ILTJ010726.84+322439.4 & 2700.0 & 2262.5  & 0.017 & --     & Too zoomed in          & Composite              & Visual inspection      & N  & Y  & N \\
3C34     &  ILTJ011018.65+314719.7 &   49.0 &   60.3  & 0.689 & 0.482* & Too zoomed in          & Flood-fill             & Visual inspection      & Y  & Y  & N \\
3C42     &  ILTJ012830.25+290259.3 &   29.0 &   44.5  & 0.395 & 0.396  & Ingest RGZL            & Flood-fill             & Ridge line code        & Y  & N  & Y \\
3C43     &  ILTJ012959.80+233820.9 &    1.3 &    5.3* & 1.470 & 1.465  & Create initial sources & Gaussian               & LR                     & Y  & -- & Y \\
3C47     &  ILTJ013624.29+205720.2 &   77.0 &   85.3  & 0.425 & 0.263* & Ingest RGZL            & Flood-fill             & Visual inspection      & Y  & Y  & N \\
3C55**   &  ILTJ015710.68+285139.3 &   72.0 &  126.7  & 0.735 & 0.892* & Too zoomed in          & Flood-fill             & Visual inspection      & Y  & N  & N \\
3C67     &  ILTJ022412.27+275011.7 &    3.0 &    7.6* & 0.310 & --     & Create initial sources & Gaussian               & LR                     & Y  & -- & N \\
3C68.2*  &  ILTJ023423.87+313417.1 &   30.0 &   37.0  & 1.575 & --     & Create initial sources & Flood-fill             & LR                     & Y  & Y  & N \\
3C186    &  ILTJ074417.47+375317.4 &    2.5 &    7.0* & 1.063 & --     & Create initial sources & Gaussian               & LR                     & Y  & -- & N \\
3C196    &  ILTJ081336.06+481302.2 &    6.0 &   18.1  & 0.871 & 0.870  & Too zoomed in          & Composite              & Visual inspection      & Y  & N  & Y \\
3C200    &  ILTJ082725.43+291845.2 &   24.5 &   38.6  & 0.458 & 0.456  & Too zoomed in          & Flood-fill             & Visual inspection      & Y  & N  & Y \\
3C204    &  ILTJ083744.99+651335.2 &   37.0 &   48.6  & 1.112 & --     & Too zoomed in          & Flood-fill             & Visual inspection      & Y  & N  & N \\
3C205    &  ILTJ083906.53+575414.0 &   19.0 &   31.3  & 1.534 & --     & Too zoomed in          & Flood-fill             & Visual inspection      & Y  & N  & N \\
3C217    &  ILTJ090850.67+374819.2 &   14.0 &   27.5  & 0.897 & 0.763* & Too zoomed in          & Composite              & Visual inspection      & Y  & N  & N \\
3C216    &  ILTJ090933.49+425346.6 &    5.3 &    7.6* & 0.668 & --     & Create initial sources & Gaussian               & LR                     & Y  & -- & N \\
3C219    &  ILTJ092108.34+453858.4 &  190.0 &  201.9  & 0.174 & --     & Ingest RGZL            & Composite              & Visual inspection      & N  & Y  & N \\
3C234    &  ILTJ100148.66+284708.3 &  112.0 &  197.8  & 0.185 & 0.503* & Too zoomed in          & Composite              & Visual inspection      & Y  & N  & N \\
3C236    &  ILTJ100615.47+345221.7 & 2478.0 & 2405.3  & 0.099 & 0.099  & Too zoomed in          & Composite              & Visual inspection      & N  & Y  & Y \\
3C239    &  ILTJ101145.45+462819.8 &   13.5 &   19.7  & 1.781 & 1.223* & Create initial sources & Gaussian               & LR                     & Y  & N  & N \\
3C244.1  &  ILTJ103333.94+581436.0 &   51.0 &   67.0  & 0.428 & 0.429  & Ingest RGZL            & Flood-fill             & Visual inspection      & Y  & N  & Y \\
3C247    &  ILTJ105858.75+430123.4 &   14.6 &   28.2  & 0.749 & --     & Ingest RGZL            & Gaussian               & Ridge line code        & Y  & N  & N \\
3C252    &  ILTJ111132.28+354044.3 &   57.0 &  125.9  & 1.105 & 0.938* & Too zoomed in          & Composite              & Visual inspection      & N  & N  & N \\
3C254    &  ILTJ111438.56+403719.8 &   15.0 &   29.5  & 0.734 & 0.857* & Too zoomed in          & Flood-fill             & Visual inspection      & Y  & N  & N \\
3C263*   &  ILTJ113957.80+654748.2 &   51.0 &   79.7  & 0.652 & --     & Ingest RGZL            & Gaussian               & Visual inspection      & Y  & N  & N \\
3C265    &  ILTJ114529.19+313344.0 &   79.0 &   90.7  & 0.811 & 0.319* & Too zoomed in          & Flood-fill             & Ridge line code        & Y  & Y  & N \\
3C266    &  ILTJ114543.38+494608.1 &    5.5 &   13.2  & 1.272 & 0.926* & Create initial sources & Gaussian               & LR                     & Y  & N  & N \\
3C268.3  &  ILTJ120624.71+641336.8 &    1.3 &    6.6* & 0.371 & 0.372  & Create initial sources & Gaussian               & LR                     & Y  & -- & Y \\
3C268.4  &  ILTJ120913.61+433919.3 &   10.4 &   22.7  & 1.400 & --     & Create initial sources & Gaussian               & LR                     & Y  & N  & N \\
3C270.1  &  ILTJ122033.80+334310.2 &   11.0 &   23.7  & 1.519 & 1.209* & Ingest RGZL            & Flood-fill             & Visual inspection      & Y  & N  & N \\
3C280    &  ILTJ125657.50+472020.2 &   13.7 &   32.1  & 0.996 & 0.954* & Ingest RGZL            & Flood-fill             & Visual inspection      & Y  & N  & Y \\
3C284    &  ILTJ131104.39+272807.6 &  178.1 &  317.6  & 0.239 & 0.240  & Too zoomed in          & Composite              & Visual inspection      & Y  & N  & Y \\
3C285    &  ILTJ132120.00+423513.1 &  180.0 &  193.7  & 0.079 & 0.079  & Ingest RGZL            & Composite              & Visual inspection      & N  & Y  & Y \\
3C287    &  ILTJ133037.69+250910.9 &    0.1 &    4.7* & 1.055 & --     & Create initial sources & Gaussian               & LR                     & N  & -- & N \\
3C286    &  ILTJ133108.27+303032.8 &    4.0 &    6.4* & 0.849 & 0.850  & Create initial sources & Gaussian               & LR                     & N  & -- & Y \\
3C288    &  ILTJ133849.67+385111.3 &   36.2 &   39.0  & 0.246 & --     & Ingest RGZL            & Flood-fill             & Ridge line code        & Y  & Y  & N \\
3C289    &  ILTJ134526.38+494632.4 &   11.8 &   20.3  & 0.967 & 0.848* & Create initial sources & Gaussian               & LR                     & Y  & N  & N \\
3C292    &  ILTJ135042.00+642931.6 &  140.0 &  148.5  & 0.710 & --     & Ingest RGZL            & Flood-fill             & Visual inspection      & N  & Y  & N \\
3C293    &  ILTJ135216.93+312655.3 &  256.0 &  271.6  & 0.045 & 0.045  & Too zoomed in          & Composite              & Visual inspection      & N  & Y  & Y \\
3C294**  &  ILTJ140644.03+341125.0 &   16.2 &   32.9  & 1.786 & --     & Ingest RGZL            & Flood-fill             & Visual inspection      & Y  & N  & N \\
3C295    &  ILTJ141120.58+521208.4 &    6.0 &   12.8  & 0.461 & 0.462  & Create initial sources & Gaussian               & LR                     & N  & N  & Y \\
3C299    &  ILTJ142105.83+414449.6 &   11.3 &   11.9* & 0.367 & --     & Create initial sources & Gaussian               & LR                     & Y  & -- & N \\
3C303    &  ILTJ144301.55+520137.5 &   47.0 &   47.1  & 0.141 & 0.141  & Ingest RGZL            & Flood-fill             & Visual inspection      & Y  & Y  & Y \\
3C305    &  ILTJ144921.73+631614.1 &   13.6 &   11.2  & 0.042 & 0.042  & Ingest RGZL            & Gaussian               & Visual inspection      & Y  & N  & Y \\
3C319    &  ILTJ152405.35+542813.8 &  105.2 &  114.9  & 0.192 & 0.188* & Too zoomed in          & Flood-fill             & Visual inspection      & Y  & Y  & Y \\
3C322    &  ILTJ153501.20+553649.2 &   37.0 &   49.9  & 1.681 & 1.459* & Ingest RGZL            & Flood-fill             & Ridge line code        & Y  & N  & N \\
3C325    &  ILTJ154958.52+624121.2 &   17.5 &   33.2  & 0.860 & --     & Too zoomed in          & Flood-fill             & Visual inspection      & Y  & N  & N \\
3C330    &  ILTJ160935.79+655640.3 &   60.0 &   85.7  & 0.549 & 0.366* & Deduplicate            & Flood-fill             & LR                     & Y  & N  & N \\
NGC6109  &  ILTJ161734.28+350206.5 &  890.0 &  847.7  & 0.030 & 0.030  & Too zoomed in          & Manual                 & Visual inspection      & N  & Y  & Y \\
3C338    &  ILTJ162839.06+393259.1 &  117.0 &  150.3  & 0.030 & --     & Too zoomed in          & Flood-fill             & Visual inspection      & N  & N  & N \\
3C337**  &  ILTJ162852.85+441904.8 &   45.5 &   56.0  & 0.635 & --     & Too zoomed in          & Flood-fill             & Visual inspection      & Y  & Y  & N \\
3C343    &  ILTJ163433.80+624535.9 &    0.2 &    6.6* & 0.988 & 0.445* & Create initial sources & Gaussian               & LR                     & Y  & -- & N \\
3C343.1  &  ILTJ163828.20+623444.3 &    0.2 &    4.8* & 0.750 & 0.484* & Create initial sources & Gaussian               & LR                     & Y  & -- & N \\
3C345    &  ILTJ164258.70+394837.4 &   20.0 &   15.5  & 0.594 & 0.593  & Create initial sources & Gaussian               & LR                     & Y  & N  & Y \\
3C351    &  ILTJ170442.40+604445.0 &   74.0 &   43.0  & 0.371 & --     & Too zoomed in          & Flood-fill             & Visual inspection      & Y  & N  & N \\
3C352    &  ILTJ171044.08+460129.8 &   15.0 &   22.5  & 0.806 & --     & Create initial sources & Gaussian               & LR                     & Y  & N  & N \\
3C441    &  ILTJ220604.96+292919.4 &   36.5 &   47.9  & 0.708 & 0.686* & Ingest RGZL            & Flood-fill             & Ridge line code        & Y  & N  & Y \\
3C454    &  ILTJ225134.74+184840.4 &    1.2 &    8.1* & 1.757 & 1.763  & Create initial sources & Gaussian               & LR                     & Y  & -- & Y \\
3C457    &  ILTJ231207.36+184533.3 &  210.0 &  215.4  & 0.428 & 0.427  & Ingest RGZL            & Flood-fill             & Ridge line code        & Y  & Y  & Y \\
3C465    &  ILTJ233832.62+265822.5 &  603.0 &  632.6  & 0.029 & --     & Too zoomed in          & Manual                 & Visual inspection      & N  & Y  & N \\
\hline
  \end{tabular}
  \vspace{6pt}
  
  Notes: In column 1, a star next to the name denotes that the LoTSS
  source has no optical ID. Two stars indicate that
  the LoTSS catalogue has the wrong optical ID for the source,
  compared to 3CRR. In column 4, a star indicates that the source is
  not resolved in the LoTSS catalogue and so no accurate size
  measurement is available. In column 6, a star indicates a
  photometric redshift, otherwise the LoTSS redshift is spectroscopic. In columns 10 and 11, the flux density and largest
  angular size are said to match if they agree to within 20\% of the
  3CRR catalogue value. In column 12, the redshift is said to match if the
  redshifts agree to within 10\%.
\end{table*}

There are few large fully optically identified radio catalogues in the
northern sky with which we can compare our new catalogue. One
instructive comparison is with the flux-complete 3CRR catalogue \citep{Laing+83} which
includes full optical identifications and spectroscopic redshifts.
Largest angular size (LAS) measurements from high-resolution radio maps are also
available\footnote{We use the compilation of data at
\url{https://3crr.extragalactic.info/}.}. Because the 3CRR sources are
selected to have a flux density $>10.9$ Jy at 178 MHz (on the scale of
\citealt{Roger+73}) they should all be detected by LoTSS: they are
typically large, bright sources and so we would expect
(Fig.\ \ref{fig:fractional_id}) that many of them will
have been associated and identified by visual inspection.

There are 62 3CRR sources in our sky area (Table \ref{tab:3crr}) and
all can be identified in the radio catalogue. We crossmatched by first
searching for an optical ID matching the 3CRR position within 5
arcsec, and secondly looking for bright ($>10$ Jy) sources close to
the 3CRR catalogued radio position in the LoTSS catalogue. Of the
matches, two have no optical ID in the LoTSS catalogue (these are the
high-$z$ source 3C\,68.2 where an ID might very well not have been
detectable given our data, and the quasar 3C\,263 where presumably the
host was mistaken for a star by some volunteers in RGZ(L)) and three
have the wrong ID, all from visual inspection. Given that the
optical IDs for the 3CRR sources benefit from high-resolution,
high-frequency observations, a correct optical ID fraction of 57/62
(92\%) is good; it is noteworthy that all eight IDs derived from the ridge
line code are correct. The flux density in LoTSS matches with the
extrapolation of the 3CRR flux density to 144 MHz to within 20\% in
49/62 (79\%) of cases: the sources where there is not a good match
tend to be large sources where presumably some components of the radio
source were either not detected by {\sc pybdsf} or were not correctly
associated. Only a minority of sources (19/62) have LAS measurements
in the LoTSS catalogue that match the 3CRR values to within 20\%. This is partly
because some (11) 3CRR objects are not resolved by LoTSS, but
generally the LoTSS sizes, while being correlated with the 3CRR ones,
tend to be systematically higher. Reasons for this will include the
lower resolution of LoTSS compared to the VLA maps used to measure the
3CRR sizes, which tends to make flood-fill sizes an overestimate,
issues with the composite source size discussed in Section
\ref{sec:angsize}, and possibly in some cases some physical effect
where more extended emission is seen at low frequencies.

The LoTSS catalogue includes a redshift accurate to 10\% in only 23/62
cases, almost all spectroscopic redshifts from SDSS. The redshift is
clearly not expected to be correct in the 5/62 sources that have no or
the wrong optical ID, In some cases we have no redshift at all in
LoTSS (18/62) --- many at high $z$ where Legacy photometry may not be
available, but also including low-$z$ sources like 3C\,31, 3C\,338, and
3C\,465 where we might have expected to have a SDSS spectroscopic
redshift\footnote{3C\,31 and 3C\,465's hosts are simply missing from the SDSS main
galaxy sample, presumably due to the existence of close companions
which prevented a fibre being placed on the galaxies \citep{Strauss+02}.
3C\,338's host position in the catalogue is 2.2 arcsec away from the
corresponding SDSS catalogue position and therefore it is not picked
up by our 1.5-arcsec crossmatch.}. 16/62 sources have an inaccurate
photometric redshift, failing to match within 10\%. The photometric
redshifts are only badly wrong in a few cases (the worst is 3C\,265
where the true redshift is 0.811 and the photo-$z$ 0.319), and the
3CRR sources contain a large fraction of quasars, as well as galaxies
with extremely strong emission lines, where photometric redshifts are
likely to be more challenging. Nevertheless this does illustrate the
value of a targeted spectroscopic survey of the LoTSS sources, as will
be provided by the WEAVE-LOFAR project.

Finally, we confirmed that there are no sources in the LoTSS catalogue
that should have been in the 3CRR catalogue but are not, either
because of errors in the 3CRR selection or in our association. There are
a number of unmatched sources in the overlapping sky area with 144-MHz
flux densities $>12$ Jy but all are 4C sources with catalogued 178-MHz flux
density just below the 3CRR cutoff.

\begin{figure*}
\includegraphics[width=\columnwidth]{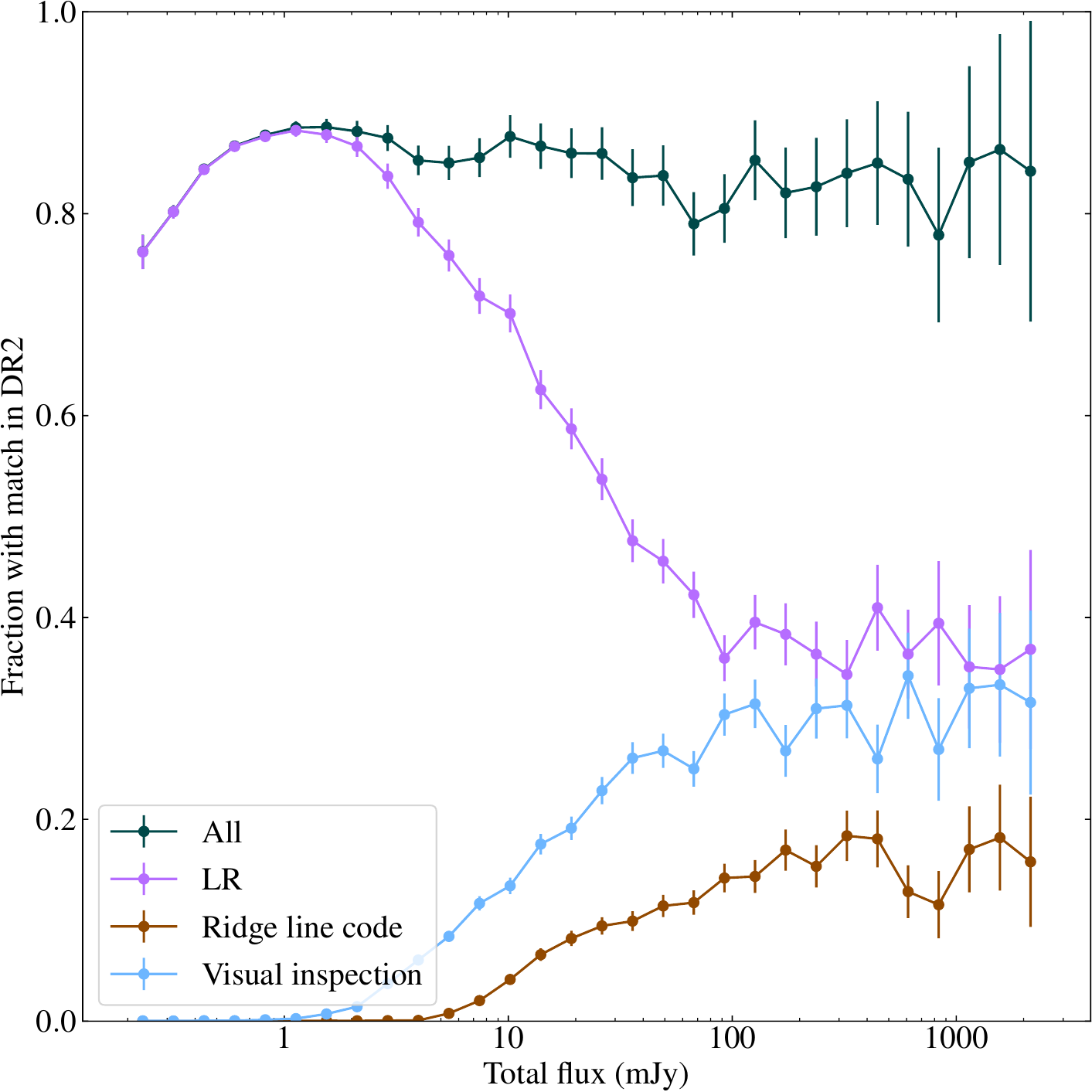}
\includegraphics[width=\columnwidth]{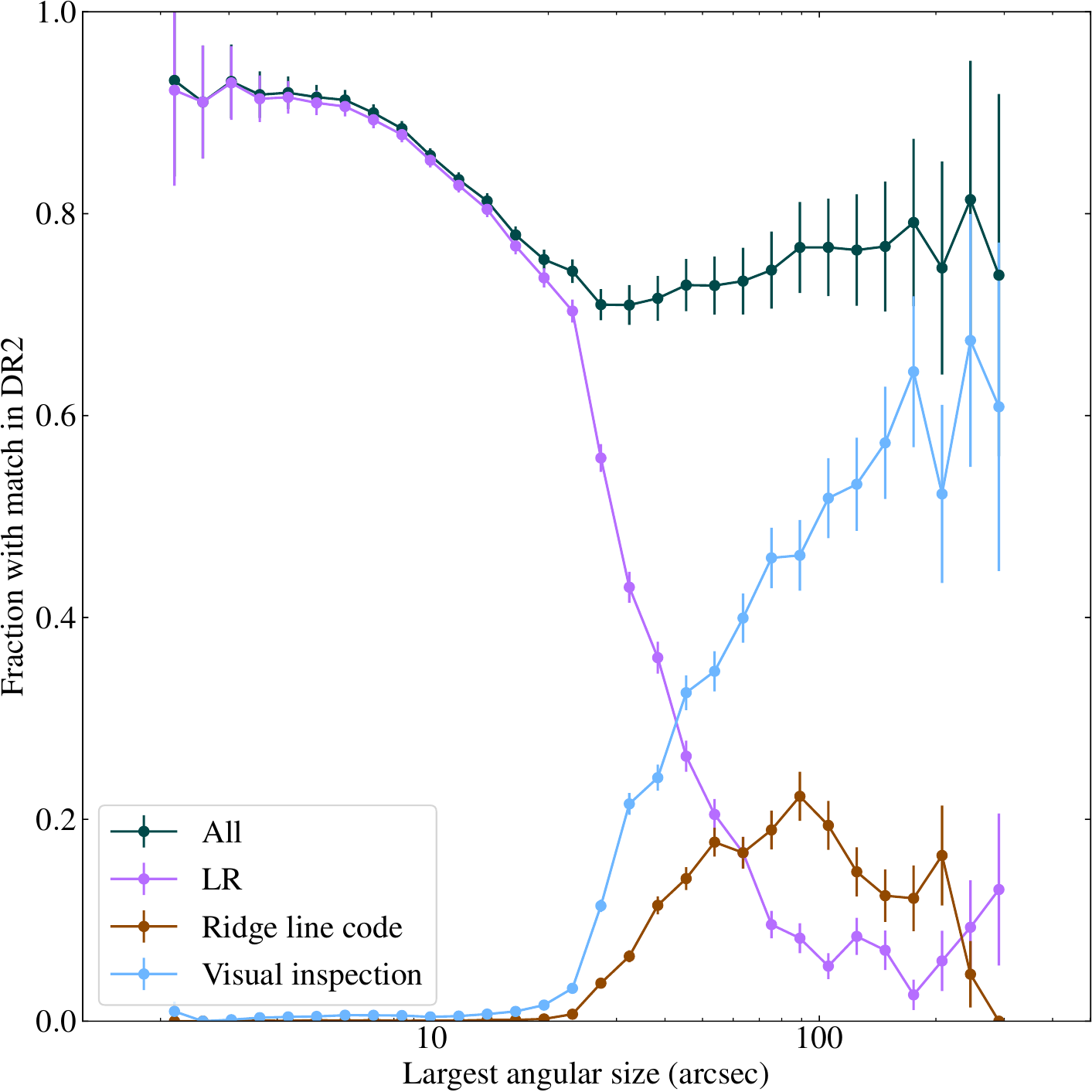}
\caption{As Fig.\ \ref{fig:fractional_id} but here the fraction of LoTSS DR1
objects that have an optical ID matching the one in DR2 are shown
broken down by their origin in DR2. Flux and angular size values here
are from DR1.}
\label{fig:fractional_id_dr12}
\end{figure*}

A further useful quality comparison is with the sources in LoTSS DR1
\citep{Williams+19}. Compared to DR2, DR1 benefited from the first
stage of visual inspection for association and identification being
done by astronomers who were able to inspect a wider range of data
(including {\it WISE} images and FIRST contours) but relied on poorer
LOFAR images with a higher noise level and used shallower PanSTARRS
optical data. As noted above, the rate of optical identification is
substantially higher in DR2. We matched DR1 and DR2 as closely as
possible by restricting a comparison to sky areas where the density of
DR1 sources is $>500$ deg$^{-2}$, which gave a matching area of 353
deg$^2$. DR1 includes 291,758 sources in this sky area, while DR2 has
401,890: the optical ID fraction is 73.0\% in DR1 and 86.8\% in DR2,
while the fraction of these IDs with redshift estimates is 70.0\% in
both datasets. Of the 212,949 sources in DR1 with IDs, 183,064
(86.0\%) have an optical positional crossmatch within 1.5 arcsec with
the IDs of DR2 sources, and these are overwhelmingly clearly the same
LOFAR source when their DR1 and DR2 total flux density is compared --
they show no obvious difference on a scatter plot comparing the total
flux densities from sources that are simply crossmatched in radio
position (of which there are 267,368, or 91.6\% of DR1, within a 3
arcsec match radius). Fig.\ \ref{fig:fractional_id_dr12} shows that
the match fraction is lowest for faint and large sources, and best for
bright and compact ones, as expected since we would hope that the LR
algorithm would select the same sources in both DR1 and DR2. There is
no evidence, comparing Fig.\ \ref{fig:fractional_id} and
Fig.\ \ref{fig:fractional_id_dr12}, that any of our optical ID methods
from DR2 performs noticeably better or worse relative to DR1 than the
others. There is no significant dip in the matching identification
fraction at flux levels of a few mJy, suggesting that these sources
are not any worse identified in DR2 than in DR1. The non-matching
sources are also not uniformly distributed on the sky, which may
suggest that per-mosaic astrometric uncertainties or a
position-dependent higher fraction of spurious sources in one or other
catalogue are responsible for some of the unmatched sources. Overall
we view the good agreement between the two independently identified
catalogues as positive, but the discrepancies illustrate that we have
not yet converged on a process that gives identical optical IDs for a
given region of radio and optical sky.

We cross-checked the optical IDs in our catalogue against those
derived by \cite{OSullivan+23} in their study of 2,461 polarized sources in
LoTSS DR2 as part of the Magnetism Key Science Project (MKSP): their
IDs were derived using a separate private Zooniverse project using the
same radio data as us, but carried out by astronomers rather than the
public, and using both Legacy and {\it WISE} data for optical IDs. The
polarized sources on which the catalogue is based are bright and often
extended relative to DR2 sources as a whole. \cite{OSullivan+23}
obtained an 88\% optical ID rate, similar to ours overall, but only 76\%
of their sources with optical IDs have the same IDs in our catalogue.
The sizes and flux densities of the MKSP sources place them in the
regime where we have the lowest optical ID rates
(Fig.\ \ref{fig:fractional_id}) and so the discrepancy is not
surprising: as noted above, professional astronomers seem to give
significantly higher optical ID rates than citizen scientists for
extended sources, and {\it WISE} images are often better than the
Legacy survey for high-$z$ host galaxies. This is a further indication
that it might be possible to obtain more IDs with more targeted visual
inspection, although at considerably increased cost in time.

Finally, we compared with the results for the first data release (DR1)
of the original Radio Galaxy Zoo (RGZ) project \citep{Wong+23}, which
provides a catalogue of 99,624 sources derived from the FIRST survey,
of which 56\% have a {\it WISE} counterpart ID derived from visual
inspection by citizen scientists. Taking the overlapping sky area (all
in the `Spring' field of LoTSS DR2) and cutting regions where there is
a low density of LoTSS DR2 sources to avoid edge effects, we have
around 3,000 deg$^2$ in common, with our catalogue containing
2,787,742 sources while the RGZ DR1 catalogue contains 40,690. Of the
23,964 sources from the RGZ DR1 sample that have {\it WISE} positional
IDs, 20,411 (85.2\%) have a match to an optical position ID in the
LoTSS DR2 optical catalogue within 3 arcsec --- these are
overwhelmingly true matches as can be verified from comparing their
{\it WISE} magnitudes and radio properties. As with the LoTSS DR1
comparison, this gives confidence in our catalogue, since our hybrid
process involving LR matching, heuristics and visual inspection is
giving results comparable in quality to a pure visual inspection
approach. \cite{Wong+23} estimate the reliability of the RGZ DR1
catalogue to be $\sim 70$--80\%, so our agreement here is as good as
would be expected, bearing in mind that for some of these sources RGZ
and our catalogue may agree on a common but incorrect source ID. It is
interesting to note that the raw ID fraction of RGZ is significantly
higher than for our Zooniverse results (Section \ref{sec:zooniverse}):
we speculate that this is partly due to the RGZ input catalogue being
composed of simpler, brighter radio sources derived from FIRST, and
partly due to the use of {\it WISE} for the optical identification.

\subsection{Properties of the sources with optical IDs}

\begin{figure*}
  \includegraphics[width=1.0\columnwidth]{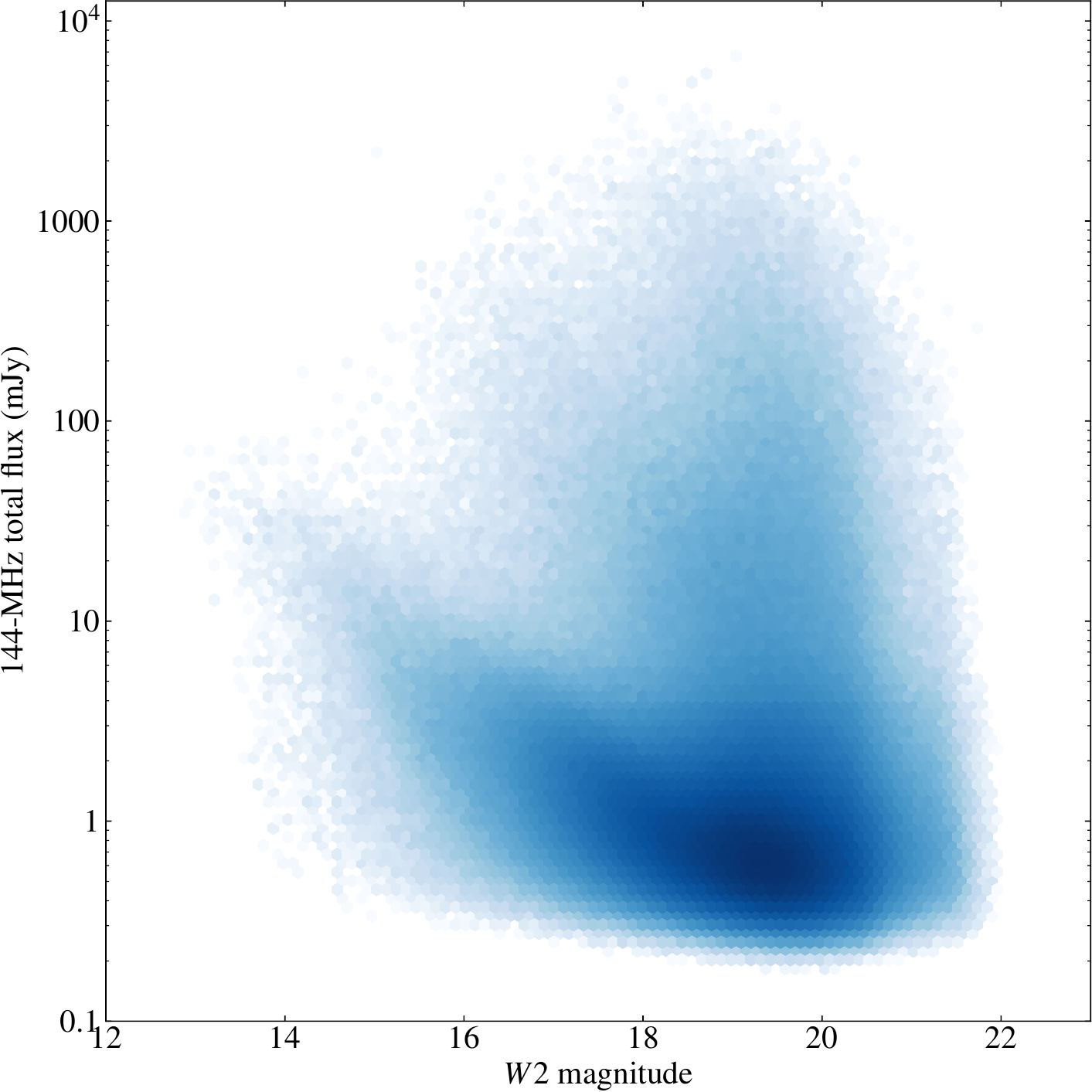}
  \includegraphics[width=1.0\columnwidth]{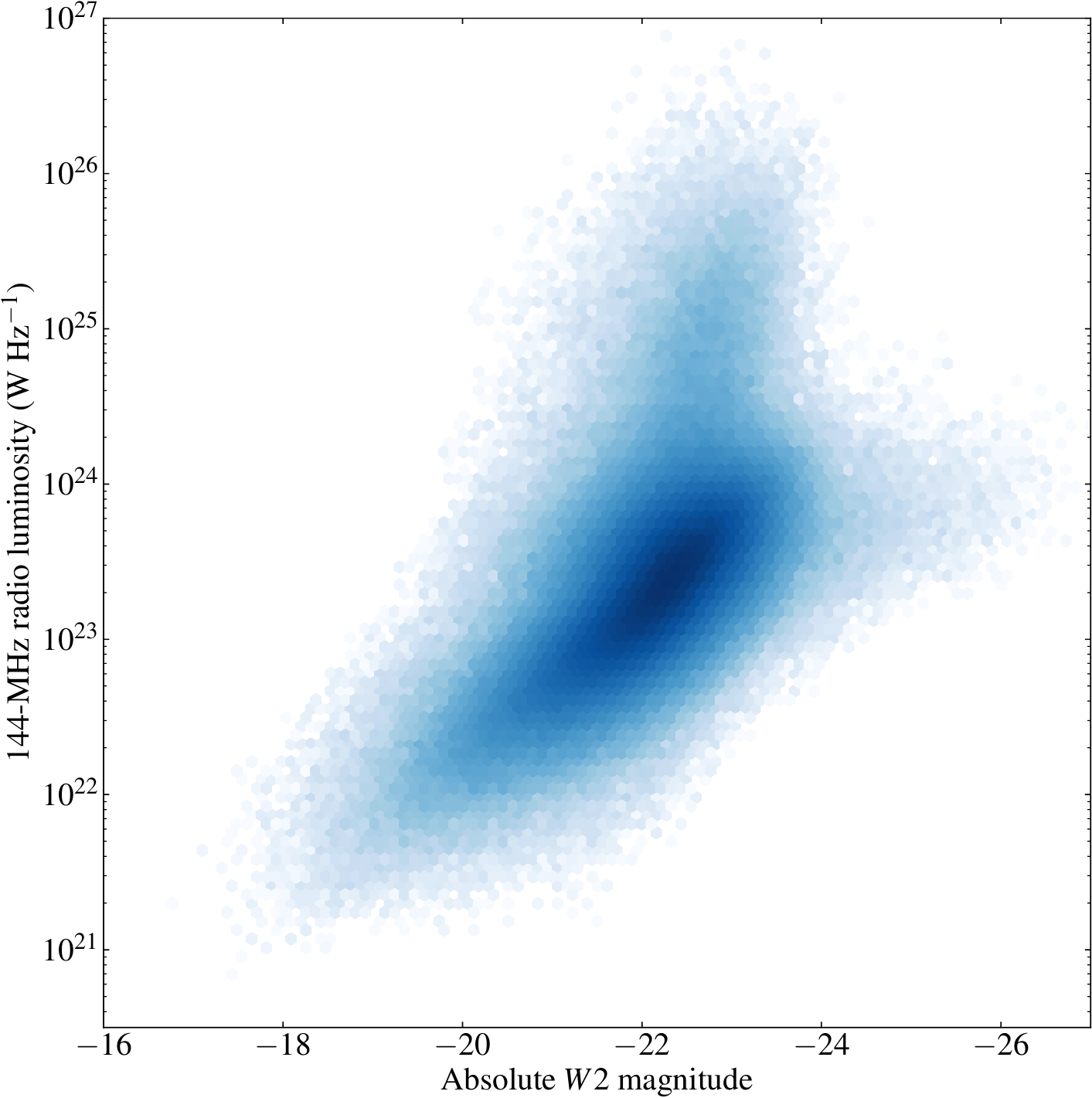}
  \caption{Relations between radio and optical properties of objects
    in the catalogue. Left: logarithmic density plot of 3,226,797 {\it
      WISE}-detected DR2 sources
    showing total LOFAR flux
    density against {\it WISE} band 2 AB magnitude. Right: logarithmic
    density plot of 851,356 DR2 sources with good {\it WISE}
    magnitudes and $z_{\rm best} <0.5$ showing total 144-MHz radio
    luminosity against {\it WISE} band 2 absolute AB magnitude, with
    approximate K-correction using the $W1$-$W2$ colour.}
  \label{fig:teapot}
\end{figure*}

Fig. \ref{fig:teapot} (left panel) shows an example of the relation
between optical or IR apparent magnitude and 144-MHz flux density for the
nearly three million sources with usable photometry in {\it WISE} bands 1
and 2. This `teapot plot' (which has a counterpart in the far-IR, e.g.
\citealt{Hardcastle+16}) exhibits two distinct branches, one which
shows a clear and close to linear correlation between radio and mid-IR
flux for bright galaxies (due to star-forming galaxies on the main
sequence of star formation) and one branch with brighter radio sources
and fainter IR galaxies, with no clear relationship between radio flux
and IR properties, which represents the AGN population. (A less
clearly defined branch to brighter magnitudes above flux densities of
10 mJy represents the quasar population.) These relationships would
not appear in a flux-flux plot unless the bulk of our optical
identifications were correct. Using the good redshifts available for a
subset of the sample, we can see the same relation in physical
quantities in the right-hand panel of Fig.\ \ref{fig:teapot}, where
the main sequence of star formation is seen as a diagonal line with a
plume of luminous points above it representing the RLAGN
population: radio-quiet quasars occupy the right-hand side of the
plot. The relatively narrow optical magnitude range occupied by RLAGN is
a consequence of the fact that they are much more common in the most
massive galaxies \citep[e.g.][]{Sabater+19}. The relation between radio
luminosity and absolute magnitude in this plot appears quite
tight (with around half a decade of scatter) and persists over $\sim
5$ magnitudes. Care would need
to be taken in selecting RLAGN using this plot alone, although it is
clear that a luminosity cut $>10^{25}$ W Hz$^{-1}$, as used in part by
\cite{Hardcastle+19}, would efficiently select true radio-excess AGN. We will return
to the question of RLAGN selection in this sample in a future paper.

\begin{figure*}
\includegraphics[width=0.915\columnwidth]{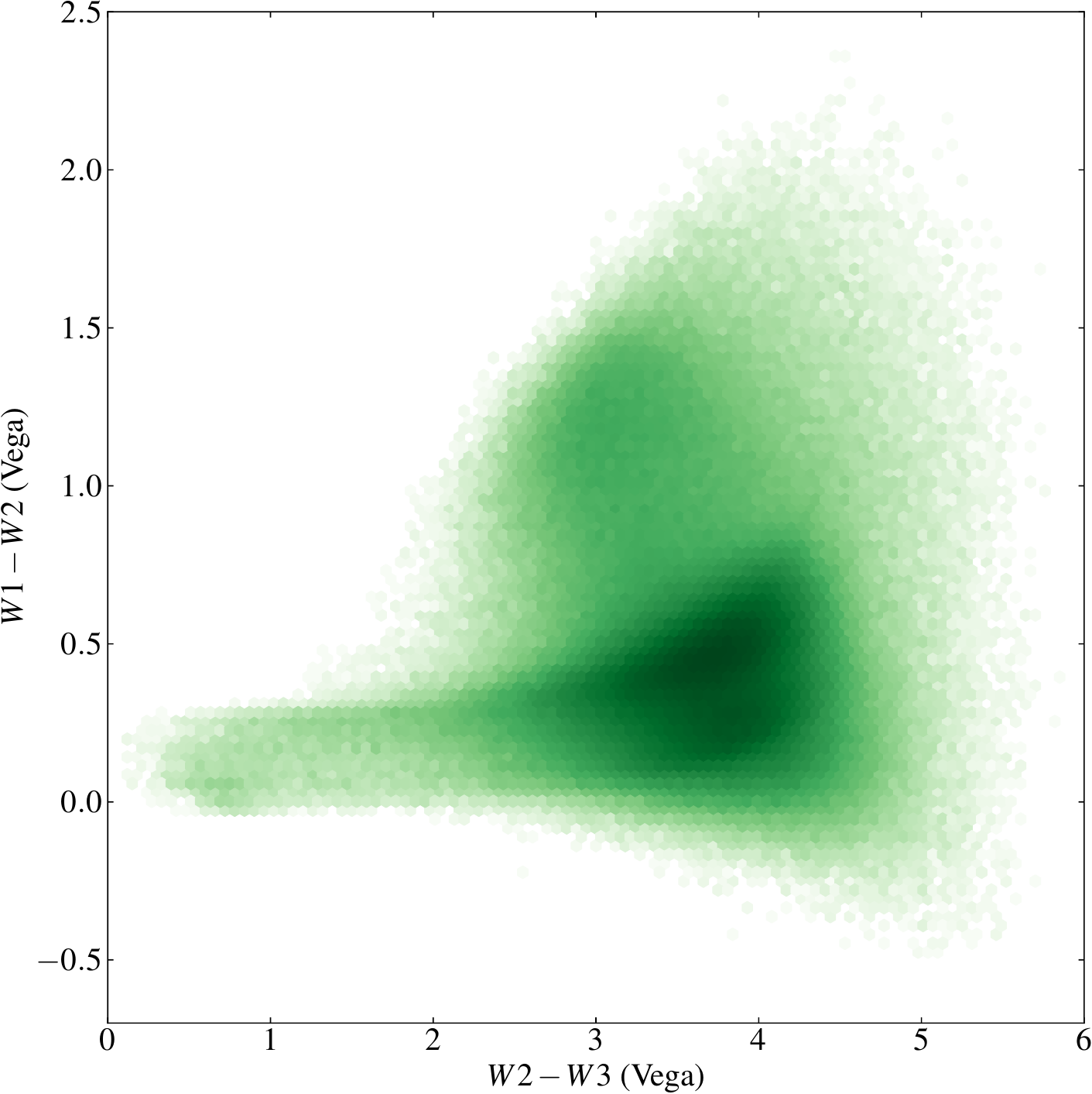}
\includegraphics[width=1.081\columnwidth]{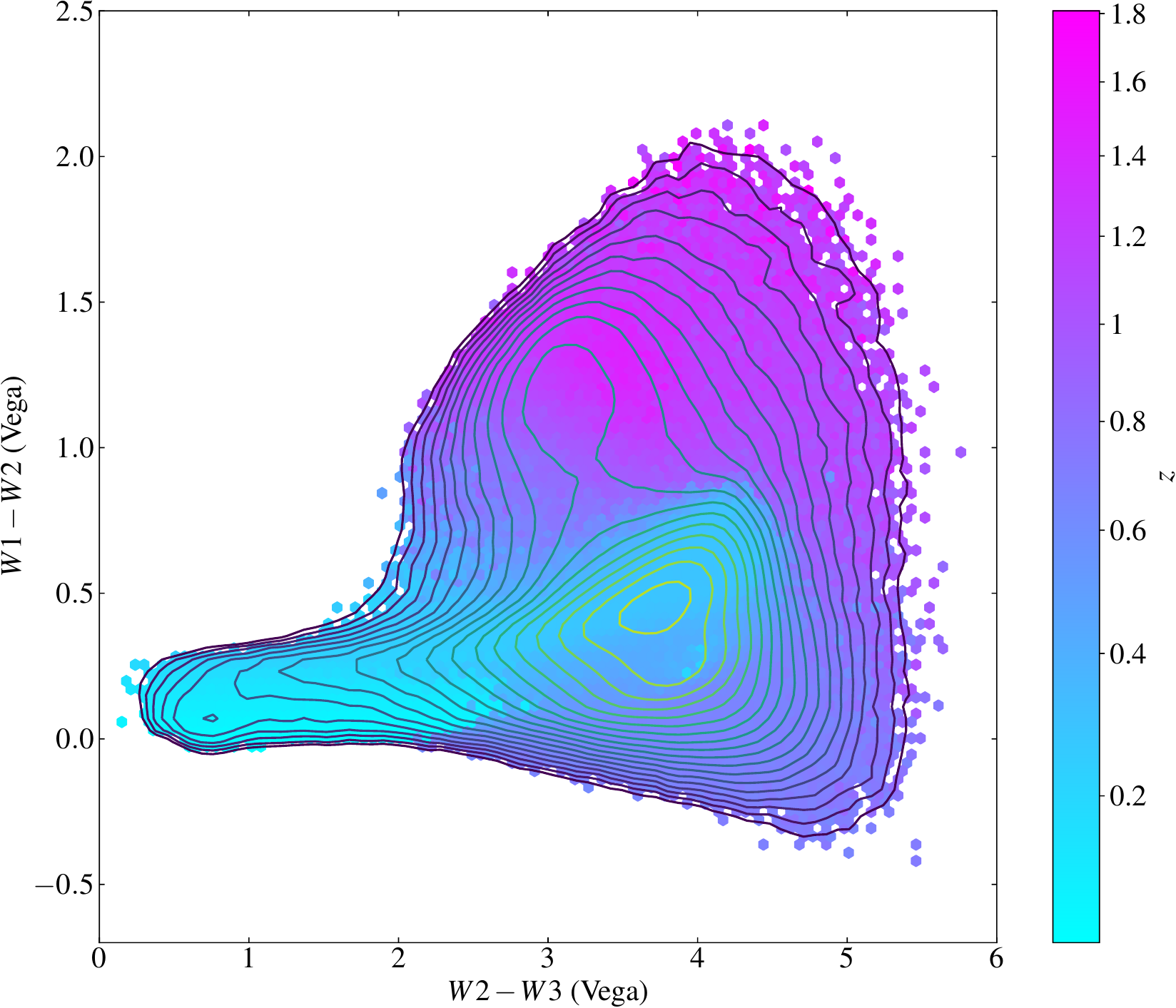}
\caption{Catalogued objects in the {\it WISE} color-colour space. Left: {\it WISE} colour-colour logarithmic density plot for 1,273,882 objects with
  good {\it WISE} photometry in the radio catalogue. Right: the same
  plot but showing the median redshift in each bin for the subset of 1,119,991 objects
with good {\it WISE} photometry and also well-defined $z_{\rm best} <
4$, overlaid with KDE-estimated logarithmic density contours.}
\label{fig:wisecc}
\end{figure*}

{\it WISE} colour-colour plots are widely used to classify optical
sources \citep[e.g.,][]{Assef+10,Stern+12,Gurkan+14}. Fig.\ \ref{fig:wisecc} shows the colour-colour plot for
1.3 million radio-source counterparts with good {\it WISE}
photometry (by which we mean sources that are detected and have magnitude errors
$<0.3$ in all three bands). Radio source counterparts are widespread across this plot
but normal galaxies occupy a curved locus with a relatively narrow
range of $W1-W2$ colours but considerable spread in $W2-W3$.
Star-forming galaxies, which lie on the main sequence lines in
Fig.\ \ref{fig:teapot}, are concentrated in a relatively small colour
space. Away from the normal galaxy locus, we see that the upper part
of the plot (with red $W1-W2$ colours) are mostly high-$z$ objects
and therefore largely quasars. Intermediate-$z$ objects lying below
the normal galaxy locus with blue $W1-W2$ colours are non-quasar AGN hosts.

\begin{figure*}
\includegraphics[width=0.915\columnwidth]{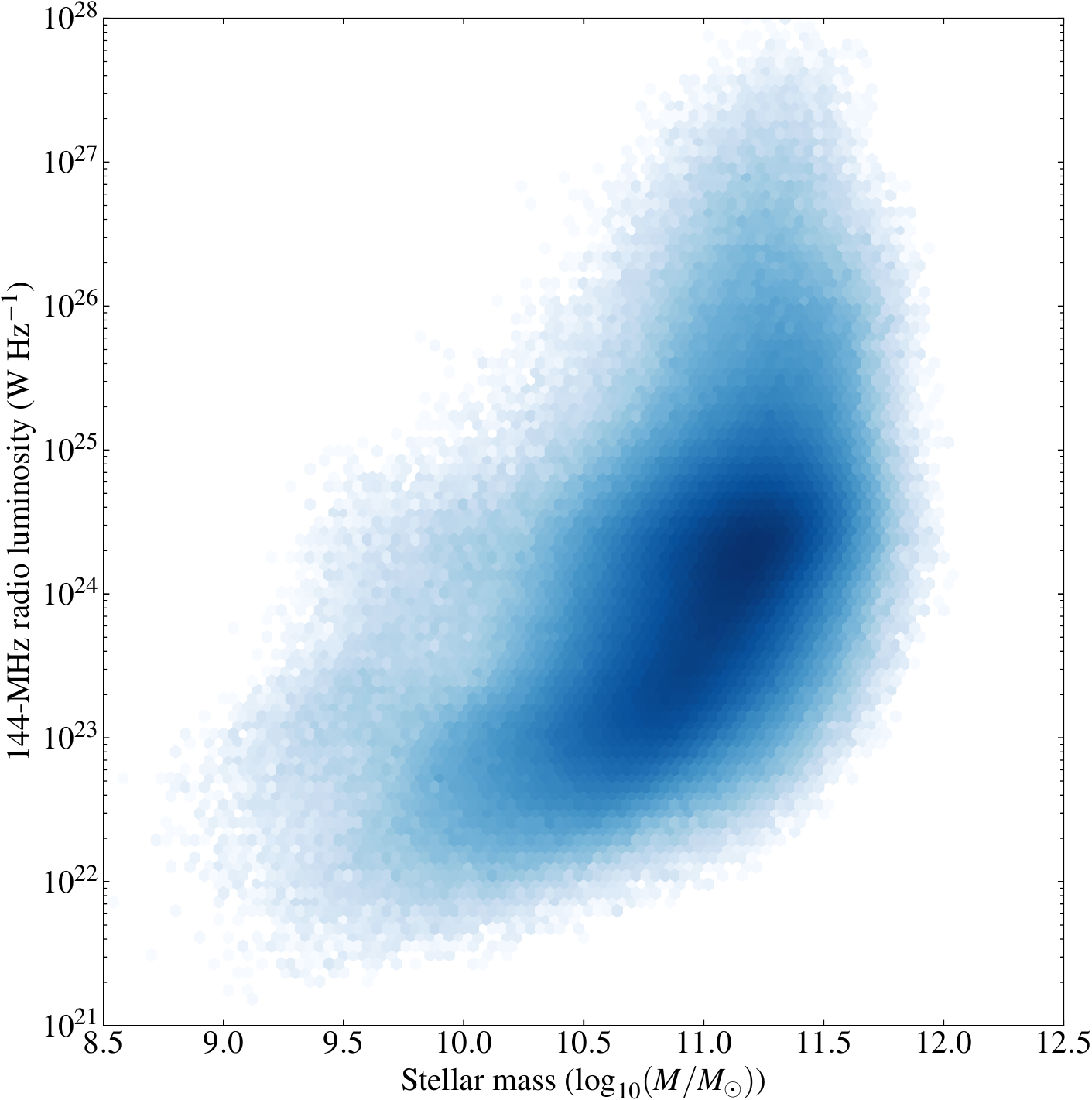}
\includegraphics[width=1.081\columnwidth]{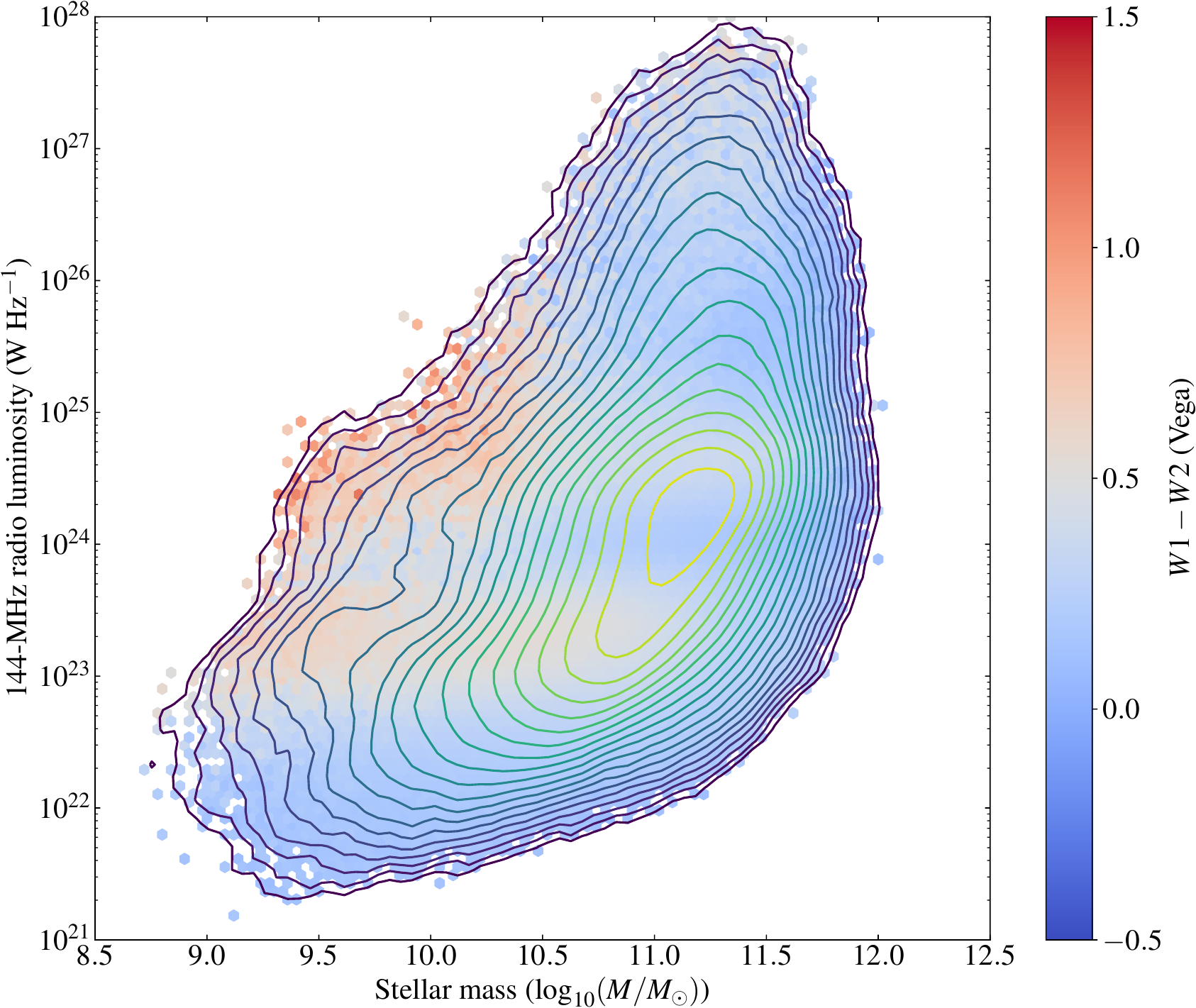}
\caption{The relationship between mass and radio luminosity for
  objects in the catalogue. Left: Logarithmic density plot of radio luminosity as a
  function of mass for 1,737,778
  radio sources with good mass estimates and usable $W1$ and $W2$ magnitudes. Right: the same plot but
  showing the median $W1-W2$ colours for each cell, overlaid with KDE-estimated logarithmic density contours.}
\label{fig:masslum}
\end{figure*}

Finally, our mass estimates allow us to look at the relationship
between physical quantities such as mass and radio luminosity.
Fig.\ \ref{fig:masslum} shows a plot of the relation between those two
for sources with mass estimates flagged as reliable in the catalogue. 
Again the main sequence of star formation can be seen as a
luminosity-mass relation in the lower part of the plot, while RLAGN
have radio luminosity independent of mass. A visible horizontal
scatter between luminosities of $10^{24}$ and $10^{25}$ W Hz$^{-1}$ is
the result of contamination by quasars, which do not have accurate
mass estimates, as can be seen by considering their {\it WISE}
colours, but overall this plot shows the expected behaviour and
we clearly have the statistics for more detailed studies of the
relationship between mass and radio properties in future papers.

\subsection{Extreme sources}

Another way of investigating the quality of the catalogue is to sort
by measured or inferred physical quantities to search for sources with extreme
properties, which could be present in error. The brightest radio sources in the
catalogue, as discussed above (Section \ref{sec:compare}) are the 3CRR
objects, and these are on the whole correctly identified with their
host galaxies. The largest sources in terms of angular size include
the degree-scale radio galaxy NGC 315, the giant radio galaxies
3C\,236 and 3C\,31, various other less well-known large FRI sources,
and the spiral galaxy M101: none of the largest ten objects in the
catalogue appear to be spurious associations, with the least plausible
being ILTJ010331.88+230426.1, a putative large FRII source. There are
89 sources in the catalogue with angular sizes $>10$ arcmin. Because
our catalogue is based on the 6-arcsec imaging, the images used for
visual inspection had limited sensitivity to extended structure, and
so we do not expect to see all the large sources found in visual inspection
of lower-resolution images \citep{Oei+22,Oei+23}.

Turning to physical quantities, the highest-redshift radio source in
our catalogue is at $z=6.6$: as discussed above, all sources with
$z>5$ come from the high-redshift quasar catalogue of
\cite{Gloudemans+22}. For objects with reliable redshifts we can look
at radio luminosity estimates. The most luminous object in our
catalogue is 3C\,196 at $z=0.870$ (Table \ref{tab:3crr}), followed by
the $z=3.03$ object ILTJ142921.88+540611.2 (6C B142744.1+541929), both
at around $2 \times 10^{29}$ W Hz$^{-1}$. In total there are 25
objects with radio luminosity $>10^{29}$ W Hz$^{-1}$. This is the
level that is reached by the most powerful 3CRR sources and
corresponds to jet powers around $10^{40}$ W \citep{Hardcastle+19}.
The vast majority of these powerful sources have redshifts that come
from the SDSS quasar catalogue and so are as reliable as the SDSS
redshifts: most are unresolved in the radio so their optical IDs are
not in doubt. It is noteworthy that there are none of these very
powerful sources at $z \ga 4$, presumably because the nature of AGN
accretion or environments and/or the very high radiative losses to
inverse-Compton emission prevent them from occurring, since we could
certainly detect them if they were optically identified.

Finally we can look at the physically largest sources. Our largest
object, ILTJ152932.89+601538.1, has a nominal size of 6.9 Mpc, though
we caution that this relies on a photometric redshift of $0.916$, a
slightly too large estimated angular size, and an uncertain
identification. However, even if identified instead with the $z=0.798$ galaxy
associated with ILTJ152933.05+601552.6 and given a hand-measured size
of 827 arcsec, its reduced computed size of 6.2 Mpc would still
make it the largest radio galaxy known to date. Including this object,
which will be discussed further by Oei et al (in prep), and
ILTJ110838.03+291731.4, which at 5.7 Mpc becomes the second largest
giant candidate discovered,
there are 13 candidate sources with projected size $>4$ Mpc, all of
which are convincing FRII radio galaxies on inspection, and four of
which have spectroscopic redshifts. However, the
optical IDs for these large angular size sources should be treated
with caution as there could be multiple candidate hosts for each
source. In total in the catalogue there are 8,541 sources with
estimated physical size $>700$ kpc, the standard threshold for a
`giant' radio source \citep{Machalski+06} in a modern cosmology, though
careful size measurements will be necessary to confirm whether they
meet this threshold value.

\subsection{Caveats and user advice}

There are a number of potential issues affecting the scientific
  use of a catalogues of optical IDs like this one. Here we outline a
  few points that users of the catalogue should be aware of.

The first and most obvious issue is that the catalogue is not
complete, in the technical sense that we do not have optical IDs for
all the radio sources in the catalogue; moreover, we do not have
redshifts for all the sources that have optical IDs. This limitation
comes primarily from the optical and IR data available: the optical ID
catalogues for the LoTSS deep fields \citep{Kondapally+21} demonstrate
that it is possible to get much closer to completeness, even for a
radio survey significantly deeper than DR2, if one has substantially
deeper optical and IR data than we have over the whole northern sky.
When using the wide-area catalogue, though, the incompleteness means
that one cannot, for example, select on radio properties such as radio
luminosity and be certain that one has selected all the sources that
physically should have been selected. Given that only 57\% of sources
have a good spectroscopic or photometric redshift, the bias
introduced by incompleteness could be substantial, though it is likely
to affect predominantly low-mass and/or high-redshift objects. So, for
example, it is reasonable to expect that the catalogue is close to complete for
low-$z$ massive galaxies, but the catalogue user needs to conduct
their own tests to quantify and account for the effects of this
incompleteness for their science use case. For example, standard completeness correction techniques
for the construction of luminosity functions for populations need to
take account of the non-trivial selection functions for both optical
ID and redshift incompleteness.

A more subtle issue is that the catalogue only lists objects that
  are detected in the original DR2 radio catalogue
  \citep{Shimwell+22}. Flux densities measured for detected objects
  should be reasonably secure, though it is important to consider the
  effects of detection incompleteness, Eddington bias and the possibility that a source
  might not be fully deconvolved at the faint end. Most of these
  issues can be avoided by applying a higher flux density cut to the
  catalogue, such as the 1.1 mJy flux density reported by
  \cite{Shimwell+22} to be the 95\% completeness limit. The flux density
  scale for DR2 is accurate to the 5--10\% level
  \citep{Hardcastle+21}. However, if a catalogue user wishes to
  measure the LoTSS maximum-likelihood flux density for a known pre-existing sample of optical
  objects, they should proceed in two stages. First they should
  cross-match their sample on optical position
  with the present catalogue, which will almost always give the best
  estimate of the radio properties of a given optical galaxy that
  appears here, including the effects of extended or multi-component
  sources. Secondly, they should return to the LoTSS images to
  estimate the flux measurements (or, if desired and appropriate to
  the analysis being conducted, upper limits) for objects that do not
  appear in this catalogue, which takes account both of the
  non-uniform noise in the LoTSS images and of sources that may be
  genuinely detected but are missing from the LoTSS radio catalogue.
  Neglecting the second step and considering only objects found in the
  present catalogue is likely to lead to a significantly
  biased analysis.

\section{Summary and conclusions}
\label{sec:summary}

We have found optical IDs and associations for 4.1
million radio sources in the LoTSS DR2 area. At more than
an order of magnitude larger than our previous work in DR1
\citep{Williams+19}, this is by far the
largest optically identified radio survey yet carried out. In addition to the extensive use of
likelihood-ratio (LR) cross-matching, including the ridge-line
analysis of \cite{Barkus+21}, we made use of $\sim 950,000$ visual
inspections by citizen scientist volunteers and $\sim 150,000$ by astronomers,
including filtering, too-zoomed-in, and blend workflows as well as the
internal Zooniverse project. We roughly estimate the human time cost
of these inspections, based on a notional 30 seconds per object and a
standard working pattern, at around six person-years.

We achieve an 85.0\% optical ID rate, and the science-ready catalogue
that we generate includes high-quality photometric redshifts for the
optical IDs, spectroscopic redshifts from SDSS and HETDEX where
possible, and, for the 58\% of sources with a good redshift estimate,
derived quantities including radio luminosity and physical size
estimates. Galaxy mass estimates are also provided as a by-product of
the photometric redshift process. A comparison with the bright,
extended sources in the 3CRR catalogue \citep{Laing+83} shows that the
quality of our optical identifications and redshift estimates is
generally good for this class of object. Followup with WEAVE-LOFAR
\citep{Smith+16} will obtain spectroscopic redshifts for most of the $\sim 330,000$ bright
sources in the sample with flux density $>8$ mJy, which may include
many high-$z$ radio galaxies. This is the first work to combine (at
scale) statistical, citizen science, and expert matching based on
homogeneous radio source extraction parameters and multi-wavelength
ancillary data, paving the way toward incorporating more advanced
matching techniques that will prove crucial to work using SKA and LSST
surveys.

The use of a citizen science project for work such as this, while
immensely rewarding to the participants and the science team alike, is time-consuming and, as discussed in Section
\ref{sec:zooniverse}, gives relatively low optical identification
rates which have to be supplemented by expert visual inspection and/or
additional algorithms. For the still larger task of generating optical
IDs for the remainder of the full LoTSS northern sky survey, and for
future surveys with the SKA, it will be essential to learn from the
results of this work. While human visual inspection seems hard to
avoid for the most complex sources, algorithms for association
\citep{Mostert+22} and optical identification \citep{Barkus+21} may
soon be able to deal with a much larger fraction of radio sources. The
associations and identifications that we have generated may be used to
train future generations of machine-learning algorithms.

Our publicly released catalogue should provide a resource for a vast
number of scientific projects based on the radio properties of active
and star-forming galaxies. We expect to make future releases of the
catalogues incorporating improved optical IDs, further spectroscopic
redshifts including those from HETDEX, WEAVE, and DESI, and
environmental and radio spectral information.

Although LoTSS is currently largely
generating images using only the Dutch baselines of LOFAR, with a
typical resolution of 6 arcsec, a stretch goal of the project is to
exploit the much higher resolution provided by the full International
LOFAR Telescope (ILT), which can be $\sim 0.3$ arcsec at 144 MHz
\citep{Morabito+21}, over large areas of the sky. Exploitation of all-sky high-resolution imaging, when available, should significantly improve the optical identification rate.
    
\bibliographystyle{aa}
\renewcommand{\refname}{REFERENCES}
\bibliography{mjh,kjd,cards}
\clearpage
\onecolumn
\appendix
\section{Table descriptions}
\label{sec:tables}
Table \ref{tab:sourcecat} gives a description of the columns in the
source catalogue and Table \ref{tab:compcat} gives a description of
the columns in the component catalogue.

\LTcapwidth=\textwidth

\begin{longtable}{lllp{11cm}}
\caption{Columns for the main catalogue. `Type' gives the Python data
  type and its length in bits.\label{tab:sourcecat}}\\
\hline\hline
Column name&Type&Units&Description\\
\hline
\endfirsthead
\caption{continued. Magnitudes are AB magnitudes. Notes: (1) {\tt type} is  "PSF"=stellar,
  "REX"="round exponential galaxy", "DEV"=deVauc, "EXP"=exponential,
  "COMP"=composite, "DUP"=Gaia source fit by different model; (2) {\tt
    flag$\_$qual} selects sources with reliable redshifts, with
  reasonable uncertainty, minimal contamination from nearby sources,
  low star-likelihood and free from imaging artefacts based on {\tt
    maskbits} (3) Non-blank {\tt z$\_$best} combines SDSS spec-$z$
  with reliable photo-$z$ (4) Cosmology is the standard cosmology for
  this paper. (5) See Tables \ref{tab:id_flags_final} and \ref{tab:id_flags_additional}.}\\
\hline\hline
Column name&Type&Units&Description\\
\hline
\endhead
\hline
\endfoot
{\tt Source$\_$Name}&bytes184&&Object identifier (ILT name)\\
{\tt RA}&float64&deg&Radio right ascension (mean position )\\
{\tt DEC}&float64&deg&Radio declination (mean position)\\
{\tt E$\_$RA}&float64&arcsec&Error on radio right ascension\\
{\tt E$\_$DEC}&float64&arcsec&Error on radio declination\\
{\tt Total$\_$flux}&float64&mJy&144-MHz total flux density\\
{\tt E$\_$Total$\_$flux}&float64&mJy&Error on total flux density\\
{\tt Peak$\_$flux}&float64&mJy/beam&144-MHz peak flux density\\
{\tt E$\_$Peak$\_$flux}&float64&mJy/beam&Error on peak flux density\\
{\tt S$\_$Code}&bytes8&&PyBDSF source code or Z for composite source\\
{\tt Mosaic$\_$ID}&bytes88&&LoTSS mosaic of source image\\
{\tt Maj}&float64&arcsec&Major axis of fitted Gaussian\\
{\tt Min}&float64&arcsec&Minor axis of fitted Gaussian\\
{\tt PA}&float64&deg&Position angle of fitted Gaussian\\
{\tt E$\_$Maj}&float64&arcsec&Error on major axis\\
{\tt E$\_$Min}&float64&arcsec&Error on minor axis\\
{\tt E$\_$PA}&float64&deg&Error on position angle\\
{\tt DC$\_$Maj}&float64&arcsec&Deconvolved major axis of fitted Gaussian\\
{\tt DC$\_$Min}&float64&arcsec&Deconvolved minor axis of fitted Gaussian\\
{\tt DC$\_$PA}&float64&deg&Deconvolved position angle of fitted Gaussian\\
{\tt Isl$\_$rms}&float64&mJy/beam&rms noise in island\\
{\tt FLAG$\_$WORKFLOW}&int64&&Flag for workflow status (internal)\\
{\tt ID$\_$flag}&int64&&Flag for workflow status (internal) (5)\\
{\tt Prefilter}&int64&&Prefilter status (internal)\\
{\tt Postfilter}&int64&&Postfilter status (internal)\\
{\tt lr$\_$fin}&float64&&Final likelihood ratio value (internal)\\
{\tt optRA}&float64&deg&Optical right ascension (see Position$\_$from)\\
{\tt optDec}&float64&deg&Optical declination (see Position$\_$from)\\
{\tt Composite$\_$Size}&float64&arcsec&Max size of convex hull surrounding components for composite sources\\
{\tt Composite$\_$Width}&float64&arcsec&Transverse size of convex hull surrounding components for composite sources\\
{\tt Composite$\_$PA}&float64&deg&Position angle on the sky of longest axis of convex hull\\
{\tt Assoc}&int64&&Number of components used to form composite source\\
{\tt ID$\_$Qual}&float64&&Quality of association from RGZ(L)\\
{\tt Assoc$\_$Qual}&float64&&Quality of association from RGZ(L)\\
{\tt Blend$\_$prob}&float64&&Blend probability from RGZ(L) or manual flagging)\\
{\tt Other$\_$prob}&float64&&Other problem probability from RGZ(L)\\
{\tt Created}&bytes192&&Origin of radio component assignment\\
{\tt Position$\_$from}&bytes136&&Origin of {\tt{optRA}}, {\tt{optDec}}\\
{\tt Renamed$\_$from}&bytes184&&Original name e.g. in RGZ if a composite source\\
{\tt ID$\_$RA}&float64&deg&Right ascension of positional match in Legacy/WISE crossmatch catalogue\\
{\tt ID$\_$DEC}&float64&deg&Declination of positional match in Legacy/WISE crossmatch catalogue\\
{\tt UID$\_$L}&bytes128&&Legacy ID if any\\
{\tt UNWISE$\_$OBJID}&bytes128&&UNWISE ID if any\\
{\tt ID$\_$NAME}&bytes128&&Legacy ID if present else WISE ID else blank if no ID exists\\
{\tt Separation}&float64&&Offset between {\tt{optRA}}, {\tt{optDec}} and {\tt{ID$\_$RA}}, {\tt{ID$\_$DEC}} (non-zero only for visual inspection)\\
{\tt Legacy$\_$ID}&int64&&Unique source ID combining release, brick ID and objid\\
{\tt HPX}&int64&&Healpix of Legacy brick (internal)\\
{\tt release}&int64&&Legacy release number\\
{\tt brickid}&int64&&Legacy brick ID\\
{\tt objid}&int64&&Legacy object ID\\
{\tt maskbits}&int64&&bitwise mask indicating that an object touches a pixel in the `coadd/*/*/*maskbits*` maps, as catalogued on the DR8 bitmasks page\\
{\tt fracflux$\_$g}&float64&&Profile-weighted fraction of the flux from other sources divided by the total flux in g (typically [0,1])\\
{\tt fracflux$\_$r}&float64&&Profile-weighted fraction of the flux from other sources divided by the total flux in r (typically [0,1])\\
{\tt fracflux$\_$z}&float64&&Profile-weighted fraction of the flux from other sources divided by the total flux in z (typically [0,1])\\
{\tt type}&bytes32&&Morphological model (1)\\
{\tt ra}&float64&deg&Right ascension of match in Legacy catalogue\\
{\tt dec}&float64&deg&Declination of match in Legacy catalogue\\
{\tt pstar}&float64&&Star likelihood based on GMM modelling (type='PSF' sources only)\\
{\tt star}&bytes40&&Likely star based on {\tt{pstar}} or proper motion (deprecated), blank if no match\\
{\tt ANYMASK$\_$OPT}&bytes40&&Bitwise mask set if the central pixel from any image satisfies each condition in any of g, r or z as catalogued on the DR8 bitmasks page\\
{\tt gmmcomp}&bytes16&&Gaussian Mixture Model component to which source belongs (and hence the gpz++ class used for prediction)\\
{\tt zphot}&float64&&Photo-z estimate\\
{\tt zphot$\_$err}&float64&&Predicted 1-sigma uncertainty on photometric redshift (after magnitude calibration)\\
{\tt var.density}&float64&&gpz++ predicted variance from density of training set\\
{\tt var.tr.noise}&float64&&gpz++ predicted variance from noise in training set\\
{\tt var.in.noise}&float64&&gpz++ predicted variance from noise in fluxes used in prediction\\
{\tt flag$\_$qual}&int64&&Predicted photo-z quality flag, 0 if bad, 1 if good (2)\\
{\tt zspec$\_$sdss}&float32&&SDSS spectroscopic redshift if available\\
{\tt zwarning$\_$sdss}&int32&&0 if SDSS redshift is good, 1 if bad\\
{\tt plate$\_$sdss}&int32&&SDSS plate number\\
{\tt mjd$\_$sdss}&int32&&SDSS MJD\\
{\tt fiberid$\_$sdss}&int32&&SDSS fibre ID\\
{\tt z$\_$hetdex}&float32&&HETDEX spectroscopic redshift if available\\
{\tt z$\_$hetdex$\_$conf}&float32&&HETDEX spectroscopic redshift confidence\\
{\tt hetdex$\_$sourceid}&int64&&HETDEX source ID\\
{\tt z$\_$desi}&float64&&DESI spectroscopic redshift if available\\
{\tt z$\_$desi$\_$err}&float64&&DESI spectroscopic redshift error\\
{\tt desi$\_$sourceid}&int64&&DESI source ID\\
{\tt 2RXS$\_$ID}&bytes168&&ID in 2RXS\\
{\tt XMMSL2$\_$ID}&bytes184&&ID in XMM source catalogue\\
{\tt Resolved}&bool&&Boolean flag to indicate whether source is resolved\\
{\tt LAS}&float64&arcsec&Estimate of angular size, only valid for sources with {\tt{Resolved}} == True\\
{\tt LAS$\_$from}&bytes80&&Source for the {\tt{LAS}} column\\
{\tt z$\_$best}&float64&&Spec-z if available and good, else photo-z if available and good, else blank (3)\\
{\tt z$\_$source}&bytes48&&String describing origin of {\tt{z$\_$best}}\\
{\tt Size}&float64&kpc&LAS times angular size distance (4)\\
{\tt L$\_$144}&float64&W/Hz&Radio luminosity in W/Hz for alpha=0.7 (4)\\
{\tt LM$\_$size}&float64&arcsec&Size from LoMorph code\\
{\tt LM$\_$flux}&float64&mJy&Flux density from LoMorph code\\
{\tt Bad$\_$LM$\_$flux}&bool&&Flag to say that LoMorph flux is bad\\
{\tt Bad$\_$LM$\_$image}&bool&&Flag to say that LoMorph mask is bad\\
{\tt Field}&bytes48&&Which of the two fields the data come from\\
{\tt Legacy$\_$Coverage}&bool&&Flag to say whether source is in the DESI Legacy sky area\\
{\tt mag$\_$g}&float32&mag&Magnitude in g-band\\
{\tt magerr$\_$g}&float32&mag&Magnitude error in g-band, or blank for upper limit\\
{\tt mag$\_$r}&float32&mag&Magnitude in r-band\\
{\tt magerr$\_$r}&float32&mag&Magnitude error in r-band, or blank for upper limit\\
{\tt mag$\_$z}&float32&mag&Magnitude in z-band\\
{\tt magerr$\_$z}&float32&mag&Magnitude error in z-band, or blank for upper limit\\
{\tt mag$\_$w1}&float32&mag&Magnitude in WISE band 1\\
{\tt magerr$\_$w1}&float32&mag&Magnitude error in WISE band 1, or blank for upper limit\\
{\tt mag$\_$w2}&float32&mag&Magnitude in WISE band 2\\
{\tt magerr$\_$w2}&float32&mag&Magnitude error in WISE band 2, or blank for upper limit\\
{\tt mag$\_$w3}&float32&mag&Magnitude in WISE band 3\\
{\tt magerr$\_$w3}&float32&mag&Magnitude error in WISE band 3, or blank for upper limit\\
{\tt mag$\_$w4}&float32&mag&Magnitude in WISE band 4\\
{\tt magerr$\_$w4}&float32&mag&Magnitude error in WISE band 4, or blank for upper limit\\
{\tt WISE$\_$Src}&bytes80&&Origin of the WISE measurements\\
{\tt Mass$\_$median}&float64&dex(solMass)&Mass estimate\\
{\tt Mass$\_$l68}&float64&dex(solMass)&68\% lower confidence bound on mass\\
{\tt Mass$\_$u68}&float64&dex(solMass)&68\% upper confidence bound on mass\\
{\tt g$\_$rest}&float64&mag&Rest-frame g-band magnitude from SED fit\\
{\tt r$\_$rest}&float64&mag&Rest-frame r-band magnitude from SED fit\\
{\tt z$\_$rest}&float64&mag&Rest-frame z-band magnitude from SED fit\\
{\tt U$\_$rest}&float64&mag&Rest-frame U-band magnitude from SED fit\\
{\tt V$\_$rest}&float64&mag&Rest-frame V-band magnitude from SED fit\\
{\tt J$\_$rest}&float64&mag&Rest-frame J-band magnitude from SED fit\\
{\tt K$\_$rest}&float64&mag&Rest-frame K-band magnitude from SED fit\\
{\tt w1$\_$rest}&float64&mag&Rest-frame WISE band-1 magnitude from SED fit\\
{\tt w2$\_$rest}&float64&mag&Rest-frame WISE band-1 magnitude from SED fit\\
{\tt flag$\_$mass}&bool&&True if a mass is measured and reliable\\
{\tt r$\_$50}&float32&arcsec&Half-light {\tt{ra}}dius of Legacy optical exponential/DeVaucouleurs/composite model\\
{\tt r$\_$50$\_$err}&float32&arcsec&1-sigma uncertainty on {\tt{r$\_$50}}\\

\end{longtable}

\begin{longtable}{lllp{11cm}}
\caption{Columns for the component catalogue. Description as for
  Table \ref{tab:sourcecat}.\label{tab:compcat}}\\
\hline\hline
Column name&Type&Units&Description\\
\hline
\endfirsthead
\caption{continued.]}\\
\hline\hline
Column name&Type&Units&Description\\
\hline
\endhead
\hline
\endfoot
{\tt Component$\_$Name}&bytes184&&Object identifier (ILT name)\\
{\tt RA}&float64&deg&Radio right ascension (mean position )\\
{\tt DEC}&float64&deg&Radio declination (mean position)\\
{\tt E$\_$RA}&float64&arcsec&Error on radio right ascension\\
{\tt E$\_$DEC}&float64&arcsec&Error on radio declination\\
{\tt Total$\_$flux}&float64&mJy&144-MHz total flux density\\
{\tt E$\_$Total$\_$flux}&float64&mJy&Error on total flux density\\
{\tt Peak$\_$flux}&float64&mJy/beam&144-MHz peak flux density\\
{\tt E$\_$Peak$\_$flux}&float64&mJy/beam&Error on peak flux density\\
{\tt S$\_$Code}&bytes8&&PyBDSF source code\\
{\tt Mosaic$\_$ID}&bytes88&&LoTSS mosaic of source image\\
{\tt Maj}&float64&arcsec&Major axis of fitted Gaussian\\
{\tt Min}&float64&arcsec&Minor axis of fitted Gaussian\\
{\tt PA}&float64&deg&Position angle of fitted Gaussian\\
{\tt E$\_$Maj}&float64&arcsec&Error on major axis\\
{\tt E$\_$Min}&float64&arcsec&Error on minor axis\\
{\tt E$\_$PA}&float64&deg&Error on position angle\\
{\tt DC$\_$Maj}&float64&arcsec&Deconvolved major axis of fitted Gaussian\\
{\tt DC$\_$Min}&float64&arcsec&Deconvolved minor axis of fitted Gaussian\\
{\tt DC$\_$PA}&float64&deg&Deconvolved position angle of fitted Gaussian\\
{\tt Created}&bytes232&&Origin of radio component\\
{\tt Deblended$\_$from}&bytes176&&If the component was created by deblending, the name of the original catalogued source from which it was deblended\\
{\tt Parent$\_$Source}&bytes184&&The source in the source catalogue of which this component is part\\
{\tt Field}&bytes48&&Which of the two fields the component is taken from\\

\end{longtable}

\section{Acknowledgements}

We thank an anonymous referee for comments that allowed us
  to improve the paper. MJH, DJBS, and JCSP acknowledge support from the UK STFC [ST/V000624/1]. MAH
acknowledges support from STFC grant [ST/X002543/1]. 
KJD acknowledges funding from the European Union's Horizon 2020 research and innovation programme under the Marie Sk\l{}odowska-Curie grant agreement No. 892117 (HIZRAD) and support from the STFC through an Ernest Rutherford Fellowship (grant number ST/W003120/1).
LA is grateful for support from the STFC via CDT studentship grant ST/P006809/1. 
BB is grateful for support from the STFC [ST/S505614/1]. 
JHC acknowledges support from the STFC [ST/T000295/1 and ST/X001164/1]. EO acknowledges
support from the VIDI research programme with project number
639.042.729, which is financed by the Netherlands Organisation for
Scientific Research (NWO). PNB and RK are grateful for support from
STFC via grant ST/V000594/1. AD acknowledges support by the BMBF
Verbundforschung under the grant 05A20STA. CLH acknowledges support
from the Leverhulme Trust through an Early Career Research Fellowship.
MJ acknowledges support from the National Science Centre, Poland under
grant UMO-2018/29/B/ST9/01793. MKB acknowledges support from the
National Science Centre, Poland under grant no. 2017/26/E/ST9/00216.
MH thanks the Ministry of Education and Science of the Republic of
Poland for support and granting funds for the Polish contribution to
the International LOFAR Telescope (arrangement no. 2021/WK/02) and for
maintenance of the LOFAR PL-612 Baldy station (MSHE decision no.
28/530020/SPUB/SP/2022). FdG acknowledges the support of the ERC CoG
grant number 101086378. SPO acknowledges support from the Comunidad de
Madrid Atracción de Talento program via grant 2022-T1/TIC-23797. IP acknowledges support from INAF under the
Large Grant 2022 funding scheme (project ``MeerKAT and LOFAR Team up:
a Unique Radio Window on Galaxy/AGN co-Evolution)''. DJS acknowledges
support by the project ``NRW-Cluster for data intensive radio
astronomy: Big Bang to Big Data (B3D)'' funded through the programme
``Profilbildung 2020'', an initiative of the Ministry of Culture and
Science of the State of North Rhine-Westphalia. HT gratefully
acknowledges the support from the Shuimu Tsinghua Scholar Program of
Tsinghua University, the fellowship of China Postdoctoral Science
Foundation 2022M721875, and long lasting support from JBCA machine
learning group and Doa Tsinghua machine learning group. MV acknowledges financial support from the Inter-University Institute for Data Intensive Astronomy (IDIA), a partnership of the University of Cape Town, the University of Pretoria, the University of the Western Cape and the South African Radio Astronomy Observatory, and from the South African Department of Science and Innovation's National Research Foundation under the ISARP RADIOSKY2020 Joint Research Scheme (DSI-NRF Grant Number 113121) and the CSUR HIPPO Project (DSI-NRF Grant Number 121291).

LOFAR is the Low Frequency Array, designed and constructed by ASTRON. It has observing, data processing, and data storage facilities in several countries, which are owned by various parties (each with their own funding sources), and which are collectively operated by the ILT foundation under a joint scientific policy. The ILT resources have benefited from the following recent major funding sources: CNRS-INSU, Observatoire de Paris and Université d'Orléans, France; BMBF, MIWF-NRW, MPG, Germany; Science Foundation Ireland (SFI), Department of Business, Enterprise and Innovation (DBEI), Ireland; NWO, The Netherlands; The Science and Technology Facilities Council, UK; Ministry of Science and Higher Education, Poland; The Istituto Nazionale di Astrofisica (INAF), Italy.

This research made use of the Dutch national e-infrastructure with support of the SURF Cooperative (e-infra 180169) and the LOFAR e-infra group. The Jülich LOFAR Long Term Archive and the German LOFAR network are both coordinated and operated by the Jülich Supercomputing Centre (JSC), and computing resources on the supercomputer JUWELS at JSC were provided by the Gauss Centre for Supercomputing e.V. (grant CHTB00) through the John von Neumann Institute for Computing (NIC).

This research made use of the University of Hertfordshire
high-performance computing facility and the LOFAR-UK computing
facility located at the University of Hertfordshire (\url{https://uhhpc.herts.ac.uk}) and supported by
STFC [ST/P000096/1], and of the Italian LOFAR IT computing
infrastructure supported and operated by INAF, and by the Physics
Department of Turin University (under an agreement with Consorzio
Interuniversitario per la Fisica Spaziale) at the C3S Supercomputing
Centre, Italy.

This research made use of {\sc Astropy}, a
community-developed core Python package for astronomy
\citep{AstropyCollaboration13} hosted at
\url{http://www.astropy.org/}, of {\sc Matplotlib} \citep{Hunter07},
of {\sc APLpy}, an open-source astronomical plotting package for
Python hosted at \url{http://aplpy.github.com/}, and of {\sc topcat}
and {\sc stilts} \citep{Taylor05}.

The Legacy Surveys consist of three individual and complementary projects: the Dark Energy Camera Legacy Survey (DECaLS; Proposal ID \#2014B-0404; PIs: David Schlegel and Arjun Dey), the Beijing-Arizona Sky Survey (BASS; NOAO Prop. ID \#2015A-0801; PIs: Zhou Xu and Xiaohui Fan), and the Mayall z-band Legacy Survey (MzLS; Prop. ID \#2016A-0453; PI: Arjun Dey). DECaLS, BASS and MzLS together include data obtained, respectively, at the Blanco telescope, Cerro Tololo Inter-American Observatory, NSF’s NOIRLab; the Bok telescope, Steward Observatory, University of Arizona; and the Mayall telescope, Kitt Peak National Observatory, NOIRLab. The Legacy Surveys project is honoured to be permitted to conduct astronomical research on Iolkam Du'ag (Kitt Peak), a mountain with particular significance to the Tohono O'odham Nation.

NOIRLab is operated by the Association of Universities for Research in Astronomy (AURA) under a cooperative agreement with the National Science Foundation.

This project used data obtained with the Dark Energy Camera (DECam), which was constructed by the Dark Energy Survey (DES) collaboration. Funding for the DES Projects has been provided by the U.S. Department of Energy, the U.S. National Science Foundation, the Ministry of Science and Education of Spain, the Science and Technology Facilities Council of the United Kingdom, the Higher Education Funding Council for England, the National Center for Supercomputing Applications at the University of Illinois at Urbana-Champaign, the Kavli Institute of Cosmological Physics at the University of Chicago, Center for Cosmology and Astro-Particle Physics at the Ohio State University, the Mitchell Institute for Fundamental Physics and Astronomy at Texas A\&M University, Financiadora de Estudos e Projetos, Fundacao Carlos Chagas Filho de Amparo, Financiadora de Estudos e Projetos, Fundacao Carlos Chagas Filho de Amparo a Pesquisa do Estado do Rio de Janeiro, Conselho Nacional de Desenvolvimento Cientifico e Tecnologico and the Ministerio da Ciencia, Tecnologia e Inovacao, the Deutsche Forschungsgemeinschaft and the Collaborating Institutions in the Dark Energy Survey. The Collaborating Institutions are Argonne National Laboratory, the University of California at Santa Cruz, the University of Cambridge, Centro de Investigaciones Energeticas, Medioambientales y Tecnologicas-Madrid, the University of Chicago, University College London, the DES-Brazil Consortium, the University of Edinburgh, the Eidgen\"ossische Technische Hochschule (ETH) Zurich, Fermi National Accelerator Laboratory, the University of Illinois at Urbana-Champaign, the Institut de Ciencies de l'Espai (IEEC/CSIC), the Institut de Fisica d'Altes Energies, Lawrence Berkeley National Laboratory, the Ludwig Maximilians Universit\"at Munchen and the associated Excellence Cluster Universe, the University of Michigan, NSF's NOIRLab, the University of Nottingham, the Ohio State University, the University of Pennsylvania, the University of Portsmouth, SLAC National Accelerator Laboratory, Stanford University, the University of Sussex, and Texas A\&M University.

BASS is a key project of the Telescope Access Program (TAP), which has been funded by the National Astronomical Observatories of China, the Chinese Academy of Sciences (the Strategic Priority Research Program “The Emergence of Cosmological Structures” Grant \# XDB09000000), and the Special Fund for Astronomy from the Ministry of Finance. The BASS is also supported by the External Cooperation Program of Chinese Academy of Sciences (Grant \# 114A11KYSB20160057), and Chinese National Natural Science Foundation (Grant \# 11433005).

This project, and the the Legacy Survey project, makes use of data
products from the Near-Earth Object Wide-field Infrared Survey
Explorer ({\it NEOWISE}), which is a project of the Jet Propulsion
Laboratory/California Institute of Technology. {\it NEOWISE} is funded by the National Aeronautics and Space Administration.

The Legacy Surveys imaging of the DESI footprint is supported by the Director, Office of Science, Office of High Energy Physics of the U.S. Department of Energy under Contract No. DE-AC02-05CH1123, by the National Energy Research Scientific Computing Center, a DOE Office of Science User Facility under the same contract; and by the U.S. National Science Foundation, Division of Astronomical Sciences under Contract No. AST-0950945 to NOAO.

    HETDEX is led by the University of Texas at Austin McDonald Observatory and Department of Astronomy with participation from the Ludwig-Maximilians-Universität München, Max-Planck-Institut für Extraterrestrische Physik (MPE), Leibniz-Institut für Astrophysik Potsdam (AIP), Texas A\&M University, Pennsylvania State University, Institut für Astrophysik Göttingen, The University of Oxford, Max-Planck-Institut für Astrophysik (MPA), The University of Tokyo and Missouri University of Science and Technology.

    Observations for HETDEX were obtained with the Hobby-Eberly
    Telescope (HET), which is a joint project of the University of
    Texas at Austin, the Pennsylvania State University,
    Ludwig-Maximilians-Universität München, and
    Georg-August-Universität Göttingen. The HET is named in honor of
    its principal benefactors, William P. Hobby and Robert E. Eberly.
    The Visible Integral-field Replicable Unit Spectrograph (VIRUS)
    was used for HETDEX observations. VIRUS is a joint project of the
    University of Texas at Austin, Leibniz-Institut für Astrophysik
    Potsdam (AIP), Texas A\&M University, Max-Planck-Institut für Extraterrestrische Physik (MPE), Ludwig-Maximilians-Universität München, Pennsylvania State University, Institut für Astrophysik Göttingen, University of Oxford, and the Max-Planck-Institut f\"ur Astrophysik (MPA).

    Funding for HETDEX has been provided by the partner institutions, the National Science Foundation, the State of Texas, the US Air Force, and by generous support from private individuals and foundations.

This research used data obtained with the Dark Energy Spectroscopic Instrument (DESI). DESI construction and operations is managed by the Lawrence Berkeley National Laboratory. This material is based upon work supported by the U.S. Department of Energy, Office of Science, Office of High-Energy Physics, under Contract No. DE–AC02–05CH11231, and by the National Energy Research Scientific Computing Center, a DOE Office of Science User Facility under the same contract. Additional support for DESI was provided by the U.S. National Science Foundation (NSF), Division of Astronomical Sciences under Contract No. AST-0950945 to the NSF’s National Optical-Infrared Astronomy Research Laboratory; the Science and Technologies Facilities Council of the United Kingdom; the Gordon and Betty Moore Foundation; the Heising-Simons Foundation; the French Alternative Energies and Atomic Energy Commission (CEA); the National Council of Science and Technology of Mexico (CONACYT); the Ministry of Science and Innovation of Spain (MICINN), and by the DESI Member Institutions: \url{https://www.desi.lbl.gov/collaborating-institutions}. The DESI collaboration is honored to be permitted to conduct scientific research on Iolkam Du'ag (Kitt Peak), a mountain with particular significance to the Tohono O'odham Nation. Any opinions, findings, and conclusions or recommendations expressed in this material are those of the author(s) and do not necessarily reflect the views of the U.S. National Science Foundation, the U.S. Department of Energy, or any of the listed funding agencies.

\end{document}